\documentclass{emulateapj} 
\usepackage{lscape}
\usepackage{color}

\shorttitle{The IR Nuclear Emission of Seyferts on Parsec Scales}
\shortauthors{Ramos Almeida et al.}

\begin{document}

\title{The Infrared Nuclear Emission of Seyfert Galaxies on Parsec Scales: Testing the Clumpy Torus models}

\author{C. Ramos Almeida\altaffilmark{1}, N. A. Levenson\altaffilmark{2}, J. M. Rodr\'\i guez Espinosa\altaffilmark{1}, 
A. Alonso-Herrero\altaffilmark{3}, A. Asensio Ramos\altaffilmark{1}, J. T. Radomski\altaffilmark{4}, C. Packham\altaffilmark{5}, R. S. Fisher\altaffilmark{6}, 
and C. M. Telesco\altaffilmark{5}}

\altaffiltext{1}{Instituto de Astrof\'\i sica de Canarias (IAC), C/V\'\i a L\'{a}ctea, s/n, E-38205, La Laguna, Tenerife, Spain; cra@iac.es}
\altaffiltext{2}{Department of Physics and Astronomy, University of Kentucky, Lexington, KY 40506.}
\altaffiltext{3}{Instituto de Estructura de la Materia, CSIC, E-28006 Madrid, Spain.}
\altaffiltext{4}{Gemini South Observatory, Casilla 603, La Serena, Chile.}
\altaffiltext{5}{Astronomy Department, University of Florida, 211 Bryant Space Science Center, P.O. Box 112055, Gainesville, FL 32611-2055.}
\altaffiltext{6}{Gemini Observatory, Northern Operations Center, 670 North A\`{o}hoku Place, Hilo, HI 96720.}

\begin{abstract}

We present subarcsecond resolution mid-infrared (mid-IR) photometry
in the wavelength range from 8 to 20 \micron~of eighteen Seyfert
galaxies, reporting high spatial resolution nuclear fluxes for the entire
sample. 
We construct spectral energy distributions (SEDs) that the AGN
dominates, relatively uncontaminated by starlight, adding near-IR measurements
from the literature at similar angular resolution.
We find that the IR SEDs of intermediate-type Seyferts are flatter and present higher 10
to 18 \micron~ratios than those of Seyfert 2 galaxies.  
We fit the individual SEDs with 
clumpy dusty torus models using the in-house-developed BayesClumpy tool.
We find that the clumpy models reproduce the high spatial resolution measurements.
Regardless of the Seyfert type, even with high spatial  
resolution data, near- to mid-IR SED fitting poorly constrains the radial extent of the torus. 
For the Seyfert 2 galaxies, 
we find that edge-on geometries are more probable than face-on views, with a number of clouds along equatorial rays
of $N_0 = 5$--15. The 10 \micron~silicate feature is generally modeled in shallow absorption. 
For the intermediate-type Seyferts, $N_0$ and the inclination angle of
the torus are lower than those of the Seyfert 2 nuclei, 
with the silicate feature appearing in weak emission or absent. 
The columns of material responsible for the X-ray absorption 
are larger than those inferred from the model fits for most of the galaxies,
which is consistent with X-ray absorbing gas being located within the dust sublimation radius
whereas the mid-IR flux arises from an area farther from the accretion disc.
The fits yield both the bolometric luminosity of the intrinsic AGN and the torus integrated luminosity, 
from which we derive the reprocessing efficiency of the torus.
In the models, the outer radial extent of the torus scales with the
AGN luminosity, and we find the tori to be confined to scales less than 5 pc.

\end{abstract}


\section{Introduction}
\label{intro}

The unified model for active galaxies \citep{Antonucci93,Urry95} explains the observed differences between 
Type-1 and Type-2 objects as due to orientation. Although the unification model may not be universally 
applicable, the most accepted scheme is that there is dust surrounding 
the central region of AGN distributed in a toroidal geometry \citep{Pier93,Granato94,Efstathiou95,Granato97}. 
Thus, the central engines of Type-1 active galactic nuclei (AGN) 
can be seen directly, resulting in typical spectra with both narrow and broad emission lines coming from the 
Narrow-Line Region (NLR) and the Broad-Line Region (BLR), respectively. In Type-2 AGN, the active nucleus and 
the BLR are obscured by an optically and geometrically thick dusty torus, 
which  explains the lack of broad lines in their observed spectra. 
The torus dust grains absorb ultraviolet photons 
from the central engine and re-radiate them in the IR, peaking at mid-IR wavelengths (7 - 26 \micron).

Pioneering work in modelling dusty tori was performed by \citet{Pier92}. They assumed a uniform dust density distribution
of a few parsecs radial extent. However, in order to better reproduce the observations, 
\citet{Pier93} enlarged the torus, e.g., to account for the far-IR emission. Thus, the coolest material in the torus must be located 
far from the central engine, implying  $\sim$100 pc scale tori. Further investigations with smooth dust distributions 
were carried out by \citet{Granato94}, \citet{Efstathiou95}, \citet{Granato97}, and \citet{Siebenmorgen04}.

However, recent ground-based mid-IR 
observations of nearby Seyferts reveal that the torus size is likely restricted to a few parsecs.
For example, \cite{Packham05} and \cite{Radomski08} established upper limits of 2 and 1.6 pc for the outer radii of the Circinus galaxy and 
Centaurus A tori, respectively using data from the Gemini Telescopes. 
Additionally, interferometric observations obtained with the MID-infrared Interferometric Instrument (MIDI) at the Very Large Telescope Interferometer (VLTI) 
of Circinus, NGC 1068, and Centaurus A suggest a 
scenario where the torus emission would only extend out to R = 1 pc \citep{Tristram07}, 
R = 1.7 - 2 pc \citep{Jaffe04,Raban09}, and R = 0.3 pc \citep{Meisenheimer07}, respectively.

Thus, during the last decade, an intensive search for an alternative geometry to the early homogeneous torus models has been carried 
out in order to explain the observations. Before \citet{Pier92,Pier93} publications, \citet{Krolik88} claimed
that smooth dust distributions cannot survive within the AGN vicinity. They proposed that the material in the torus must be 
distributed in a clumpy structure, in order to prevent the dust grains from being destroyed by the hot surrounding gas. 
The first results of radiative transfer calculations of a clumpy medium were reported by \citet{Nenkova02}, and 
further work was done by \citet{Dullemond05}, \citet{Fritz06}, \citet{Elitzur06}, and \citet{Ballantyne06}.

Thus, the clumpy dusty torus models
\citep{Nenkova02,Nenkova08a,Nenkova08b,Honig06,Schartmann08}
propose that the dust is distributed in clumps, instead of
homogeneously filling the torus volume. These models are making
significant progress in accounting for the mid-IR emission of AGNs. 
A fundamental difference between clumpy and smooth density distributions
of dust is that the former implies that both directly-illuminated and shadowed cloud faces may
exist at different distances from the central engine.  Thus, the dust
temperature is not a function of the radius only, as was the case in
the homogeneous models, since illuminated and shadowed clouds
contribute 
to the IR emission from all viewing angles.  Homogeneous
torus models had serious difficulties in predicting the observed
differences in the variety of SEDs \citep{Spinoglio89,Fadda98,Kuraszkiewicz03,Alonso03,Rigby04} and the
strengths of the 10 \micron~silicate feature in both Seyfert 1 and
Seyfert 2 nuclei \citep{Roche91,Granato97,Mason06}, and also the large optical depths along
the line of sight indicated by X-ray observations \citep{Simpson94,Tozzi06}.
Inhomogeneous torus models solve these problems. The clumpy models
have been employed in several observational studies, e.g., the first
Spitzer analysis of AGN in the GOODS fields \citep{Treister04}, and in
the analysis of the 10 \micron~spatially resolved spectra of NGC 1068
and NGC 2110 nuclei \citep{Mason06,Mason09}.

Since the reprocessed radiation from the torus is re-emitted in the
IR, this range (especially the mid-IR) 
is key to put constraints
on the clumpy dusty torus models.  In comparing the predictions of any
torus model with observations,
the small-scale torus emission must be isolated. 
 Large aperture data (i.e.,
images with angular resolutions larger than 1-1.5\arcsec, such as those
provided by ISO, IRAS, or Spitzer) are strongly contaminated with
emission from the host galaxy. For example, starburst emission is an important
source of IR flux in the majority of AGN \citep{Netzer07,Barmby06,Polletta07,Ramos09}.
\citet{Mason06} found for NGC 1068 that all flux measurements within
apertures smaller than 0.5\arcsec~were well fitted by the models,
whereas larger aperture fluxes were much higher and presented 
a different spectral shape in the mid-IR. This excess of emission could
be due to nearby dust outside the torus, that is also emitting in the
IR (see e.g., \citealt{Roche06}). \citet{Mason06} concluded for NGC
1068 that the torus contributes 
less than 30\% of the 10
\micron~flux within apertures larger than 1\arcsec. The bulk of this
large-aperture flux comes from dust in the ionization cone, which 
occupies a larger volume than the torus dust.  

At shorter IR
wavelengths ($\sim$ 1-2.2 \micron) the flux contained in apertures of a few arcseconds partly
comes from the torus 
and partly from stellar
emission. \citet{Alonso96} found that for Seyfert 2 galaxies (Sy2), the fluxes
obtained within a 3\arcsec~aperture are dominated by stellar
emission in the near-IR, out to $\sim$ 2~\micron.  Indeed,
\citet{Kotilainen92} found for a complete sample of Seyfert galaxies,
most of them Seyfert 1 (Sy1), that starlight contributes significantly
to the nuclear emission in the near-IR (J, H, and K bands), even within a small projected aperture (of 3\arcsec).
Consequently, high angular resolution observations at these
wavelengths are crucial to separate the pure nuclear emission from 
that of the host galaxy, which 
contaminates or hides the AGN emission.

The optimal way to estimate near-IR nuclear fluxes uncontaminated by star formation, in as much as possible, is the
use of either subarcsecond spatial resolution data (e.g., from the HST or adaptive optics) or high spatial resolution 
ground-based data, extracting the nuclear emission in a proper manner (e.g., using PSF subtraction).
In the mid-IR, it is necessary to isolate the torus emission from that produced by dust in the ionization cones 
or dust heated by intense star formation. 
With various high-resolution SEDs covering the near- and mid-IR ranges, 
it is possible to identify SED features that are attributable to the torus.

It is worth mentioning that mid-IR interferometry constitutes the most precise 
way of characterizing the torus structure using observations and  comparing them with the torus models (e.g., \citealt{Honig06}). However, 
the signal-to-noise required for VLTI observations strongly restricts the number of observable sources to the nearest AGN only.

In this work, we report new high-resolution mid-IR imaging data for 18 nearby Seyfert galaxies, for which
we have estimated nuclear mid-IR fluxes\footnote{The mid-IR images of Circinus, NGC 4151, and Centaurus A
are already published by \citet{Packham05} and \citet{Radomski03,Radomski08}, respectively. However, we present  new 
calculations of the nuclear mid-IR fluxes in this work.}.
We also compiled near-IR high spatial resolution
fluxes from the literature to construct pure-nuclear SEDs. We have fitted these AGN SEDs with the 
clumpy dusty torus models of \citet{Nenkova08a,Nenkova08b} to constrain the parameters 
that describe the clumpy models when applied to Seyfert galaxies. 
Table \ref{sources} summarizes key observational properties of the sources in the sample. 
Section 2 describes the observations, data reduction, and compilation
of near-IR fluxes. Section 3 presents the main observational results, and  \S 4  presents the modelling results.
We draw conclusions about the clumpy torus models and AGN obscuration
in general  in \S 5.  Finally, Section 6 summarizes the main conclusions of this work.

\begin{deluxetable*}{lccccc}
\tablewidth{0pt}
\tabletypesize{\footnotesize}
\tablecaption{Basic Galaxy Data}
\tablehead{
\colhead{Galaxy} & \colhead{Seyfert Type} & \colhead{Morphology} & \colhead{$z$} & \colhead{Distance} & \colhead{Scale} \\
 & & & & \colhead{(Mpc)} & \colhead{(pc~arcsec$^{-1}$)}} 
\startdata
Centaurus A &  Sy2			& S0 pec	    & 0.0018 &   3.5 \tablenotemark{a}   & 17	\\
Circinus    &  Sy2			& SA(s)b:	    & 0.0014 &   4 \tablenotemark{a}     & 20	\\
IC 5063	    &  Sy2			& SA(s)0+:	    & 0.0113 &   45    & 219	\\
Mrk 573	    &  Sy2 \tablenotemark{b}	& (R)SAB(rs)0+:     & 0.0172 &   69    & 334	\\
NGC 1386    &  Sy2	    		& SB(s)0+	    & 0.0029 &   11 \tablenotemark{a}    & 56	\\  
NGC 1808    &  Sy2			& (R'$_{1}$)SAB(s:)b& 0.0033 &   11 \tablenotemark{a}    & 64	\\
NGC 3081    &  Sy2	    		& (R$_{1}$)SAB(r)0/a& 0.0080 &   32    & 155    \\
NGC 3281    &  Sy2	    		& SAB(rs+)a 	    & 0.0107 &   43    & 208	\\  
NGC 4388    &  Sy2	    		& SA(s)b: sp	    & 0.0084 &   34    & 163	\\
NGC 5728    &  HII/Sy2    		& (R$_{1}$)SAB(r)a  & 0.0094 &   38    & 182    \\  
NGC 7172    &  HII/Sy2    		& Sa pec sp	    & 0.0087 &   35    & 169	\\  
NGC 7582    &  Sy2	    		& (R'$_{1}$)SB(s)ab & 0.0053 &   21    & 103    \\  
\hline
NGC 1365    &  HII/Sy1.8  		& (R')SBb(s)b       & 0.0055 &   18 \tablenotemark{a}    & 107	\\  
NGC 2992    &  Sy1.9	    		& Sa pec	    & 0.0077 &   31    & 149	\\  
NGC 5506    &  Sy1.9 \tablenotemark{b}	& Sa pec sp 	    & 0.0062 &   25    & 120	\\  
\hline
NGC 3227    &  Sy1.5	    		& SAB(s) pec	    & 0.0039 &   17 \tablenotemark{a}    & 82	\\  
NGC 4151    &  Sy1.5	    		& (R')SAB(rs)ab:    & 0.0033 &   13 \tablenotemark{a}    & 64	\\
\hline
NGC 1566    &  Sy1	    		& (R'$_{1}$)SAB(rs) & 0.0050 &   20    & 97     \\  
\enddata
\tablecomments{\footnotesize{Classification, morphological class, and spectroscopic redshift of our targets taken from the NASA/IPAC Extragalactic 
Database (NED). The distance to the sources and the physical scale have been obtained using H$_{0}$=75 km~s$^{-1}$~Mpc$^{-1}$.}}
\tablenotetext{a}{\footnotesize{For the cases of the most nearby objects Centaurus A, Circinus, NGC 1386, NGC 1808, NGC 1365, 
NGC 3227, and NGC 4151 the values of the distance to the galaxies have been taken 
from \citet{Radomski08}, \citet{Packham05}, \citet{Bennert06}, \citet{Jimenez05}, \citet{Silbermann99}, \citet{Davies06}, and \citet{Radomski03}, respectively.}}
\tablenotetext{b}{\footnotesize{Mrk 573 and NGC 5506 have recently been classified as obscured Narrow-line Sy1 galaxies 
\citep{Nagar02,Dewangan05,Ramos08}.}}
\label{sources}
\end{deluxetable*}

\section{Observations and Data Reduction}
\label{observations}

\subsection{Mid-IR Imaging Observations}

Ground-based mid-IR high-angular resolution observations of 18 nearby active galaxies were carried out
over the past years for a variety of science drivers. We make use of this archive of  data in this paper. 
Most of these sources are Type-2 Seyferts, but the sample also includes two Seyfert 1.9 (Sy1.9), one Seyfert 1.8 (Sy1.8), 
two Seyfert 1.5 (Sy1.5), and one Sy1 galaxy (see Table \ref{sources}).

The first set of observations was obtained with the University of
Florida mid-IR camera/spectrometer OSCIR, in 1998 December at the 4 m
Blanco Telescope at Cerro Tololo Inter-American Observatory (CTIO) and
in 2001 May at the Gemini North Telescope. This instrument employs a
Rockwell 128x128 pixel Si:As blocked impunity band (BIB) detector
array, optimized for the wavelength range between 8 and 25 \micron. On
CTIO, the pixel scale is 0.183\arcsec, with a field of view (FOV) of
23.4\arcsec~x 23.4\arcsec, and the resolutions obtained were $\sim$1\arcsec~at both 10.7 and 18.2 \micron.  
On Gemini North, OSCIR has a plate scale
of 0.089\arcsec~pixel$^{-1}$ and a total FOV of 11\arcsec~x 11\arcsec. The 
achieved resolutions were $\sim$0.5\arcsec~at 10.7 \micron~and $\sim$0.6\arcsec~at 18.2 \micron.
Another set of observations was performed with the mid-IR
camera/spectrograph T-ReCS (Thermal-Region Camera Spectrograph;
\citealt{Telesco98}) on the Gemini-South telescope.  T-ReCS uses a
Raytheon 320x240 pixel Si:As IBC array, providing a plate scale of
0.089\arcsec~pixel$^{-1}$, corresponding to a FOV of 28.5\arcsec~x
21.4\arcsec. The resolutions obtained were 0.3-0.4\arcsec~at 8.8 \micron, $\sim$0.5\arcsec~at 
10.4 \micron, and 0.5-0.6\arcsec~at 18.3 \micron. 
The last set of observations was made with the mid-IR
camera/spectrograph Michelle \citep{Glasse97} on the Gemini North
Telescope. The Michelle detector is a Si:As IBC array with a format of
320x240 pixels. Configured in its imaging mode, Michelle has a
0.10\arcsec~pixel$^{-1}$ plate scale, 
and the resolutions of our observations were $\sim$0.4\arcsec~at 11.3 \micron~and 
$\sim$0.5\arcsec~at 18.1 \micron. This plate scale translates to
a FOV of 32\arcsec~x 24\arcsec.  A summary of the observations is
reported in Table \ref{log}.

\begin{deluxetable*}{llcccccccc}
\tabletypesize{\scriptsize}
\tablewidth{0pt}
\tablecaption{Summary of Mid-IR Observations}
\tablehead{
\colhead{Galaxy} & \colhead{Filters} & \colhead{Instrument} & \colhead{Telescope} & \colhead{Observation} & \multicolumn{2}{c}{On-Source Time (s)}  & \multicolumn{2}{c}{PSF FWHM}   \\
& & & & \colhead{epoch} & \colhead{N band} & \colhead{Q band} & \colhead{N band} & \colhead{Q band}}
\startdata
Centaurus A  	    & Si2, Qa       & T-ReCS   & Gemini S &  Jan 2004   &  2000 & 1550  & 0.30\arcsec & 0.53\arcsec      \\
Circinus            & Si2, Qa       & T-ReCS   & Gemini S &  Feb 2004   &  109  & 109   & 0.33\arcsec & 0.55\arcsec      \\
IC 5063	            & Si2, Qa       & T-ReCS   & Gemini S &  Jul 2005   &  130  & 304   & 0.40\arcsec & 0.62\arcsec      \\
Mrk 573	            &  N,  Qa       & T-ReCS   & Gemini S &  Dec 2003   &  217  & 217   & 0.36\arcsec & 0.54\arcsec      \\
NGC 1386	    &  N,  Qa       & T-ReCS   & Gemini S &  Dec 2003   &  217  & 217   & 0.31\arcsec & 0.54\arcsec      \\  
NGC 1808	    &  N, IHW18     & OSCIR    &  CTIO 4m &  Dec 1998   &  300  & 300   & 0.94\arcsec & 1.02\arcsec      \\
NGC 3081	    &  Si2, Qa      & T-ReCS   & Gemini S &  Jan 2006   &  130  & 304   & 0.30\arcsec & 0.56\arcsec      \\
NGC 3281	    &  N,   Qa      & T-ReCS   & Gemini S &  Jan 2004   &  260  & 455   & 0.34\arcsec & 0.58\arcsec      \\
NGC 4388	    &  N',  Qa      & Michelle & Gemini N &  May 2006   &  549  & 733   & 0.34\arcsec & 0.50\arcsec      \\
NGC 5728            &  Si2, Qa      & T-ReCS   & Gemini S &  Jul 2005   &  130  & 304   & 0.35\arcsec & 0.56\arcsec      \\  
NGC 7172	    &  N 	    & T-ReCS   & Gemini S &  May 2004   &  305  & \dots & 0.51\arcsec & \dots     	 \\  
NGC 7582            &  N, IHW18     & OSCIR    & CTIO 4m  &  Dec 1998   &  250  & 250   & 0.76\arcsec & 0.99\arcsec      \\
\hline
NGC 1365	    &  N, IHW18     & OSCIR    & CTIO 4m  &  Dec 1998   &  482  & 482   & 0.92\arcsec & 1.03\arcsec      \\
NGC 2992	    &  N',  Qa      & Michelle & Gemini N &  May 2006   &  730  & 1095  & 0.32\arcsec & 0.53\arcsec      \\  
NGC 5506	    &  N',  Qa      & Michelle & Gemini N &  Apr 2006   &  546  & 729   & 0.36\arcsec & 0.51\arcsec      \\  
\hline
NGC 3227	    &  N'	    & Michelle & Gemini N &  Apr 2006   &  300  & \dots & 0.39\arcsec &  \dots           \\
NGC 4151	    &  N, IHW18     & OSCIR    & Gemini N &  May 2001   &  360  & 480   & 0.53\arcsec & 0.58\arcsec      \\
\hline
NGC 1566            & Si2, Qa       & T-ReCS   & Gemini S &  Sep 2005   &  152  & 304   & 0.30\arcsec & 0.53\arcsec      \\  
\enddata       
\tablecomments{\footnotesize{Images were obtained in the 8.74 \micron~(Si2, $\Delta\lambda$ = 0.78 \micron~at 50\% cut-on/off), 
10.36 \micron~(N, $\Delta\lambda$ = 5.27 \micron), and 18.33 \micron~(Qa, $\Delta\lambda$ = 1.5 \micron) filters with 
T-ReCS; in the 11.29 \micron~(N', $\Delta\lambda$ = 2.4 \micron) and 18.11 \micron~(Qa, $\Delta\lambda$ = 1.9 \micron) 
filters with Michelle; and in the 10.75 \micron~(N, $\Delta\lambda$ = 5.2 \micron) and 
18.17 \micron~(IHW18, $\Delta\lambda$ = 1.7 \micron) filters with OSCIR.}}
\label{log}
\end{deluxetable*}

The standard chopping-nodding technique was used to remove the time-variable sky background,
the telescope thermal emission, and the so-called 1/f detector noise.
The chopping throw was 15\arcsec, and the telescope was nodded every
30 s. 
All data were reduced using in-house-developed IDL routines.
The difference for each chopped pair for each given nodding set was calculated,
and the nod sets were then differenced and combined until a single image was created. 
Chopped pairs obviously compromised by cirrus, high electronic noise, or other problems were excluded 
from any further reduction process.

\begin{figure*}[!ht]
\centering
\includegraphics[width=12cm]{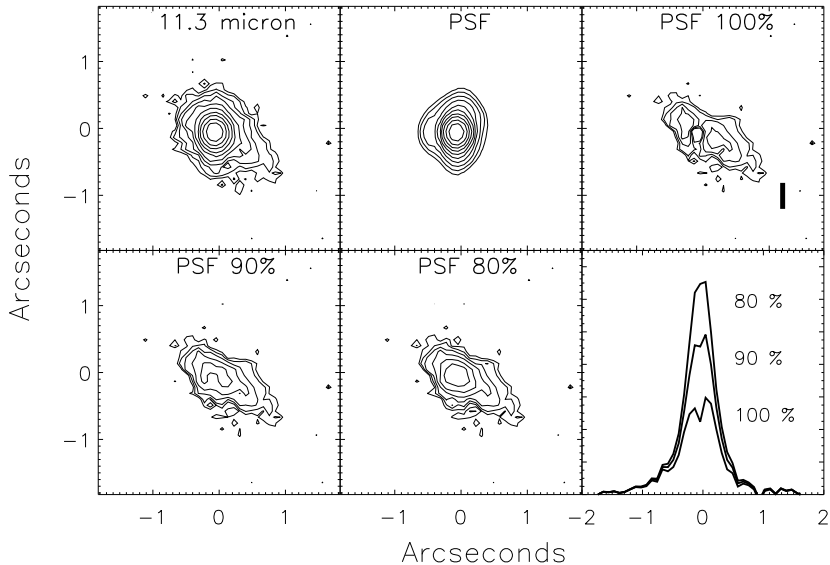}
\caption{\footnotesize{11.29 \micron~contour plots of NGC 4388, the PSF star, and the 
scaled subtraction of the PSF for this galaxy at various levels (100\%, 90\%, and 80\%).
The residuals of the subtraction in the lower right panel show the host galaxy profile, 
with  the residual at 90\% scale representing the best fit.}
\label{psf1}}
\end{figure*}

Observations of flux standard stars were made for the flux calibration of each galaxy through the 
same filters. The uncertainties in the flux calibration were found to be $\sim$5-10\% at N band 
and $\sim$15-20\% at Qa band. PSF star observations were also made immediately prior to or
after each galaxy observation to accurately sample the image quality. 
These  images  were employed to determine the unresolved (i.e., nuclear) component of each galaxy.
The PSF star, scaled to the peak of the galaxy emission, represents the maximum
contribution of the unresolved source, where we integrate flux within an aperture of
2\arcsec~for the Gemini data, and of 4\arcsec~for the CTIO data (NGC 1365, NGC 1808, and NGC 7582). 
The residual of the total emission minus the scaled PSF represents the host galaxy contribution.
A detailed study of the extended near nuclear structures of the galaxies in our sample 
will be  the subject of a forthcoming paper (J. T. Radomski et al. 2009, in preparation). 
We require a flat profile in the residual (for a realistic galaxy profile) 
and therefore reduce the scale of the PSF from matching the peak of the galaxy emission (100\%) to 
obtain the unresolved fluxes reported in Table \ref{psf}.
Figure \ref{psf1} shows an example of PSF subtraction at various levels
(in contours) for the Sy2 galaxy NGC 4388 in the N' Michelle filter, performed following 
\citet{Radomski02} and \citet{Soifer03}. 
The residual profiles from the different scales demonstrate the best-fitting result,
scaled to 90\% of the peak flux.
The uncertainty in the unresolved fluxes determination from PSF subtraction is $\sim$10-15\%. Thus, 
we estimated the errors in the flux densities reported in Table \ref{psf} by adding quadratically the 
flux calibration and PSF subtraction uncertainties, resulting in errors of $\sim$15\% at N band and 
of $\sim$25\% at Qa band. 

For the galaxies that clearly appear spatially unresolved in our mid-IR images (Mrk 573, NGC 1566, and NGC 5728), 
aperture photometry was employed to derive their nuclear density fluxes (reported in Table \ref{psf}). The rule for their calculation 
was to ensure that all the unresolved emission was collected in the chosen aperture, with the minimum contribution 
of foreground contribution as possible. Aperture correction was applied to these fluxes, employing the images of 
the corresponding standard stars in both N and Qa filters. 

\begin{deluxetable}{lcccc}
\tablewidth{0pt}
\tabletypesize{\footnotesize}
\tablecaption{Unresolved Mid-IR Fluxes}
\tablehead{
\colhead{Galaxy} & \multicolumn{2}{c}{Level of PSF subtraction} & \multicolumn{2}{c}{Flux Density (mJy)}    \\
& \colhead{N band} & \colhead{Qa band} & \colhead{N band} & \colhead{Qa band}}
\startdata
Centaurus A  	    		&  100\% &  100\%	&   710   &  2630   	\\
Circinus            		&  90\%  &  90\%	&   5620  &  12790  	\\
IC 5063	            		&  100\% &  100\%	&   399   &  2790   	\\
Mrk 573\tablenotemark{a}	&  100\% &  100\%	&   177   &  415    	\\
NGC 1386	    		&  70\%  &  70\%	&   147   &  457    	\\  
NGC 1808	    		&  40\%  &  60\%	&   227   &  560    	\\
NGC 3081	    		&  100\% &  50\%	&   83    &  231    	\\
NGC 3281	    		&  80\%  &  100\%	&   355   &  1110   	\\  
NGC 4388	    		&  90\%  &  100\%	&   195   &  803    	\\
NGC 5728\tablenotemark{a}       &  100\% &  100\%	&   25    &  184    	\\  
NGC 7172	    		&  90\%  &\nodata	&   68    &  \nodata    \\  
NGC 7582	    		&  70\%  &  50\%	&   195   &  527    	\\  
\hline
NGC 1365	    		&  90\%  &  100\%	&   321   &  640    	\\  
NGC 2992	    		&  90\%  &  90\%	&   175   &  521    	\\  
NGC 5506	    		&  100\% &  100\%	&   873   &  2200   	\\  
\hline
NGC 3227	    		&  100\% &\nodata	&   401   &  \nodata    \\  
NGC 4151	    		&  90\%  &  100\%	&   1320  &  3200   	\\
\hline
NGC 1566\tablenotemark{a}       &  100\% &  100\%	&   29    &  117    	\\  
\enddata        
\tablecomments{\footnotesize{The percentages of PSF subtraction level are reported 
in the employed filters (listed in Table \ref{log}). 
Errors in flux densities  are dominated in 
general by uncertainties in the flux calibration and PSF subtraction ($\sim$15\% at N band and $\sim$25\% at Qa band).}}
\tablenotetext{a}{\footnotesize{For the galaxies Mrk 573, NGC 1566, and NGC 5728, 
aperture photometry was performed with the following aperture radii: 0.89\arcsec, 0.45\arcsec, and 0.62\arcsec, 
respectively. These fluxes include the corresponding aperture corrections.}}
\label{psf}
\end{deluxetable}

\subsection{Compilation of Near-IR High Spatial Resolution Data}

	To construct well-sampled IR SEDs, we compiled
near-IR high spatial resolution nuclear fluxes from the literature (reported in Table \ref{literature}).  
These fluxes correspond to the observed emission from the nuclear region of the galaxies (unresolved component), with 
typical spatial scales of the same order as those of our nuclear mid-IR fluxes.
NICMOS/HST fluxes are available for 10 galaxies, having 
FWHM for an unresolved PSF $\sim$0.13\arcsec~in 
the F160W filter and $\sim$0.09\arcsec~in F110W. 
For the cases of 
Centaurus A and Circinus, there are also diffraction-limited and near-diffraction-limited adaptive optics NACO/VLT fluxes,
respectively. (The FWHM of Centaurus A unresolved component is 0.1\arcsec~in the near-IR bands and  
for Circinus 0.2\arcsec~in the near-IR bands and $< 0.13\arcsec$~in the L' and M' bands.) 
We also compiled seeing-limited near-IR nuclear fluxes 
from the literature, obtained with data from IRCAM3 on the 3.8 m United Kingdom IR Telescope (UKIRT),
with NSFCam at the 3 m NASA IRTF telescope, with the IRAC-1 IR array camera on the ESO 2.2 m telescope, 
and with ISAAC at the 8 m VLT. 
The angular resolution of these natural seeing-limited images is 0.6--0.7\arcsec~in K band and 0.6--0.9\arcsec~in L band.
These ground-based near-IR nuclear fluxes were estimated either using aperture photometry (scaling an annulus 
of the J-band images to the observed counts in H and K,~~\citealt{Simpson98,Galliano08}),
surface brightness profile deconvolution \citep{Quillen01}, or PSF scaling \citep{Alonso01,Alonso03}.
Due to the lack of high spatial resolution nuclear fluxes of Seyfert galaxies in the literature, 
the SEDs in our sample range from very well-sampled
SEDs (e.g., Circinus and Mrk 573) to galaxies for which there are no high spatial resolution near-IR nuclear fluxes 
available (NGC 1566 and NGC 5728).  

\begin{deluxetable*}{lccccclc}
\tabletypesize{\footnotesize}
\tablewidth{0pt}
\tablecaption{High Spatial Resolution Near-IR Fluxes}
\tablehead{
\colhead{Galaxy} & \multicolumn{5}{c}{Flux Density (mJy)} & \colhead{Filters} & \colhead{Reference(s)}  \\
& \colhead{J band} & \colhead{H band} & \colhead{K band} & \colhead{L band} & \colhead{M band} & &}
\startdata
Centaurus A  			& 1.3$\pm$0.1     & 4.5$\pm$0.3  & 34$\pm$2    & 200$\pm$40   & \nodata 	 &   NACO J,H,K,L           	& a \\
Circinus\tablenotemark{a}    	& 1.6$\pm$0.2     & 4.8$\pm$0.7  & 19$\pm$2    & 380$\pm$38   & 1900$\pm$190 	 &   F160W, NACO J,K,L,M    	& b \\
IC 5063	    			& \nodata         & 0.3$\pm$0.1  & \nodata     & \nodata      & \nodata 	 &   F160W                  	& c \\
Mrk 573	    			& 0.15$\pm$0.06   & 0.54$\pm$0.04& 3.2$\pm$0.6 & 18.8$\pm$3.8 & 41.3$\pm$8.3 	 &   F110W,F160W, NSFCam K',L,M & d \\ 
NGC 1386	    		& \nodata         & 0.2$\pm$0.1  & \nodata     & \nodata      & \nodata 	 &   F160W                  	& c \\
NGC 1808	    		& 15.5$\pm$4.5    & \nodata      & 30.5$\pm$8.5& 27.5$\pm$10.5& \nodata		 &  ISAAC J,Ks,L'           	& e \\
NGC 3081	    		& \nodata         & 0.22$\pm$0.13& \nodata     & \nodata      & \nodata 	 &  F160W                   	& c \\
NGC 3281	    		& \nodata         & 1.3$\pm$0.2  & 7.7$\pm$0.8 & 103$\pm$9    & 207$\pm$25       &  IRAC-1 H,K, IRCAM3 L',M 	& f  \\
NGC 4388	    		& 0.06$\pm$0.02   & 0.71$\pm$0.28& \nodata     & 40$\pm$8     & \nodata          &  F110W,F160W, NSFCam L   	& d \\ 
NGC 5728     			& \nodata         & \nodata      & \nodata     & \nodata      & \nodata 	 &  \nodata                 	& \nodata   \\
NGC 7172	    		& \nodata         & $<$0.4       & 3.4$\pm$0.7 & 30$\pm$6     & 61$\pm$12        &  IRCAM3 H,K,L',M         	& g  \\
NGC 7582			& \nodata         & 22.6$\pm$2.3 & \nodata     & 117$\pm$18   & 142$\pm$21 	 &  F160W, ISAAC L,M        	& c,h \\
\hline
NGC 1365	    		& \nodata         & 8.3$\pm$0.8  & 78$\pm$8    & 205$\pm$41   & 177$\pm$35   	 &  F160W                   	& i \\
NGC 2992	    		& \nodata         & $<$1         & 2.8$\pm$0.6 & 22.7$\pm$4.5 & 35.7$\pm$7.1     &  IRCAM3 H,K,L',M         	& g \\
NGC 5506	    		& 13.8$\pm$2.8    & 59$\pm$12    & 120$\pm$24  & 340$\pm$68   & 530$\pm$106      &  IRCAM3 J,H,K,L',M       	& g  \\
\hline
NGC 3227	    		& \nodata         & 10.6$\pm$1.1 & 22.6$\pm$4.5& 46.7$\pm$9.3 & \nodata		 &   F160W, NSFCam K',L     	& d \\ 
NGC 4151	    		& 69$\pm$7        & 104$\pm$10   & 177$\pm$35  & 325$\pm$65   & 449$\pm$34       &   F110W,F160W, NSFCam K',L, IRCAM3 M & d \\ 
\hline
NGC 1566	    		& \nodata         & \nodata      & \nodata     & \nodata      & \nodata	         &  \nodata                 & \nodata \\
\enddata
\tablenotetext{a}{\scriptsize{In the case of Circinus, we also use the 2.42 \micron~flux (31$\pm$3 mJy) from \citet{Prieto04}.}}
\tablecomments{\scriptsize{Ground-based instruments and telescopes are: NACO and ISAAC on the 8 m VLT,  NSFCam on the 3 m NASA IRTF, IRCAM3 on the  3.8 m UKIRT, and IRAC-1 on the 2.2 m ESO telescope. 
Measurements in the F110W and F160W filters are from NICMOS on HST.}}
\tablerefs{\scriptsize{(a) \citet{Meisenheimer07}; (b) \citet{Prieto04}; (c) \citet{Quillen01}; (d) \citet{Alonso03}; (e) \citet{Galliano08}; (f) \citet{Simpson98};
(g) \citet{Alonso01}; (h) \citet{Prieto02}; (i) \citet{Carollo02}}}
\label{literature}
\end{deluxetable*}

\section{SED Observational Properties}

\subsection{High versus Low Spatial Resolution Spectral Energy Distributions}

Important observational constraints for torus modelling arise from the
shape of the IR SEDs of Seyfert galaxies, as
the bulk of the torus emission is concentrated in
the IR range, peaking at mid-IR wavelengths (7 - 26
\micron).  When large aperture data (i.e., of angular
resolutions larger than 1--1.5\arcsec) such as those from ISO, Spitzer, or
IRAS, are employed for constructing SEDs, the IR fluxes are a mixture
of AGN plus host galaxy emission. In particular, for the case of Sy2
galaxies, stellar emission dominates 
the near-IR fluxes at $\lambda~\lesssim$ 2.2 \micron~even 
within 3\arcsec~apertures \citep{Alonso96}. 
Even for Sy1 galaxies, where the direct emission of
the AGN is generally the dominant contribution to the near-IR
flux, \citet{Kotilainen92} found that starlight contributes
significantly to their nuclear emission. 
The stellar contribution can also be significant in large-aperture
mid-IR measurements. For example, \citet{Ramos07} found that
the ISOCAM mid-IR fluxes (with an effective resolution in the 
diffraction limit of $\sim$4\arcsec) of a sample of $\sim$60 Seyfert galaxies
are dominated by AGN emission in Sy1, although there is a
non-negligible contribution from the host galaxy that becomes larger in Sy2. 

Using the data described in Section \ref{observations} we have constructed high spatial resolution SEDs from $\sim$ 1-18 \micron. 
In order to compare them with
large aperture measurements, we have compiled IRAS 12 and 25
\micron~fluxes for the galaxies in our sample, taken from NED (Table \ref{slopes}).  
We compare the nuclear mid-IR spectral slopes calculated from our N and Q data points
($\alpha_{MIR}$; column 3)
with the slopes from the IRAS 12 and 25 \micron~measurements
($\alpha_{IRAS}$; column 4).
If the spectral shape of the IRAS
SEDs were similar to that of our mid-IR high-spatial resolution data,
we would find $\alpha_{MIR}$ = $\alpha_{IRAS}$.  However, that is not
the case, and except for a couple of galaxies for which both slopes
are similar (Mrk 573, NGC 1365, and NGC 1808), 
the rest of sources are far from the $\alpha_{MIR}$ =
$\alpha_{IRAS}$ relationship. 

\begin{deluxetable*}{lcccccccc}
\tabletypesize{\footnotesize}
\tablewidth{0pt}
\tablecaption{Spectral Shape Information}
\tablehead{
\colhead{Galaxy} & \colhead{$\alpha_{IR}$} & \colhead{$\alpha_{NIR}$} & \colhead{$\alpha_{MIR}$} & \colhead{$\alpha_{IRAS}$} & 
\colhead{H/N} & \colhead{N/Q} & \colhead{N/IRAS 12} & \colhead{Q/IRAS 25}}
\startdata
Average     & 3.1$\pm$0.9  & 3.6$\pm$0.8 & 2.0$\pm$0.2 & \dots       &    0.003  &  0.23   & \dots   &   \dots        \\   
Centaurus A & 2.8$\pm$0.8  & 3.4$\pm$0.8 & 1.8$\pm$0.2 & 0.3$\pm$0.1 &    0.006  &  0.27   & 0.03    &   0.09	      \\
Circinus    & 3.6$\pm$1.0  & 4.5$\pm$1.0 & 1.1$\pm$0.1 & 1.8$\pm$0.2 &    0.001  &  0.44   & 0.30    &   0.19	      \\
IC5063	    & 3.8$\pm$1.5  & 4.2$\pm$1.7 & 2.7$\pm$0.4 & 1.8$\pm$0.2 &    0.001  &  0.14   & 0.36    &   0.65	      \\
Mrk 573	    & 2.9$\pm$0.8  & 3.4$\pm$0.8 & 1.7$\pm$0.1 & 1.5$\pm$0.1 &    0.003  &  0.43   & 0.63    &   0.49	      \\
NGC 1386    & 3.3$\pm$1.2  & 3.7$\pm$1.4 & 2.0$\pm$0.2 & 1.4$\pm$0.1 &    0.001  &  0.32   & 0.28    &   0.31	      \\
NGC 1808    & 1.8$\pm$0.3  & 1.7$\pm$0.2 & 1.7$\pm$0.1 & 1.6$\pm$0.1 &    \dots  &  0.41   & 0.04    &   0.03	      \\
NGC 3081    & 3.0$\pm$1.1  & 3.5$\pm$1.3 & 1.4$\pm$0.1 & \dots       &    0.003  &  0.36   & \dots   &   \dots	      \\
NGC 3281    & 2.6$\pm$0.6  & 3.1$\pm$0.7 & 2.0$\pm$0.2 & 1.4$\pm$0.1 &    0.004  &  0.32   & 0.39    &    0.42	      \\
NGC 4388    & 3.0$\pm$0.8  & 3.3$\pm$0.7 & 2.7$\pm$0.4 & 1.7$\pm$0.2 &    0.004  &  0.24   & 0.19    &    0.22	      \\
NGC 5728    & \dots        & \dots       & 2.7$\pm$0.2 & 2.2$\pm$0.3 &   \dots   &  0.14   & 0.12    &    0.18	      \\
NGC 7172    & 1.8$\pm$0.2  & 1.8$\pm$0.2 & \dots       & 1.0$\pm$0.1 &  $<$0.006 &  \dots  & 0.16    &   \dots	      \\
NGC 7582    & 1.2$\pm$0.1  & 1.1$\pm$0.1 & 1.9$\pm$0.1 & 1.6$\pm$0.1 &    0.116  &  0.37   & 0.08    &    0.07	      \\ 
\hline
NGC 1365    & 1.8$\pm$0.4  & 2.0$\pm$0.5 & 1.3$\pm$0.1 & 1.4$\pm$0.1 &    0.026  &  0.50   & 0.06    &   0.04	      \\ 
NGC 2992    & 2.4$\pm$0.5  & 2.7$\pm$0.4 & 2.1$\pm$0.2 & 1.0$\pm$0.1 & $<$0.006  &  0.34   & 0.24    &   0.33	      \\
NGC 5506    & 1.7$\pm$0.3  & 2.0$\pm$0.3 & 1.2$\pm$0.1 & 1.6$\pm$0.1 &    0.068  &  0.40   & 0.68    &   0.53	      \\
\hline
NGC 3227    & 1.8$\pm$0.2  & 1.8$\pm$0.2 & \dots       & 0.9$\pm$0.1 &    0.026  &  \dots  & 0.43    &   \dots	      \\
NGC 4151    & 1.4$\pm$0.2  & 1.3$\pm$0.1 & 1.6$\pm$0.1 & 1.2$\pm$0.1 &    0.079  &  0.41   & 0.66    &   0.66	      \\
\hline
NGC 1566    & \dots        & \dots       & 2.0$\pm$0.2 & 0.6$\pm$0.1 &   \dots   &  0.25   & 0.02    &   0.04	      \\
\enddata        
\tablecomments{\footnotesize{Fitted spectral indexes (f$_{\nu}~\alpha~\nu^{-\alpha}$) in the whole range ($\alpha_{IR}$, from $\sim$1 to 18 \micron), in
the near-IR ($\alpha_{NIR}$, from $\sim$1 to $\sim$9 \micron), and in the mid-IR ($\alpha_{MIR}$, using the N and Q band data points).
The IRAS slopes ($\alpha_{IRAS}$, determined using the 12 and 25 \micron~IRAS fluxes) are shown also for comparison. 
H/N and N/Q band ratios are reported, together with the N/12 \micron~and Q/25 \micron~IRAS ratios.}}
\label{slopes}
\end{deluxetable*}

An analogous comparison between our unresolved mid-IR fluxes and the
IRAS measurements comes from the N/IRAS 12 \micron~and
Q/IRAS 25 \micron~ratios (see Table \ref{slopes}). The low values of these ratios indicate the
large amount of emission related to stellar processes that is contributing to the IRAS fluxes
in contrast to our nuclear  mid-IR fluxes, that are representative of torus emission. 
The lowest values of the ratios correspond to Centaurus A, NGC 1365, NGC 1808, NGC 1566, and NGC 7582.

In summary, the high spatial resolution mid-IR measurements  provide a spectral shape of the SEDs 
that is different from that of large aperture data SEDs. 
The nuclear SEDs presented in this work will allow to us
characterize the torus emission 
and consequently, to use torus models to constrain the distribution of
dust in the immediate AGN vicinity.

\subsection{Average Seyfert 2 Spectral Energy Distribution}

In order to derive general properties of Sy2 galaxies, which constitute the majority of our sample, 
we constructed an average Sy2 SED, considering only the highest angular resolution data to avoid
as much as possible the stellar contamination. This average Sy2 template will be used to compare
with the individual SEDs of the Sy2 analyzed in this work, and also with the intermediate-type Seyferts. 
Out of the 18 galaxies in total (see Table \ref{sources}), there are 10 ``pure'' Type-2 Seyferts  
and two HII/Sy2 composite objects. The rest of the sample consists of other Seyfert types (Sy1.9, Sy1.8, Sy1.5, and Sy1). 
For constructing the Sy2 average SED, we first excluded the composite objects and the intermediate-type Seyferts. 
Second, to  ensure that we are collecting the best-measured 
non-stellar nuclear emission, 
we exclude galaxies with mid-IR data from the 4 m CTIO, because 
of their lower spatial resolution compared with that of the 8 m Gemini telescopes.
This restriction excludes NGC 1808 and NGC 7582. Indeed, these two galaxies are undergoing intense episodes of nuclear star formation, 
making it more difficult to isolate the AGN emission from that of the surrounding host galaxy.
Finally, in the construction  of the average template, we use only 
 HST (NICMOS) and VLT (NACO) adaptive optics fluxes in the near-IR. This excludes NGC 3281, for which all the compiled data are ground-based, 
and a few points in the SEDs of some of the other galaxies (e.g., Mrk 573 and NGC 4388).

\begin{figure*}[!ht]
\centering
\includegraphics[width=10cm,angle=90]{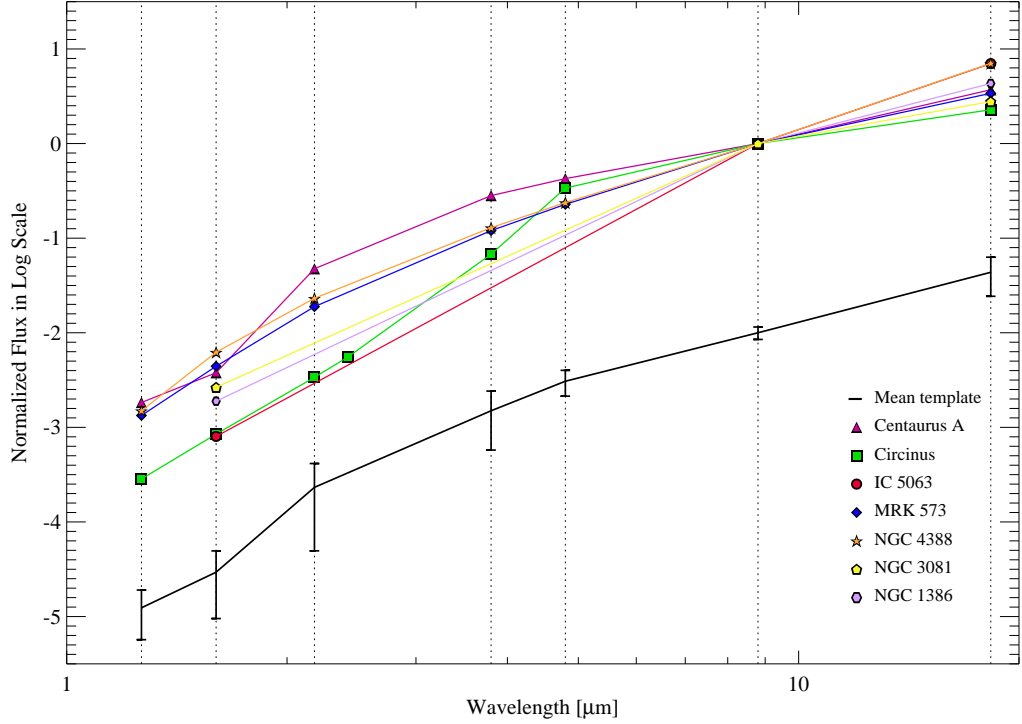}
\caption{\footnotesize{Observed IR SEDs for the seven pure Sy2 galaxies (in color and with different 
symbols) used for the construction of the 
average template (solid black). The SEDs have been normalized at 8.8 \micron, and the average SED has been shifted in the 
Y-axis for clarity.}
\label{template}}
\end{figure*}

In the end, we have seven pure Sy2 for constructing the average template: Centaurus A, Circinus, 
IC 5063, Mrk 573\footnote{Note that Mrk 573  has been recently reclassified 
by \citet{Ramos08} as an obscured NLSy1, but the IR SED
remains similar to that of the ordinary Sy2 galaxies.},
NGC 1386, NGC 3081, and NGC 4388.
Their IR SEDs are shown in Figure \ref{template}. We normalize 
the SEDs at 8.8 \micron~in order to emphasize the comparison between the N- and both the H- and Q-band data points, that
provides useful information about the obscuration (see Section \ref{nirflux}).
The measured N/Q ratio for 
the average template is 0.23$\pm$0.14, which is 
representative of the Sy2 in our sample. 

Thus, the average template is contructed with near- and mid-IR data having resolutions $\lesssim 0.6\arcsec$,
to ensure the isolation of the torus emission.
The best-sampled SED, Circinus, defines the wavelength grid,
and we interpolated nearby measurements of the other six galaxies onto this scale.
We did not interpolate the sparse observations of IC 5063 and NGC 3081. 
The interpolated fluxes  were used solely for the purpose of deriving the average Sy2 SED template, 
shown in Figure \ref{template}. The error bars correspond to the standard deviation 
of each averaged point, except for the 8.8 \micron~point (the wavelength chosen for the normalization). In this case, 
we assigned a 15\% error, the nominal percentage considered for the N-band flux measurements.

\subsection{Type-2 Seyfert SEDs}
\label{sy2:seds}

The shape of the Sy2 SEDs in our sample is generally very steep, compared with the intermediate-type Seyferts. 
In general, Sy2 have steeper 1--10 \micron{} SEDs than Sy1 \citep{Rieke78,Edelson87,Ward87,Fadda98,Alonso01,Alonso03}.
Nevertheless, we find a variety of spectral shapes for Sy2, in agreement with \citet{Alonso03}.

We measured the 1--18 \micron~IR slopes
(f$_{\nu}~\alpha~\nu^{-\alpha_{IR}}$) of the individual Sy2 galaxies
and of the average template (Table \ref{slopes}).  The IR slopes are
representative of the whole SED shape. We also calculated the near-IR ($\alpha_{NIR}$, from $\sim$1 to $\sim$9 \micron)
and mid-IR spectral indexes ($\alpha_{MIR}$, using the N and Q band data points) to compare them. A flat near-IR slope
indicates an  important contribution of the hot dust emission (up to $\sim$1000-1200 K;
\citealt{Rieke81,Barvainis87}) that comes from the inner part
of the torus, and consequently, must be hidden in smooth torus
descriptions of Sy2. 

The Sy2 SEDs are typically steep through the IR, with $\alpha_{IR} = 3.1\pm0.9$
for the average template.  However, the individual galaxies
do show some variety, with $\alpha_{IR}$ ranging from 1.8 to 3.8,
being even shallower in the uncertain NGC 7582.
Our results are similar to those of \citet{Alonso03}, who found
$\alpha_{IR}^{Alonso}>2.6$ for the 1 to 16  \micron{} slopes of 
the Sy2 galaxies in the expanded CfA sample.
Comparing the near- and mid-IR slopes for the Sy2 reported in Table \ref{slopes}, 
we find that $\alpha_{NIR} \ga \alpha_{MIR}$ for all of them, except for NGC 7582.
For the average template, $\alpha_{NIR}=3.6\pm0.8$ and $\alpha_{MIR}=2.0\pm0.2$.

The flatter slopes arise naturally in some configurations of clumpy torus  
models with
some near-IR contribution from the hot dust faces of directly-illuminated clouds,
even when the central engine is blocked from view.  In contrast, smooth
torus models do not generally produce a range of IR spectral slopes. 
Indeed, large opacity torus models without extended cone component predict $\alpha_{IR} > 4$ \citep{Pier93,Granato94}.
Homogeneus tori with lower optical depth can flatten the IR slopes
\citep{Granato97,Fadda98}, but they predict the 10 \micron~silicate
feature in strong emission for Type-1 nuclei, which is not observed
\citep{Roche91}.  The extended conical component included in the
\citet{Efstathiou95} torus models flattens the 3--5 \micron~continuum
of Sy2. However, the high resolution measurements considered here
should be mostly free of cone contamination.  Thus, clumpy dusty torus
models seem to be the most promising way for reproducing the observed range of Sy2 SEDs.

\subsection{Intermediate-Type Seyfert SEDs} 
\label{sy1p8}

The Sy1.5, Sy1.8, and Sy1.9 galaxies were added by \citet{Osterbrock81} to the Seyfert classification scheme to account for 
those galaxies with weaker
featureless continua, larger broad Balmer decrements, and weaker broad components of the permitted lines than Type-1 Seyferts. 
\citet{Osterbrock81} suggested that the observed characteristics of these intermediate Seyferts could be due
to the reddening of both the continuum and the BLR by dust. \citet{Alonso03} found for the CfA Seyfert sample, 
that the majority 
of galaxies optically classified as Sy1.8 and Sy1.9 display either IR spectral indexes and SEDs similar to those of 
Sy1 or intermediate between Sy1 and Sy2. For the Sy1.5, \citet{Alonso03} found the SEDs to be practically identical 
to those of Sy1.


\begin{figure*}[!ht]
\centering
\includegraphics[width=10cm,angle=90]{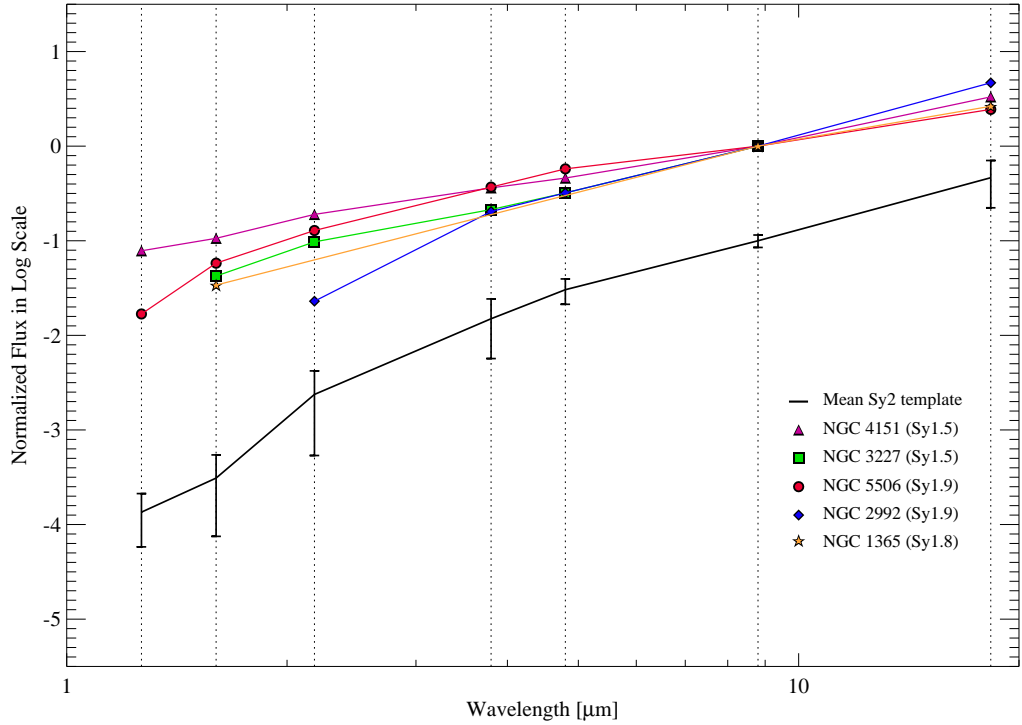}
\caption{\footnotesize{Observed IR SEDs normalized at 8.8 \micron~for the Sy1.5, Sy1.8, and Sy1.9 galaxies in our sample (in color and different symbols). 
The average Sy2 SED is displayed for comparison, displaced in the Y-axis for clarity. 
Note the difference in steepness between the Sy2 mean
SED and the intermediate types, with the exception of the Sy1.9 NGC 2992.}
\label{intermediate}}
\end{figure*}

Figure \ref{intermediate} shows the SEDs of the five intermediate-type Seyferts considered in this work 
(two Sy1.5, one Sy1.8, and two Sy1.9), together with the mean Sy2 template for 
comparison, interpolated onto the mean Sy2 wavelength grid.
Spectral indexes are reported in Table \ref{slopes}. 
%
The spectral slopes of the intermediate-type Seyferts are shallower than those
of the Sy2 (mean slope $\alpha_{IR} = 2.0\pm0.4$ for Sy1.8 and Sy1.9 and $\alpha_{IR} = 1.6\pm0.3$ for Sy1.5, as opposed to $3.1\pm0.9$).
\citet{Alonso03} report IR spectral indices (from 1 to 16 \micron)
of types 1.8 and 1.9 similar to our measurements ($\alpha_{IR}^{Alonso} = 1.8$--2.6), and
within the  interval $\alpha_{IR}^{Alonso}$ = 1.5--1.6 for Sy1 and Sy1.5 galaxies.
The steepest SED among the intermediate-type Seyferts is in
the Sy1.9 NGC 2992, which is more similar in shape to the Sy2 SEDs.
This result could be due to an incorrect or genuinely changed 
classification of this galaxy, as \citet{Trippe08} claim (see Appendix \ref{indiv:sy2}).

The near-IR slopes of the Sy1.8 and Sy1.9  have  
intermediate values between those of Sy2 and
Sy1.5, with $\alpha_{NIR} > \alpha_{MIR}$.
For the Sy1.5 galaxy NGC 4151, $\alpha_{NIR} < \alpha_{MIR}$, indicating the importance of the hot
dust contribution in the SEDs of this type of nucleus. 


The H/N and N/Q flux ratios show comparable differences among Seyfert
types (Table \ref{slopes}).  The H/N ratio is larger for Sy1.5
(0.05$\pm$0.04) than for Sy2 (0.003$\pm$0.002 for the average
template).  The N/Q ratios of the intermediate-type Seyferts are also
larger than those of Sy2 (mean values of 0.41$\pm$0.08 for Sy1.8 and
Sy1.9 and 0.41$\pm$0.06 for Sy1.5, as opposed to 0.23$\pm$0.14).

In summary, the slope of the IR SED is in general correlated with the Seyfert
type. Sy2 show steeper SEDs, 
and intermediate-type Seyferts are
flatter. Seyferts 1.8 and 1.9 present intermediate values of the IR slope 
and H/N ratio between Sy2 and Sy1.5. 
However, we find a range of spectral shapes among the Sy2
galaxies, and some intermediate-type SEDs have the same slopes as the
Sy2.  This cannot be reconciled with the predictions of early
torus models, since large optical depth homogeneus tori strictly predict steep SEDs for Sy2 and
flat SEDs for Sy1. We do not observe such a strong dichotomy (see also \citealt{Alonso03}). Torus models using lower opacities are observationally
discarded (\S \ref{sy2:seds}) and models including an 
extended conical component do not seem adecuate for being applied to our high spatial resolution data. 
Thus, we pursue clumpy torus models of \citet{Nenkova02}  to reproduce the observed SEDs.

\section{SED Modelling} 

\subsection{Clumpy Dusty Torus Models} 

\begin{figure}[!ht]
\centering
\includegraphics[width=10cm]{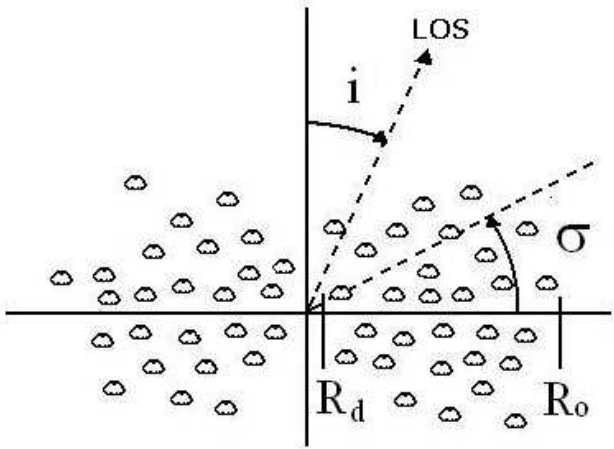}
\caption{\footnotesize{Scheme of the clumpy torus described in \citet{Nenkova08a,Nenkova08b}. The radial extent of the torus is defined by the outer radius
($R_o$) and the dust sublimation radius ($R_d$). All the clouds are supposed to have the same $\tau_{V}$, and $\sigma$ 
characterizes the width of the angular distribution. The number of cloud encounters is function of the viewing angle, $i$.}
\label{clumpy_scheme}}
\end{figure}

The clumpy dusty torus models of \citet{Nenkova02,Nenkova08a,Nenkova08b} propose that the dust
surrounding the central engine of an AGN is distributed in clumps, instead of homogeneously filling the torus volume.  
These clumps are distributed with a radial extent characterized by the parameter $Y = R_{o}/R_{d}$, where 
$R_{o}$ and $R_{d}$ are the outer and inner radius of the toroidal distribution, respectively (see Figure \ref{clumpy_scheme}). 
The inner radius is defined by the dust sublimation temperature ($T_{sub} \approx 1500$ K),
with $R_{d} = 0.4~(1500~K~T_{sub}^{-1})^{2.6}(L / 10^{45}\,\mathrm{erg ~s^{-1}})^{0.5}$ pc.  
Within this geometry, each clump has the same optical depth ($\tau_{V}$, defined at $V$).
The average number of clouds along a radial equatorial ray is $N_0$. The radial density profile is a
power-law ($\propto r^{-q}$). A width parameter, $\sigma$, characterizes the angular distribution, which has
a smooth edge.  The number of clouds along the line of sight (LOS) 
at 
an inclination angle $i$ is $N_{LOS}(i) = N_0~e^{(-(i-90)^2/\sigma^2)}$. 

The radiative transfer equations are solved for each clump using the DUSTY code
\citep{Ivezic99}, with solutions depending mainly on the location of
each clump within the torus, its optical depth, and the chosen dust
composition. We adopt a dust extinction profile corresponding to the
OHMc dust (i.e., the standard cold oxygen-rich ISM dust of
\citealt{Ossenkopf92}). DUSTY includes dust absorption, emission, and
scattering components. The total torus emission is calculated
by integrating the source function of the whole number of clumps
convolved with the radiation propagation probability along the torus
\citep{Nenkova02}. The direct AGN emission may also be included in the resulting SED,
which is appropriate for Type-1 nuclei and some intermediate types.

\subsection{BayesClumpy}
\label{bayesclumpy}

The clumpy database contains $\sim$10$^{6}$ models, calculated for a fine grid of model parameters.
The inherent degeneracy between model parameters has to be taken
into account when fitting the observables.
To this aim, \citet{Asensio09} recently developed a Bayesian inference tool (\textsc{BayesClumpy})
that extracts as much information as possible from the observations. 
They applied interpolation methods that are
able to derive models for different combinations of the six clumpy model parameters described above
even if they are not present in the original database.

The synthesis code is implemented into a Metropolis-Hastings Markov
Chain Monte Carlo  algorithm that evaluates the posterior
distribution function for the Bayesian inference. This
posterior distribution results from taking into account the a-priori
knowledge about the parameters and the information introduced by the
observations. The names and abbreviations of the six parameters that
describe the clumpy models, and the intervals considered for
the general fitting are detailed in Table \ref{parametros} (see also Figure \ref{clumpy_scheme}). 

\begin{deluxetable*}{lcl}
\tabletypesize{\footnotesize}
\tablewidth{0pt}
\tablecaption{Clumpy Model Parameters and Considered Intervals}
\tablehead{
\colhead{Parameter} & \colhead{Abbreviation} & \colhead{Interval}}
\startdata
Radial extent of the torus                             & $Y$             & 15        \\
Width of the angular distribution of clouds            & $\sigma$        & [15\degr, 75\degr]  \\
Number of clouds along the radial equatorial direction & $N_0$           & [1, 15] \\
Power-law index of the radial density profile          & $q$             & [0, 3]    \\
Inclination angle of the torus                         & $i$             & [0\degr, 90\degr]   \\
Optical depth per single cloud                         & $\tau_{V}$      & [10, 200] \\
\enddata         
\label{parametros}
\end{deluxetable*}

We introduced to the code the intervals of the parameters reported in Table \ref{parametros} as uniform distributions, thus, giving 
the same weight to all the values in each interval. 
The exception is the torus radial extent parameter $Y$, fixed as a Gaussian distribution centered at 15,
with a width of 2.5. 
We fixed this parameter after confirming that it is 
otherwise unconstrained in the fits and is not correlated with any of the other
parameters\footnote{The 
only exception is Centaurus A (which $Y$ parameter results centered in $\sim$10-15 without having 
constrained it. See Appendix \ref{indiv:sy2}).}.
The choice of  $Y$ = 15 is justified below (\S\ref{individual}). For those galaxies for which we have additional a-priori
knowledge about any of the parameters (e.g., the inclination angle of the torus
for galaxies with known masers), the priors have been set as narrow Gaussians centred at a given value.
Apart from the six parameters that characterize a
clumpy model, there is another additional parameter that accounts
for the vertical scale displacement, which we allow to vary freely, needed 
to match the fluxes of a chosen model to an observed SED. 
This vertical shift scales with 
the AGN bolometric luminosity (\S \ref{discussion}). 

In order to compare with the observations, \textsc{BayesClumpy} simulates the
effect of the employed filters on the simulated SED
by selecting the filter and introducing the observed flux and its corresponding
error (assuming Gaussian errors).
For a detailed description of the Bayesian inference applied to the clumpy models
see \citet{Asensio09}.

\subsection{Model Results} 

\subsubsection{Seyfert 2 Individual Fits}
\label{individual}

The results of the fitting process are the marginal posterior 
distributions for the six free 
parameters that describe the clumpy models and the vertical shift. These are the probability
distribution of each parameter, represented as histograms. 
Uniform priors have been employed in our analysis of the observational data.
Therefore, when the observed data introduce enough information into the fit, 
the resulting probability distributions will clearly differ from uniform distributions, 
either showing trends or being centered at certain values within the considered intervals.

We fit the individual Sy2 SEDs with the  \textsc{BayesClumpy} code,
modelling only the torus emission, assuming no direct AGN continuum is detected.
We also assume that foreground extinction from the host galaxy does not significantly 
modify the infrared SEDs of the Seyfert galaxies in the sample. For example, 
\citet{Martini99} used V-H colors of Sy2 galaxies in the CfA sample to derive 
levels of optical extinction up to A$_V\sim$ 5 mag. Based on this, 
\citet{Alonso03} applied different values of A$_V$ up to 5 magnitudes to 
a subset of torus models (those of \citealt{Efstathiou95}) to illustrate the 
effects of dust in the host galaxy, and concluded that moderate amounts 
of foreground extinction (A$_V\la$ 5 mag) will not have a significant effect
on the nuclear fluxes at $\lambda\ga$ 2 $\micron$. 
Thus, we do not consider foreground extinction in the fits with the clumpy models
for the Seyfert galaxies in our sample. The only exception is Centaurus A, since 
its core is heavily obscured by a dust lane up to A$_V\sim$ 7-8 mag \citep{Hough87,Packham96,Marconi00}.
For details on the specific fit of this galaxy see Appendix \ref{indiv:sy2}.

\begin{figure*}[!ht]
\centering
{\par
\includegraphics[width=5.3cm]{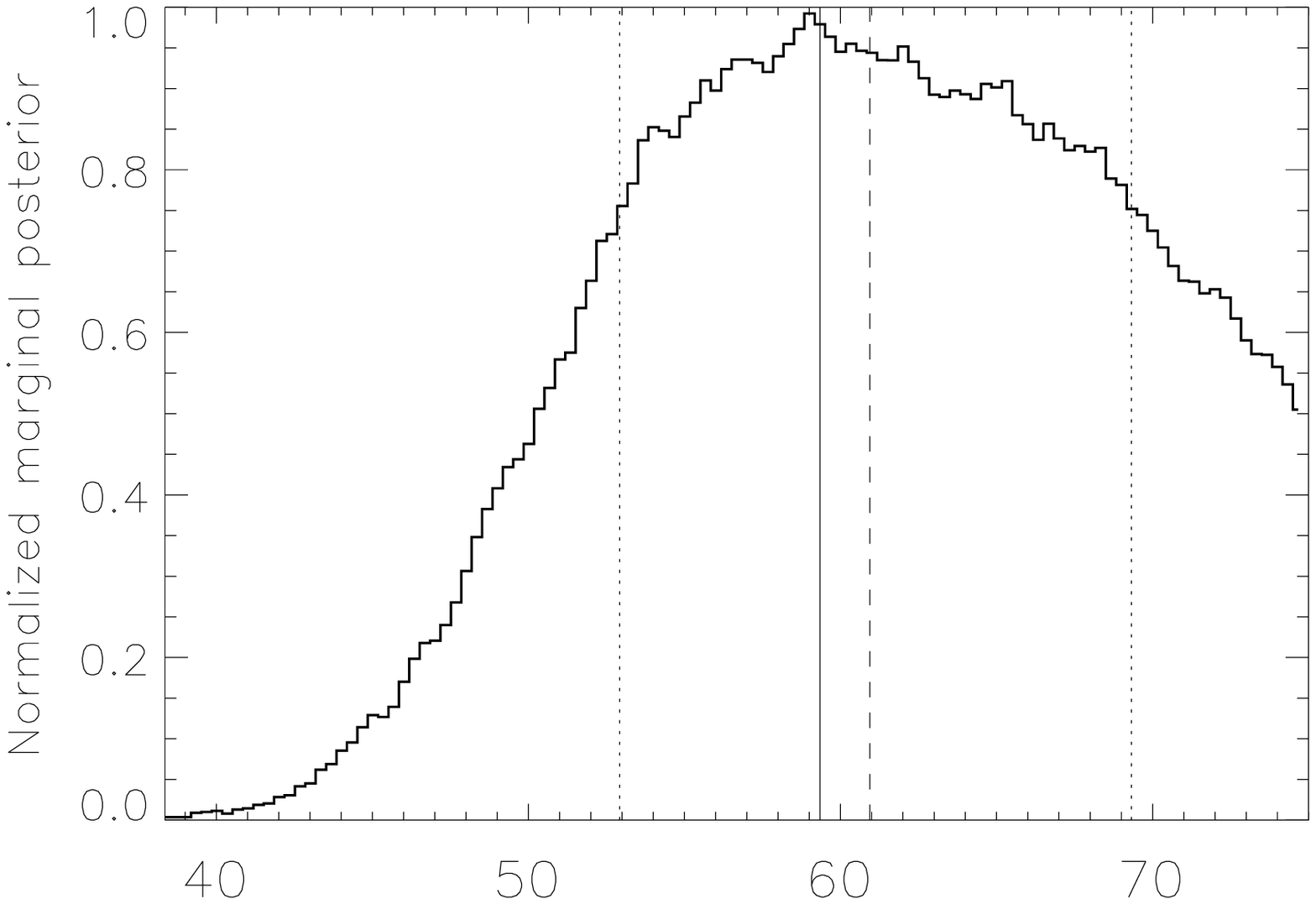}
\includegraphics[width=5.3cm]{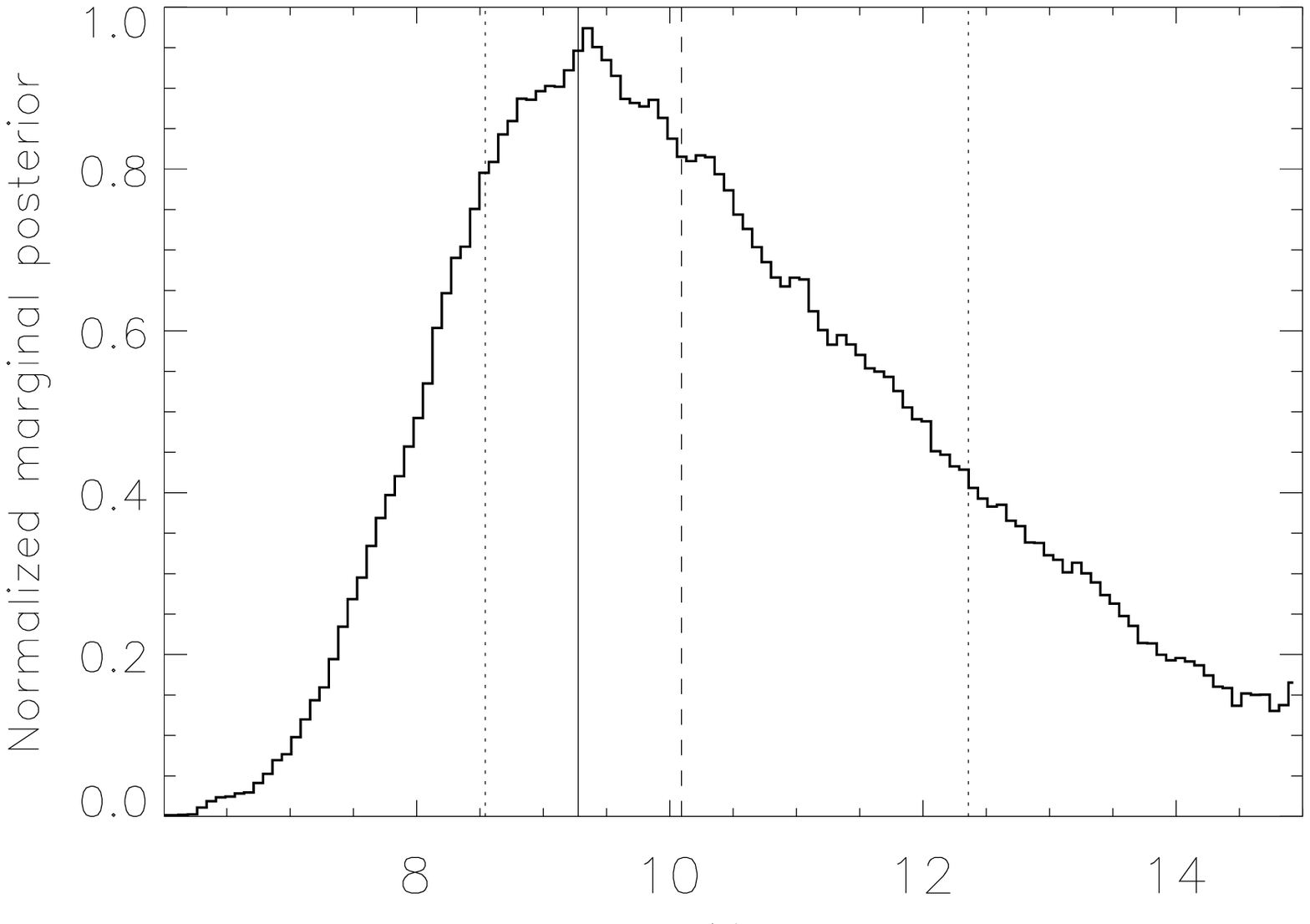}
\includegraphics[width=5.3cm]{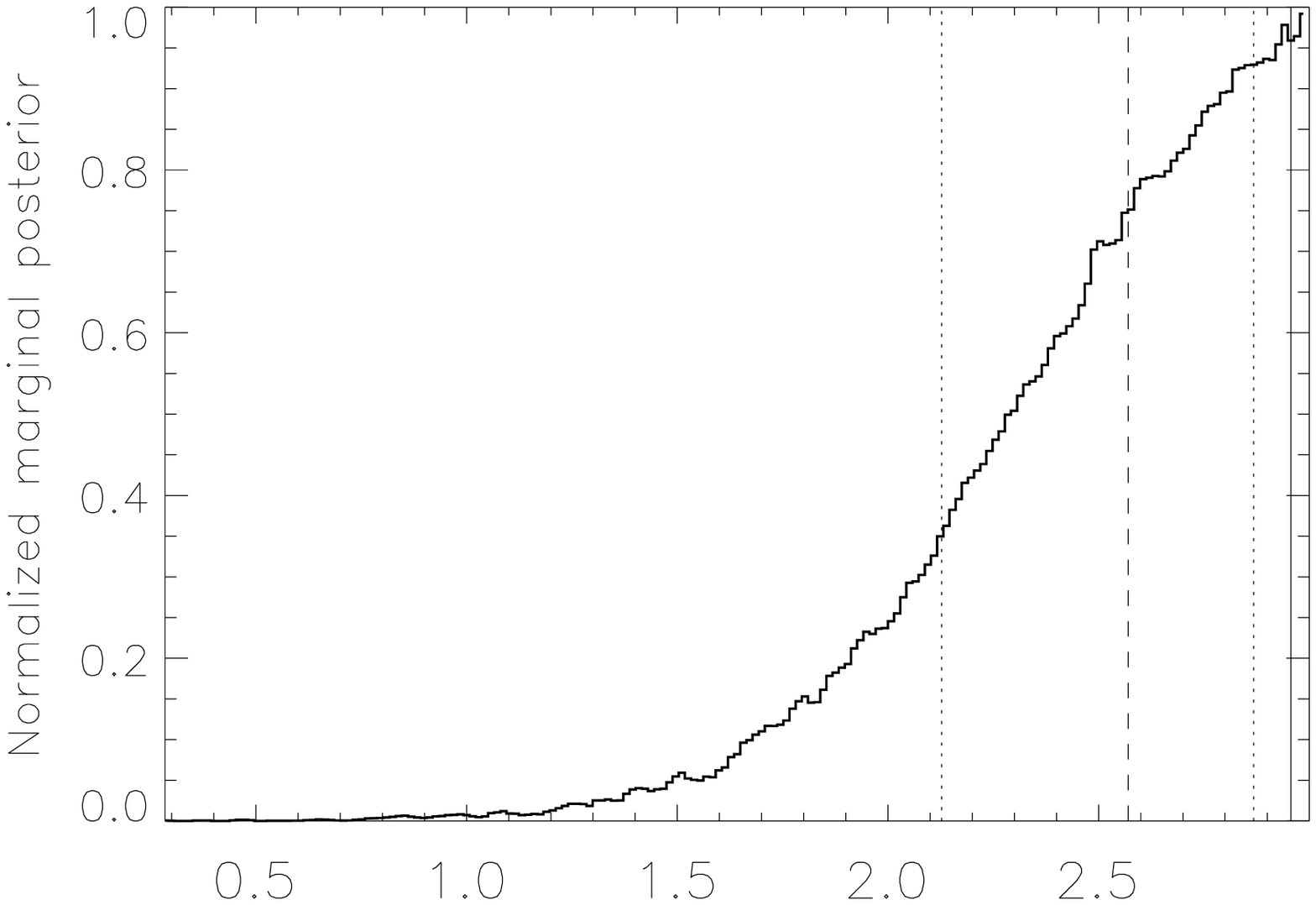}
\includegraphics[width=5.3cm]{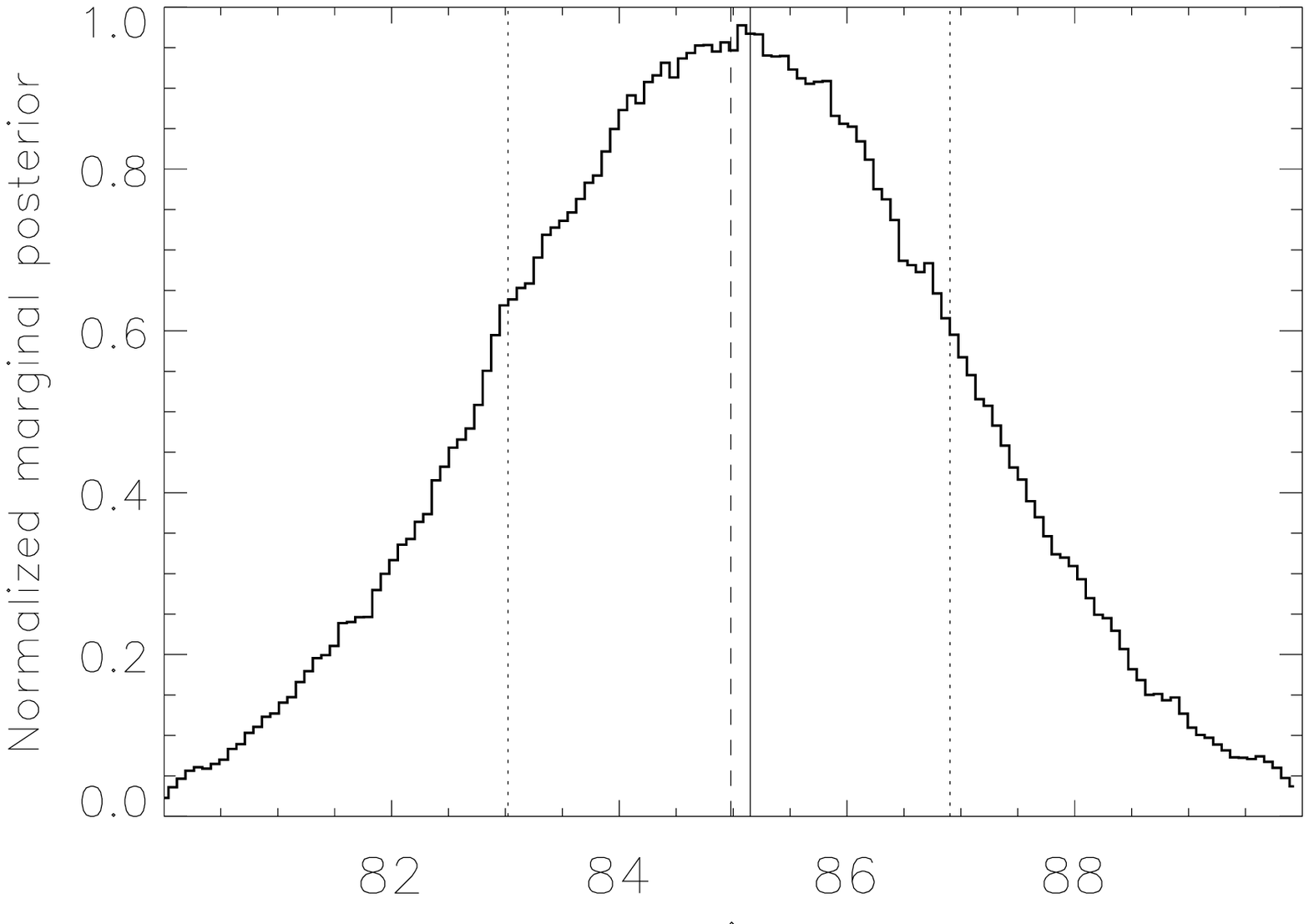}
\includegraphics[width=5.3cm]{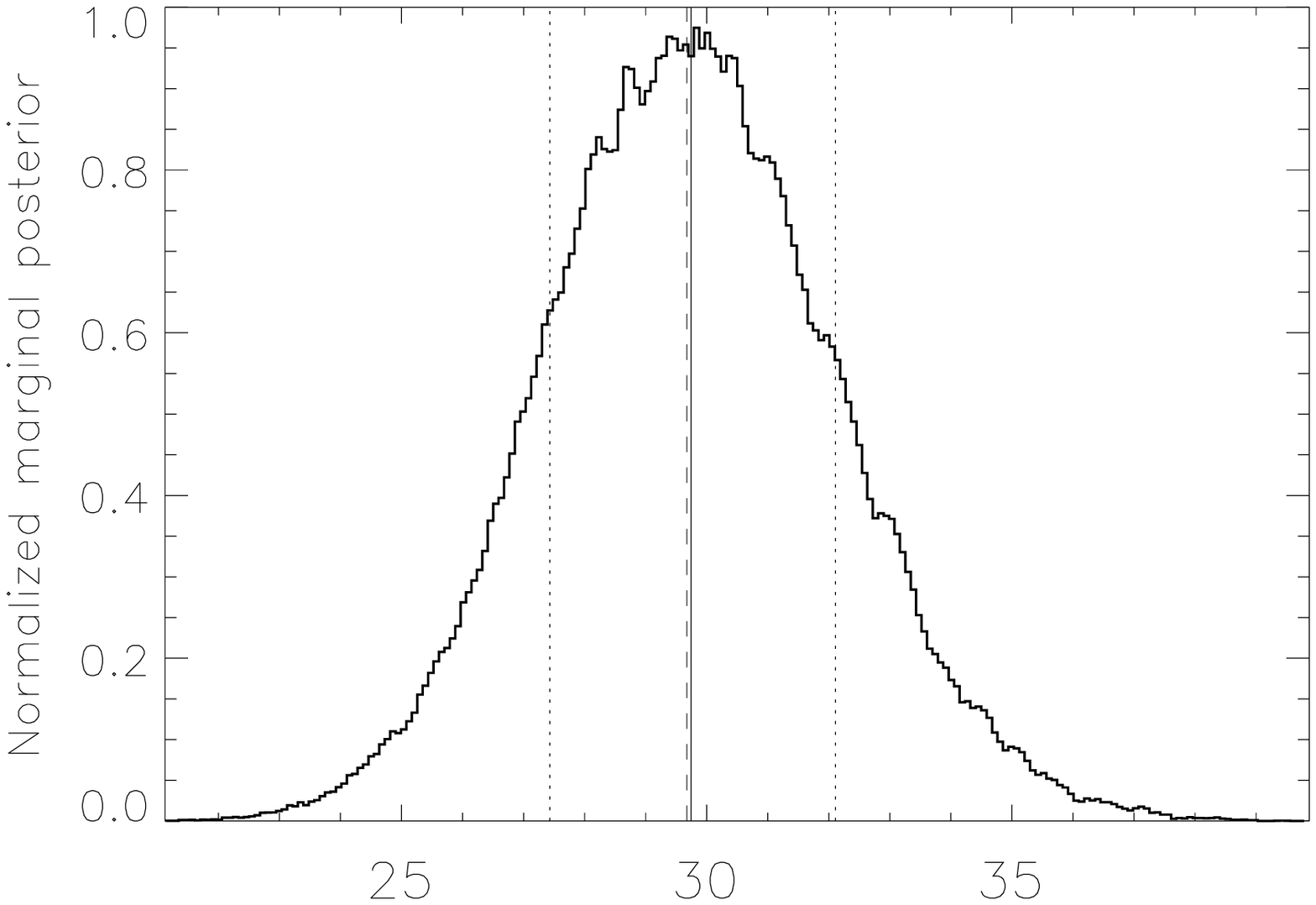}
\includegraphics[width=5.3cm]{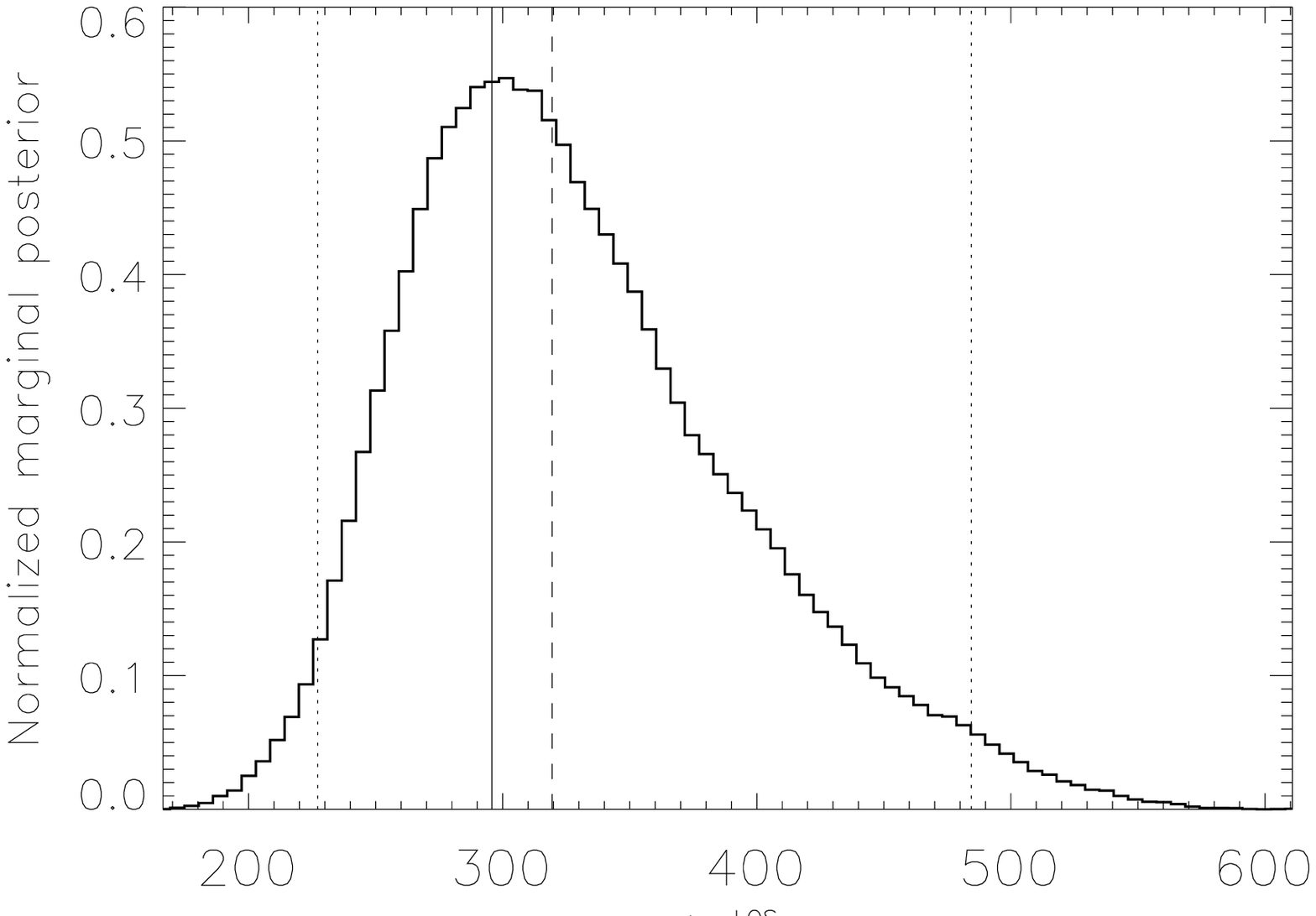}\par}
\caption{\footnotesize{Probability distributions of the free parameters that describe the clumpy models resulting from the fit 
of Circinus, together with the calculated value of the optical extinction along the LOS $A_V^{LOS}$. The vertical shift and the $Y$ parameter have been marginalized.
Solid lines represent the mode of each distribution, dashed lines correspond to the median, and dotted lines indicate 
the 68\% confidence level for each parameter around the median.}
\label{posterior}}
\end{figure*}

The histograms corresponding to the probability distributions of the free parameters are 
shown in Figure \ref{posterior} for Circinus, which constitutes the best fit in our sample. 
It also includes the histogram corresponding to the optical extinction 
produced by the torus along the LOS, computed as $A_{V}^{LOS} = 1.086~N_0~\tau_{V}~e^{(-(i-90)^{2}/\sigma^{2})}$ mag. 

We adopt a radial extent of $Y=15$ as a Gaussian prior with a width of 2.5
because this parameter is otherwise unconstrained and presents no correlation with
other parameters. 
The chosen value of $Y$=15 is consistent with all current observations of
nearby active nuclei, which place the radial extent of the torus within values smaller than $\sim$20-30
\citep{Jaffe04,Packham05,Tristram07,Meisenheimer07,Radomski08,Raban09}, 
and perhaps even as small as $\sim$5-10. The lack of restriction of the torus radial extent is in agreement with the 
findings of \citet{Nenkova08b}, who  report that IR SED fitting sets only a poor constraint on the torus size 
(see Section \ref{torus_size}).

From Figure \ref{posterior} it is clear that the Circinus data
(8 photometric data points) provide sufficient information to constrain the
model parameters. In addition to the above mentioned prior for the $Y$
parameter, we introduce a Gaussian prior on the inclination angle of the torus $i$, a Gaussian centered in
85\degr, with a width of 2\degr. We make this restriction in the $i$ parameter 
because of the detection of a water vapor megamaser \citep{Greenhill03}, which constrains 
the viewing angle to $i \sim 90 \degr$. 
The optical depth per cloud results in a narrow Gaussian centred at the median value
$\tau_{V}$ = 30$\pm$2. The number of clouds results in a
Gaussian-like distribution of median value $N_0 = 10\pm$2, and the width of the angular
distribution has a median value of $\sigma = 61\degr\pm$8. Finally,
we establish a lower limit of $q > 2.4$ at a 68\% confidence level.
The optical obscuration produced by the torus along the LOS would be of $A_V^{LOS}$ = 320$\pm_{55}^{80}$
mag.

\begin{figure*}[!ht]
\centering
\includegraphics[width=10cm,angle=90]{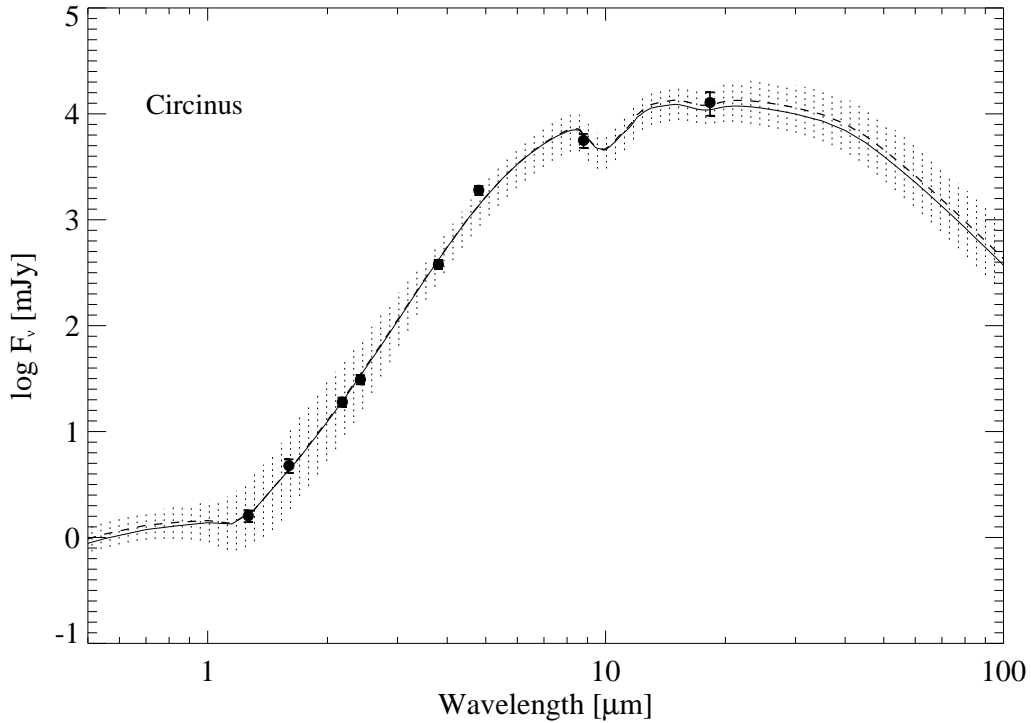}
\caption{\footnotesize{High spatial resolution Circinus SED. 
The solid line represents the ``best fit'' to the data, corresponding 
to the clumpy model described by the combination
of parameters that maximizes the probability distributions of the parameters.
The dashed line represents the model that corresponds to
the median value of the probability distribution of each parameter. The shaded region indicates the range of models compatible
with a 68\% confidence interval for each parameter around the median.}
\label{sed}}
\end{figure*}

Although the solution to the Bayesian inference problem are the
probability distributions shown in Figure \ref{posterior}, we
translate these results into corresponding spectra (Figure \ref{sed}).  The
maximum-a-posteriori (MAP) values of the parameters (i.e., the modes)
represent the ``best fit'' to the data, since the mode corresponds to the most probable value of each parameter,
and we plot the corresponding spectrum as a solid line. 
The dashed line shows the spectral model obtained using the median
value of the probability distribution of each parameter, which is
characteristic of the observed SED.  Finally, the shaded region
indicates the range of models compatible with a 68\% confidence
interval for each parameter around the median value.
The observed Circinus SED is extremelly well-fitted with the clumpy models, 
including only the reprocessed torus emission without any
stellar contribution.
The model fitting to the photometric points makes a spectroscopic prediction:
the 10 \micron~silicate feature appears in shallow absorption in
the fitted models, as expected for Sy2 galaxies
\citep{Hao07,Levenson07}. In particular, the model prediction for Circinus
is in qualitative agreement with spectroscopic observations \citep{Roche06,Tristram07}
that show the silicate feature in absorption, although deeper than the predicted by the models 
(see Appendix \ref{indiv:sy2}).

The excellent fit of the Circinus SED is the result of the combination of the 
good behaviour of the clumpy models and the accuracy of the IR 
nuclear fluxes. 
The near-diffraction-limited adaptive optics NACO/VLT near-IR data 
(providing resolution between 0.1\arcsec and 0.2\arcsec) and the  T-ReCS/Gemini South mid-IR data (resolution between 0.3\arcsec and 0.5\arcsec), 
together with the proximity of the galaxy 
provide the least contaminated (by host galaxy emission) SED of our sample, and consequently the best example of 
an AGN-dominated SED. 

We apply the same technique to all the
Sy2 individual galaxy measurements (with the exception of NGC 5728, for which we did not find near-IR nuclear fluxes in the literature, resulting then 
excluded from any further analysis), 
and plot  the spectral results in Figure \ref{sy2_1}.
The probability distributions of each parameter for all the individual galaxies are summarized 
in Table \ref{clumpy_parameters}.
Comments on the individual fits of the Sy2 galaxies in our sample are reported in Appendix \ref{indiv:sy2}. 
%
Despite the more limited flux measurements in most individual cases, the resulting
model fits are generally good.  The fitted parameters are consistent with
unified AGN schemes and  
the requirement that the central engine be hidden from direct view in Type-2 AGNs.
The only exceptional cases are NGC 1808 and NGC 7582, as both exhibit 
intense nuclear star formation and have lower spatial resolution
mid-IR measurements (obtained at the CTIO 4 m).  Thus, we consider these two model results unreliable, 
especially that of NGC 7582. 

\begin{deluxetable*}{lccccccccccccc}
\tabletypesize{\footnotesize}
\tablewidth{0pt}
\tablecaption{Parameters Derived from the Clumpy Model Fitting}
\tablehead{
\colhead{Galaxy} & \colhead{Seyfert} & \multicolumn{2}{c}{$\sigma$} & \multicolumn{2}{c}{$N_0$} & \multicolumn{2}{c}{$q$} & \multicolumn{2}{c}{$i$} &
\multicolumn{2}{c}{$\tau_{V}$} & \multicolumn{2}{c}{$A_{V}^{LOS}$}  \\
& \colhead{Type} & \colhead{Median} & \colhead{Mode} &  \colhead{Median} & \colhead{Mode} & \colhead{Median} & \colhead{Mode} & \colhead{Median} & \colhead{Mode} & \colhead{Median} & \colhead{Mode} & \colhead{Median} & \colhead{Mode}}
\startdata
Centaurus A&   2 & 58$\pm_{15}^{10}$   & 62    & 2$\pm$1	 & 1   & $<$0.2    & 0.1   & $>$85	     & 89  & 176$\pm_{25}^{16}$& 186	& 300$\pm_{80}^{100}$  & 260   \\
Circinus   &   2 & 61$\pm$8 	       & 59    & 10$\pm$2	 & 9   & $>$2.4    & 3.0   & 85 (fix)	     & 85  & 30$\pm$2	       & 30	& 320$\pm_{55}^{80}$   & 295  \\
IC 5063	   &   2 & $>$57	       & 75    & $>$11  	 & 14  & $<$1.5    & 0.4   & $>$65	     & 89  & 70$\pm_{23}^{31}$ & 66	& 780$\pm_{345}^{485}$ & 580  \\
Mrk 573	   &   2 & $<$37               & 19    & 6$\pm_{2}^{5}$  & 4   & $<$1.4    & 0.1   & 85 (fix)	     & 85  & 30$\pm_{8}^{10}$  & 27	& 185$\pm_{75}^{120}$  & 150  \\
NGC 1386   &   2 & 50$\pm_{19}^{16}$   & 51    & 11$\pm$3	 & 12  & 1.5$\pm_{1.0}^{0.9}$ & 2.1& 85 (fix)& 85  & 95$\pm_{51}^{66}$ & 51	& $<$1460& 410 \\  
NGC 1808   &   2 & $<$35	       & 16    & 8$\pm_{4}^{5}$  & 5   &1.2$\pm$0.9& 0.9   & 37$\pm_{23}^{19}$& 44 & 146$\pm_{50}^{37}$& 198	& $<$140 & 10 \\
NGC 3081   &   2 & $>$52	       & 74    & 10$\pm$3	 & 11  & $>$1.2    & 2.9   & $>$42	     & 89  & 47$\pm_{23}^{42}$ & 34	& $<$450 & 120 \\	
NGC 3281   &   2 & $>$68	       & 75    & 6$\pm_{2}^{4}$  & 5   & 1.3$\pm_{0.8}^{1.1}$&0.6& $>$49     & 88  & $<$12	       & 10	& 55$\pm_{15}^{18}$ & 50  \\
NGC 4388   &   2 & $>$53	       & 74    & 9$\pm_{3}^{4}$  & 7   & $<$1.6    & 0.2   & $>$38	     & 88  & $<$14	       & 10	& $<$80  & 50 \\
NGC 7172   &   2 & $>$54               & 74    & 5$\pm_{1}^{3}$  & 5   & $>$1.7    & 2.9   & $>$45	     & 89  & $<$12	       & 10 	& 50$\pm$20 & 50 \\
NGC 7582   &   2 & $<$29	       & 16    & $<$2		 & 1   & $>$2.5    & 3.0   & 41$\pm_{28}^{19}$& 58 & $<$27	       & 14	& $<$6      & 1 \\  
\hline
NGC 1365   &  1.8& 35$\pm_{13}^{20}$   & 31    & 7$\pm_{4}^{5}$  & 3   & $<$1.7    & 0.1   & $<$42	     & 7   & 110$\pm$60        & 88	& $<$170 & 10 \\ 
NGC 2992   &  1.9& 45$\pm_{21}^{18}$   & 52    & 7$\pm_{3}^{5}$  & 4   & $<$1.0    & 0.1   & $>$53	     & 85  & 36$\pm_{11}^{14}$ & 31	& 160$\pm_{80}^{130}$ & 130 \\
NGC 5506   &  1.9& 25$\pm_{7}^{9}$     & 15    & $<$2		 & 1   & 2.5$\pm_{0.4}^{0.3}$& 2.7 & 85 (fix)& 85  & $<$68	       & 22	& $<$90  & 30 \\ 
\hline
NGC 3227   &  1.5& 33$\pm_{12}^{20}$   & 23    & 6$\pm_{4}^{5}$  & 3   & $<$2      & 0.1      & $<$48	     & 14  &115$\pm$55         & 77	& $<$210 & 10 \\ 
NGC 4151   &  1.5& $<$32               & 16    & $<$3		 & 1   & 1.7$\pm_{0.9}^{0.8}$ & 1.9& 41$\pm_{28}^{23}$& 52 &120$\pm_{48}^{55}$& 93 & $<$65& 10 \\
\enddata          
\tablecomments{\footnotesize{The $Y$ parameter is fixed to be a Gaussian distribution centered at 15 with a width of 2.5.
For the galaxies Circinus, Mrk 573, NGC 1386, and NGC 5506 the $i$ parameter is also introduced as a Gaussian prior into the computations, centered at 85\degr with a 
width of 2\degr, based on other observations. 
Probability distributions presenting a single tail have been characterized with the mode and upper/lower limits at 68\% confidence.
Sy1.8 and Sy1.9 have been fitted with the geometry corresponding to torus emission-only.  The Sy1.5 models include the intrinsic AGN continuum emission.
The Sy2 NGC 5728 and the Sy1 NGC 1566 are not included here because of the lack
of near-IR nuclear fluxes for them, preventing  the SED fitting.}}
\label{clumpy_parameters}
\end{deluxetable*}

From the individual fits with the clumpy models, we  infer the following conclusions.
(1) For the majority of the Sy2 considered here, the clumpy models reproduce the observed SEDs, suggesting that
the high spatial resolution measurements are dominated by the reprocessed emission of the torus.
(2) The 10 \micron~silicate feature appears in absorption in the
fitted models for seven out of the eleven Sy2 (Figure \ref{sy2_1}) 
and is practically flat in the case of NGC 7172. The
unreliable NGC 1808 and NGC 7582 results, and also the Centaurus A fit
suggest silicate emission.  This is probably related with a characteristic
SED shape, characterized by low N/Q ratio plus L band excess (associated with
stellar contamination). See Section \ref{stellar_contamination} for details.
(3) The average number of clouds along an
equatorial ray is typically within the interval $N_0$ = [5, 15]. 
The exceptions are Centaurus A
($N_0$ = 2$\pm$1) and NGC 7582 ($N_0 <2$).  (4) High values of $\sigma$
are preferred from our fits ($\sigma$ = [50, 75]).  We only find values of
$\sigma < ~40$ for the galaxies Mrk 573, NGC 1808, and NGC 7582.  (5)
High values of the inclination angle of the torus are preferred, with
$i>40\degr$, in general.  (6) The combination of $i$, $\sigma$, and
$N_0$ determines the likelihood of encountering a cloud along the line
of sight to the central engine. In these cases, such encounters are
always likely, which is consistent with having the AGN blocked from
direct view, as expected in these Sy2.  (7) Relatively low values of
$\tau_{V}$ ($\leq$100) are found. The exception here are Centaurus A and NGC 1808, 
for which we find $\tau_{V} > 100$.  (8) The radial
density profile $q$ does not show any clear trend, and we find 
both low- and high-values within the considered interval.
All the above mentioned intervals or limits of the parameters correspond to  
median values. The reason for choosing median values instead of modes is that
the median gives a less biased information about the result, 
since it takes into account degeneracies, while the mode does not.

\subsubsection{Intermediate-Type Seyfert Individual Fits}
\label{individual_intermediate}

\begin{figure*}[!h]
\centering
{\par
\includegraphics[width=5.7cm,angle=90]{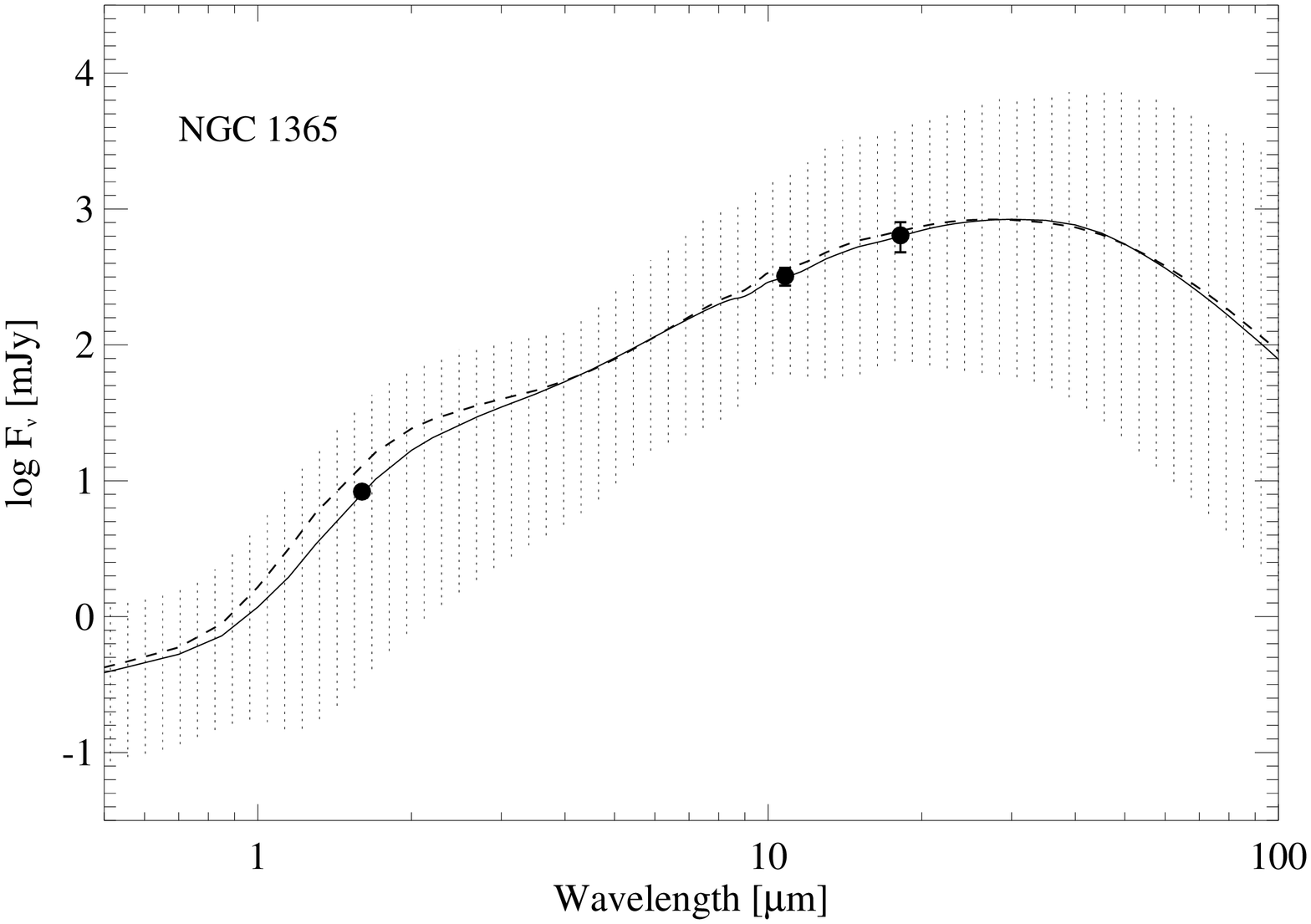}
\includegraphics[width=5.7cm,angle=90]{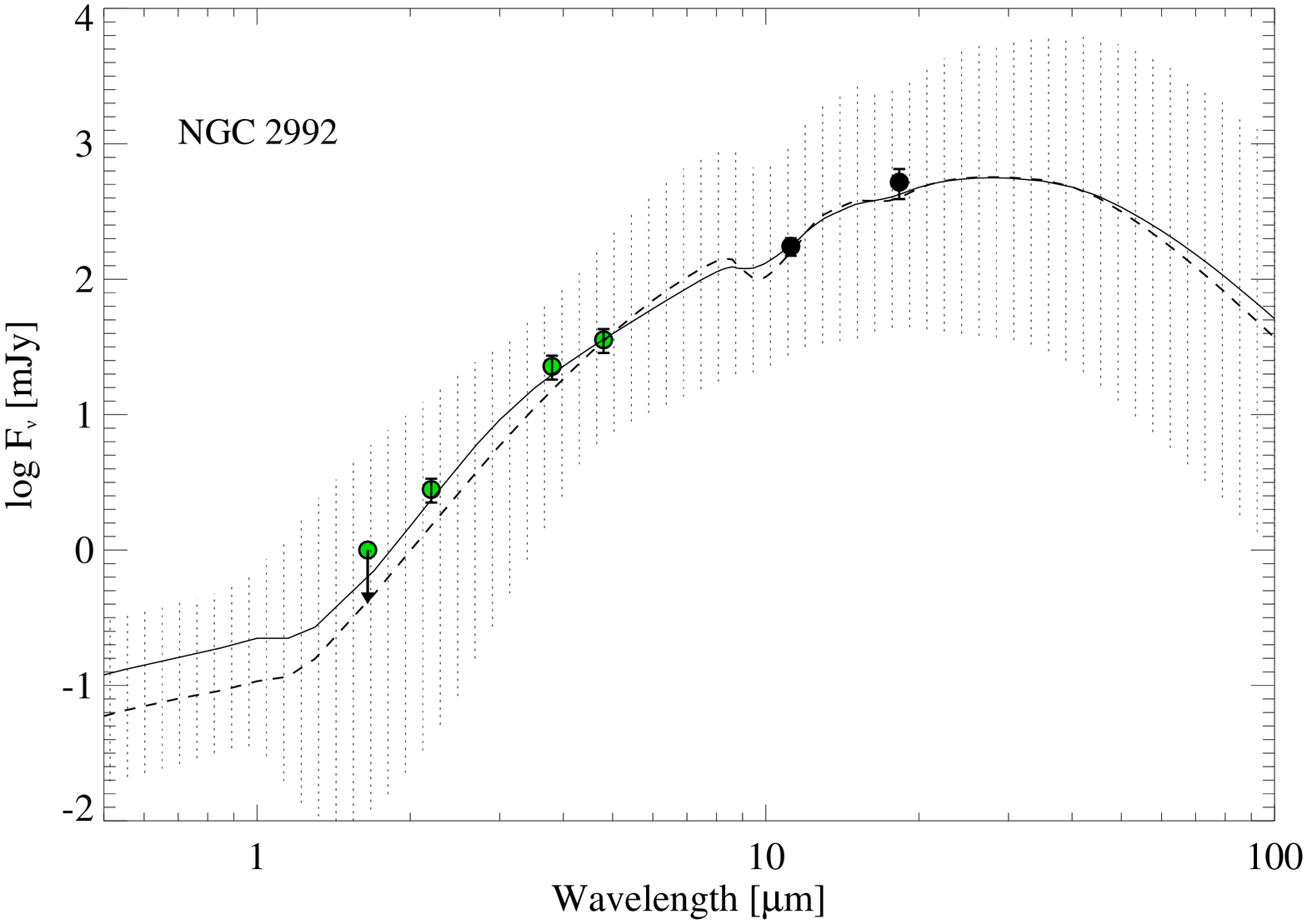}
\includegraphics[width=5.7cm,angle=90]{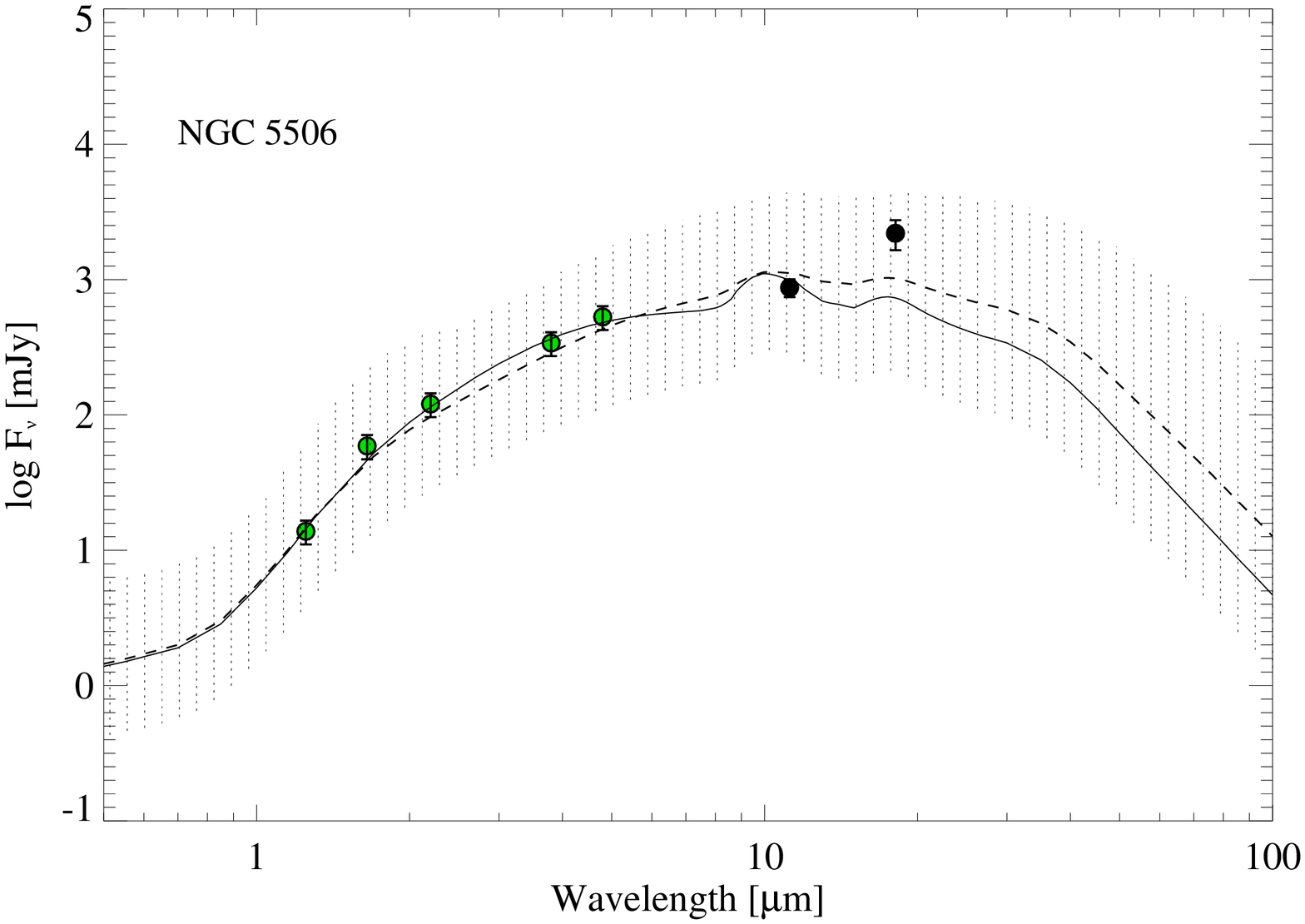}
\includegraphics[width=5.7cm,angle=90]{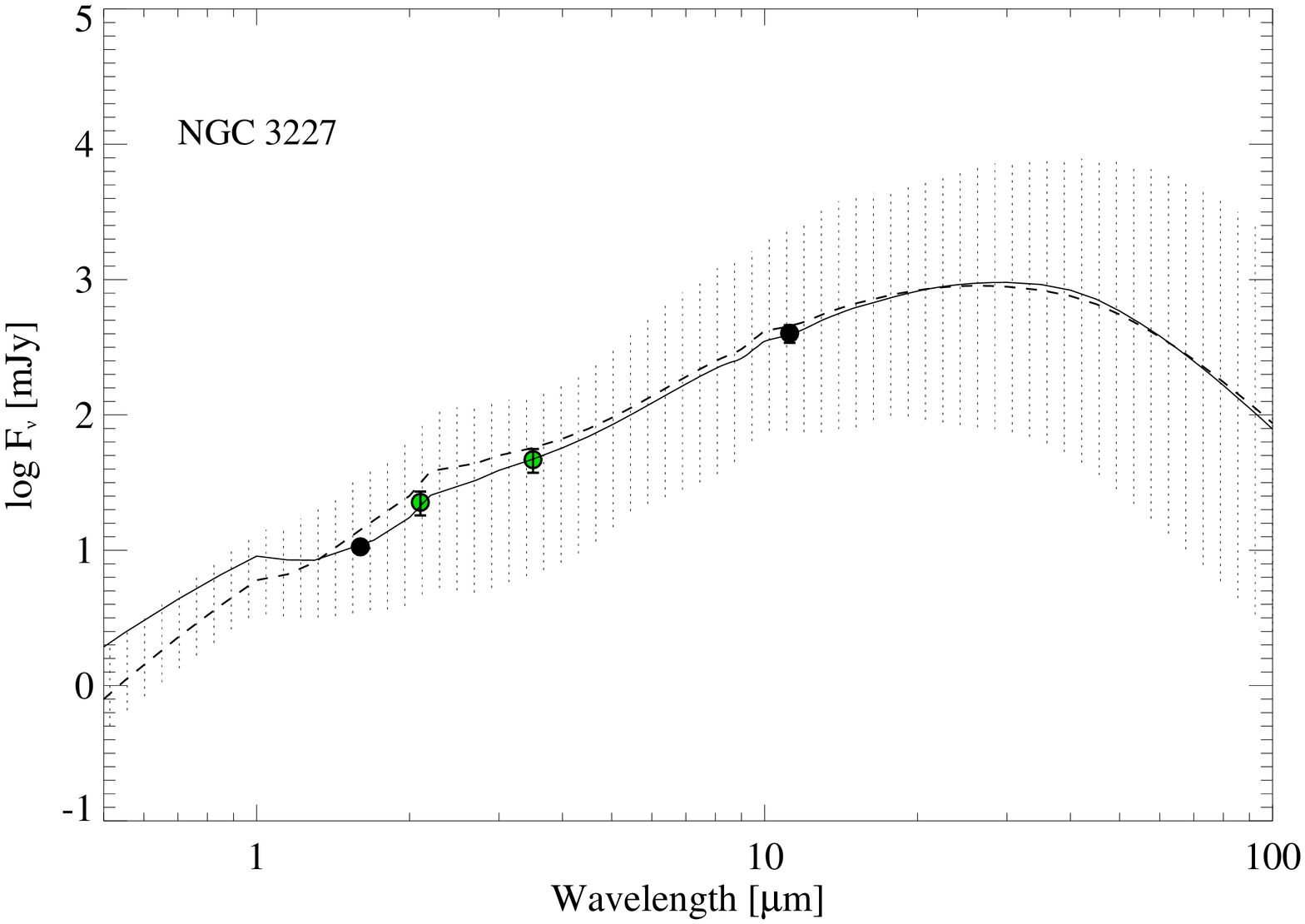}
\includegraphics[width=5.7cm,angle=90]{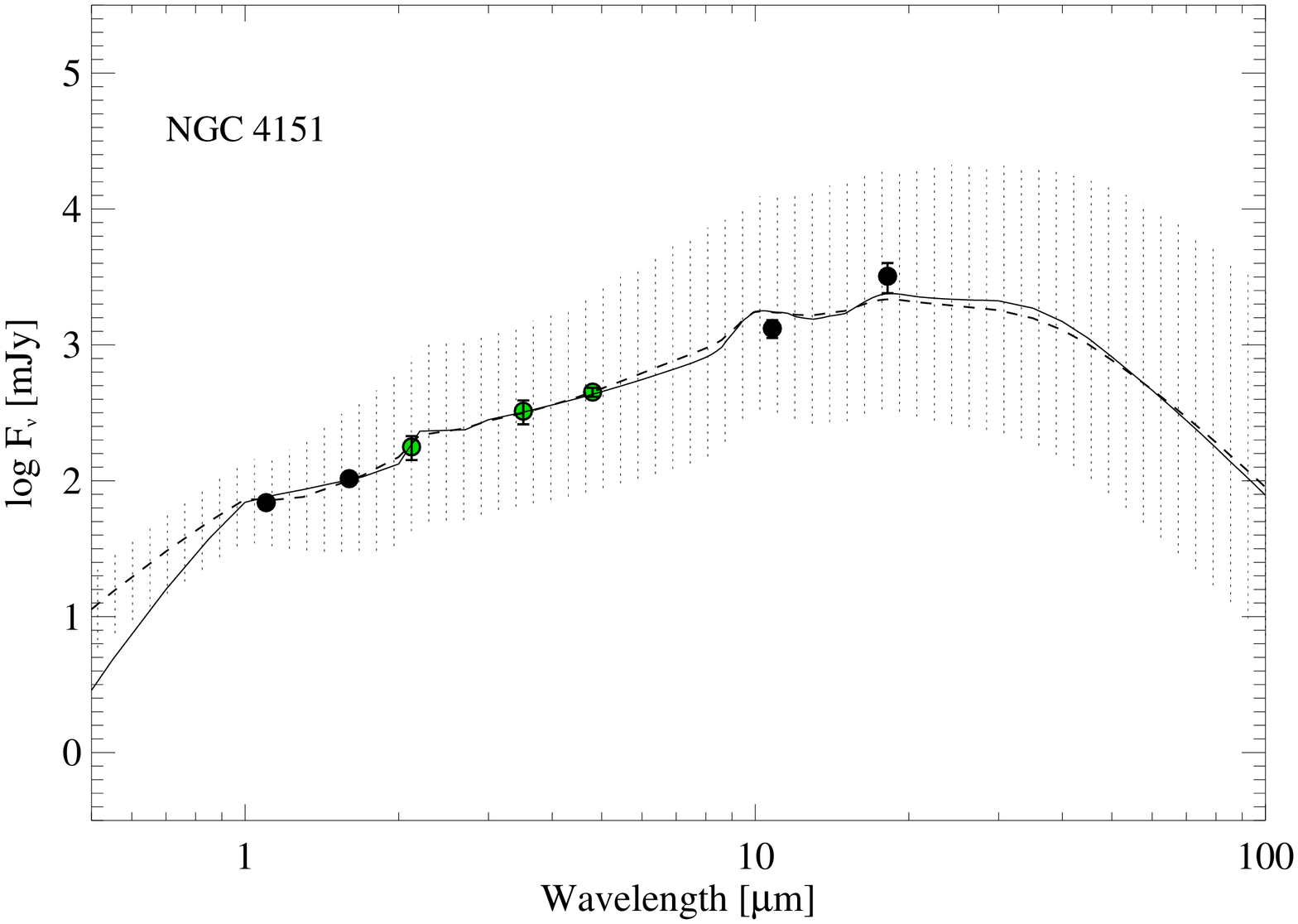}\par}
\caption{\footnotesize{Same as in Figure \ref{sy2_1} but for the intermediate-type Seyferts.}
\label{sy2_2}}
\end{figure*}

We fit the Sy1.8 and Sy1.9 nuclei the same way we fit the Sy2 SEDs,
namely considering only the reprocessed torus emission.  The optically
broad lines and the relatively strong near-IR fluxes of the Sy1.5,
however, suggest that some direct AGN contribution is also present in
these cases.  Thus, we include the intrinsic AGN emission as a broken power law
in these models. The
AGN scales self-consistently with the torus flux, and additional extinction
(separate from the clumpy torus) is a free parameter.
The resulting fits are plotted in Figure \ref{sy2_2}.
The Sy1 NGC 1566 is not included in the following because of the lack of near-IR 
nucler fluxes in the literature for it. 

From the individual fits of the intermediate-type Seyfert galaxies
with the clumpy models, we find similarities and differences with
the Sy2 results.  (1) Compared with the Sy2,
lower values of $\sigma$ are preferred in  the fits of 
both the 
Sy1.8 and Sy1.9 ($\sigma$ = [25, 45]) and Sy1.5  ($\sigma<35$).
(2) All the intermediate-type Seyferts result in a low number of clouds
along the equatorial plane ($N_0$ = [1,7]). 
(3) Low values of the inclination angle of the torus are found 
for Sy1.5 ($i<50\degr$) and also for the Sy1.8 NGC 1365. 
(4) $q$ does not show any clear trend for intermediate-type Seyferts, 
and both low- and high-values within the considered interval are found, as for Sy2.
(5) Low values of the optical depth per cloud are found for the two Sy1.9 galaxies ($\tau_{V}<70$), 
whilst for the Sy1.8 and Sy1.5, high values ($\tau_{V}>$100) are preferred.
(6) In contrast with the wide range of values of the optical extinction produced by the torus for Sy2, 
all the intermediate-type Seyferts show $A_{V}\la200$ mag.  
(7) Including some obscured AGN contribution fits the 
near-IR excess of the observed SEDs of the Sy1.5 NGC 4151 and NGC
3227.  The AGN is obscured by up to a few magnitudes at $V$ in each case. 
(8) Except for the case of NGC 2992, for which both the data and the fit are very
similar to those of Sy2 galaxies, the 10 \micron~silicate feature appears either in
weak emission or is absent in the intermediate-type Seyfert fits.
As in Section  \ref{individual}, the values of the parameters above mentioned
correspond to the median. Although the results from the modelling of Sy2 and 
intermediate-type Seyfert SEDs seem to point to different trends in their torus parameters, 
a larger set of objects is needed to clarify if these differences are really significant.

The results of our SED modelling of Seyfert galaxies are generally in good agreement with the findings of \citet{Nenkova08b},
who reproduce IR observations of both Type-1 and Type-2 Seyferts from the literature 
\citep{Sanders89,Elvis94,Alonso03,Prieto04,Mason06,Hao07,Netzer07} using clumpy torus models. They find $N_0 \sim 5$--15
clouds, $\tau_{V} \sim30$--100, $\sigma \sim30\degr$--50\degr, and the angular distribution of
the clouds having a soft edge (the only distribution we consider here), 
with  $q = 1$ or 2. They reproduce the observational data with compact torus sizes ($Y \sim 5$--10),
although they claim that the radial extent of the torus is poorly constrained from SED fitting.

\section{Interpretation of Observations and Model Fits}
\label{discussion}

\subsection{NIR Flux and Torus Inclination}
\label{nirflux}

The intermediate-type Seyferts exhibit characteristically higher H/N
ratios, and the Sy1.5 have flatter SEDs overall.  Hot dust emission
from the directly-illuminated faces of the clumps close to the central
engine contributes to the near-IR emission in all these cases.  In
addition, the Sy1.5 include a component of the direct AGN emission,
(i.e., the tail of the optical/ultraviolet power-law continuum), which
strongly flattens their IR SEDs.

In general, the relative near-IR flux depends sensitively on the torus
inclination angle.  The individual Sy2
show the same trend toward low near- to mid-IR ratios and therefore
more edge-on views ($i>40\degr$ considering the median values of the 
$i$ distributions, and $i\sim90\degr$ considering the modes). 
Only the unreliable measurements for NGC 7582
and NGC 1808 show high near- to mid-IR ratios and thus indicate more
pole-on views. 
In the context of the clumpy models, the presence of a cloud
along the line of sight, which may occur from any viewing angle,
results in Sy2 classification.
Cloud encounters are more likely at large inclination
angle, and thus we generally find 
larger values of the inclination angle for the Sy2.
However, we do not
suggest that all Sy2 are viewed exactly through the equatorial plane.
This small sample includes a number of galaxies whose tori are
known to be nearly edge-on (Circinus, Mrk 573, NGC 1386, and NGC 5506).
Hence there is a selection bias for highly inclined tori in our sample.  

The likelihood of a cloud
encounter more completely depends on the combination of viewing angle, $N_0$, and
$\sigma$. 
The preference for somewhat lower values of $i$, $N_0$, and $\sigma$ in the
intermediate-type Seyferts suggests that they present fewer clouds
along the line of sight, even allowing some direct detection of the
AGN continuum in the case of the Sy1.5.  This configuration also
increases the likelihood of unimpeded views of some 
directly-illuminated cloud faces (i.e., those on the ``back'' side
of the torus), which increases the near-IR flux.
These selection effects may account for our finding of larger values of $\sigma$ in the
Sy2 sample than \citet{Nenkova08b} report
as typical for all Seyfert types  ($\sim 30\degr$),
based on population statistics of Seyfert 1 and 2 galaxies 
\citep{Schmitt01,Hao05a}.

\subsection{Stellar Contamination in the Near-IR?}
\label{stellar_contamination}

One characteristic spectral shape the models cannot reproduce well exhibits
both strong emission toward the long-wavelength end of the near-IR 
(around L band) and simultaneously a steep N/Q ratio.
NGC 4388 is the most severe example of this problem; NGC 3281, NGC 5506,  and
NGC 7172 show similar effects. 
Because the formal uncertainties on the near-IR points are smaller than those
at Q, the  model fits tend to reproduce the near-IR
measurements while underpredicting the emission at Q in these cases.
(While we do not observe NGC 7172 at Q directly, 
the models would not accomodate the strong L band emission
combined with a rise of the SED toward 20 \micron,
assuming a typical N/Q ratio.)

We can interpret these results either as a failure of the models or as
evidence for near-IR contamination.  If the former, the models require
some modification to produce additional L band emission without otherwise
altering the spectral shapes significantly.  However, the fact that the most severe
discrepancies arise in the galaxies that have the
lowest resolution near-IR measurements
suggests instead that stellar or other extended contributions
can be significant even longwards of 2 \micron.
The L band observations of NGC 3281, NGC 4388, NGC 5506, and NGC 7172 were obtained
with natural seeing on 3 and 4 m telescopes, so even these
``nuclear'' fluxes cover scales up to 150 pc.

The possibility of stellar or other contamination of the Q band fluxes
cannot be discarded, especially those from the 4 m CTIO.  Where
strong stellar emission takes place in the nuclear region of the
galaxies, the emission of hot dust heated by this intense star
formation is an important contribution at $\sim$20 \micron~(see
e.g., \citealt{Helou04,Mason07}).  
Alternatively, the NLR can contribute more strongly at longer wavelengths, 
as the models of \citet{Groves06} show. 

For the CTIO observations the
resolution achieved in the Q band was $\sim1\arcsec$, which corresponds
to physical scales up to $\sim100$ pc 
(in NGC 1365, NGC 1808, and NGC 7582).  On the other hand, for
the Gemini data the resolutions are between 0.5\arcsec~and
0.6\arcsec~in the Q band.  Depending on the distance of each
source, this implies having resolution from $\sim10$ pc
(e.g., Centaurus A and Circinus) to $\sim150$ pc (e.g., IC 5063, NGC
3281, and Mrk 573).  Thus,  the steep N/Q ratios that the
models cannot correctly reproduce could be due to the combined effect
of stellar contamination around both the L and Q bands,
or NLR contamination that is preferentially more severe at Q.

\subsection{Torus Size}
\label{torus_size}
While the IR SEDs do not constrain the size of the torus, they are
consistent with the small torus of clumpy models, confined to scales
less than 10 pc.  Uniform density
models require the dusty torus to extend over large dimensions, to
provide cool dust that produces the IR emission (e.g., \citealt{Granato94}).  
Fundamentally, in
smooth distributions, the dust temperature is a monotonic function of
distance to the nucleus.  In contrast, in a clumpy distribution,
different dust temperatures can coexist at the same distance,
including cool dust at small radii \citep{Nenkova02}, so large tori
are not necessary.
Indeed, the flat probability distribution we found when allowing
$Y$ to be a free parameter (and as small as 5)
explicitly shows that small tori can produce the IR emission we observe.

In the clumpy models, with a steep radial density distribution
($q=2$), the SED is never sensitive to the outer torus extent because
the majority of the clouds are located very close to the nucleus for
all values of $Y$.  With flatter radial profiles, more clouds are
located farther from the central engine.  These model SEDs are then
sensitive to $Y$, but the variations are evident only at wavelengths
longer than 20 \micron. Far-IR spectral observations would provide
useful constraints \citep{Levenson07}, but high spatial resolution
measurements will remain critical because the far-IR emission from dust that
stars heat increases.

\subsection{10 \micron~Silicate Feature and Column Density Estimates}

The model fits to the photometric data yield spectral predictions.
The 10 \micron~silicate feature is of particular interest.
It is always weak 
(in emission or absorption)
in the clumpy models, unlike smooth-density
distributions in which it can appear in deep absorption
\citep{Dullemond05,Levenson07,Nenkova08b}.
Fundamentally, the weak silicate feature arises in the clumpy
model because both illuminated and dark cloud sides
contribute to the observed spectrum.
While most views of the Sy2 torus are through absorbing dark cloud faces,
silicate emission from some bright faces fills in
the feature, making it shallower.
The weak silicate absorption or
emission observed in Seyfert spectra indicate that a clumpy medium is
the dominant source of obscuration \citep{Levenson07}.
Indeed,  at high spatial resolution, where AGN emission
is practically isolated, the silicate absorption has never been observed to be deep in
Seyfert galaxies  \citep{Roche06}. 
The clumpy models reproduce the general observed trends of
silicate emission in quasars and absorption in obscured Seyfert galaxies
\citep{Roche91,Weedman05,Hao05,Siebenmorgen05,Sturm05,Levenson07}.
They can also accomodate the exceptional observations
of  silicate absorption in unobscured AGNs \citep{Roche91}
and emission in buried AGNs \citep{Mason09}, with different clump distributions.

We quantify the model silicate feature with its strength, which we define as
$\tau_{10 \micron}^{app} = \ln (F_{cont}) - \ln (F_{core})$, relative to the underlying 
continuum. 
We calculate $F_{cont}$ using a spline fit and $F_{core}$ 
allowing peak wavelength to vary, following  \citet{Sirocky08}.
The values of $\tau_{10 \micron}^{app}$ are reported in Table \ref{silicate}.
The silicate feature appears in shallow absorption in the fitted spectra
of seven out of the eleven Sy2. The remaining four sources are Centaurus A, 
NGC 7172, and the unreliable  cases of NGC 1808 and NGC 7582. 
All of them, except NGC 1808, have published mid-IR spectra that show the silicate feature
in absorption, contrary to our results (see Appendix \ref{indiv:sy2} for references). 
Excluding NGC 1808 and NGC 7582, the incorrect prediction of the silicate feature in 
very weak emission (practically flat) for NGC 7172 is probably due to 
the stellar contamination of the seeing-limited 
near-IR data described in Section \ref{stellar_contamination}. 
Indeed, the lack of Q band data for this galaxy 
limits the effectiveness of the modelling. 
The case of Centaurus A is more complicated, since the nuclear fluxes come
from a $\la$ 10 pc region, but the clumpy models reproduce the silicate feature in emission.
This could be related to the controversial nature of the near- and mid-IR emission of
this galaxy. According to several publications, synchrotron emission is significant or even dominant 
\citep{Bailey86,Turner92,Chiaberge01,Meisenheimer07}, although \citet{Radomski08} argue instead that the synchrotron contribution 
to the mid-IR flux is small.
For a detailed discussion of Centaurus A infrared emission nature and comparison with mid-IR spectroscopic data see
Appendix \ref{indiv:sy2}.

Regarding the intermediate-type Seyferts, the silicate feature appears in 
weak emission or absent, except for the case of NGC 2992 (a Sy1.9 with a SED 
very similar to those of Sy2. See Table \ref{slopes}), for which it is 
in absorption. 
In general, all the predicted silicate features, either in emission or absorption, are weak, 
which is a consequence of the contribution of both illuminated and shadowed 
cloud faces that the clumpy models predict for all the geometries considered.

For the OHMc dust extinction profile \citep{Ossenkopf92}, we can derive 
the apparent optical extinction 
using $A_V^{app} = \tau_{10 \micron}^{app} \times$ 23.6 for the galaxies with the silicate feature 
in absorption (Table \ref{silicate}; column 2).
This value can be compared with the optical extinction along the LOS derived from our fits using  
the mode values of N$_0$, $\sigma$, $\tau_V$, and $i$ parameters (A$_{V}^{LOS}$, Table \ref{clumpy_parameters}).
Modes are used here instead of medians in order to avoid lower limits that would make more complicated the comparison with $A_V^{app}$.   
As expected, we obtain $A_V^{LOS} >> A_V^{app}$ (see Table \ref{silicate}) since on the basis of the clumpy models, the true optical depth 
along LOS does not coincide with the apparent optical depth measured from the absorption silicate feature ($A_V^{app}$). 
The directly illuminated faces of the clouds seen along the LOS refill
the absortion silicate band (produced by the obscuring material), thus
producing lower apparent optical depth values. On the other hand,
A$_{V}^{LOS}$ takes into account the full model cloud distribution
along the LOS.

The columns of material implied in the X-ray absorption (Table \ref{silicate}; column 5) 
should be in principle comparable to 
or larger than those inferred from the fit of IR data with the models
(Table \ref{silicate}; column 4). Thus, we estimated $N_H^{LOS}$ from the optical depth along the LOS, using the relationship 
$N_H^{LOS}/A_V^{LOS}$ = 1.9$\times10^{21}~cm^{-2}~mag^{-1}$ from \citet{Bohlin78}.  
Larger absorbing column densities derived from X-ray measurements ($N_{H}^{X-ray}$) compared
with $N_H^{LOS}$  suggest 
that the X-ray emitting regions are affected by a larger amount of extinction than the mid-IR ones. 
As it was firstly noted by \citet{Granato97}, dust-free absorption unrelated to the torus
could be the main source of X-ray obscuration. Thus, the X-ray absorber would be gas located
inside the dust sublimation radius.
This is the case for most of the sample, with $N_{H}^{X-ray} / N_H^{LOS}$ ranging from 2 to 25 
(excluding the unrealistic NGC 7582 result). 
On the contrary, comparable values of  $N_H^{LOS}$ and $N_{H}^{X-ray}$ (those of Centaurus A, NGC 7172, and NGC 5506)
mean that the dust-free absorption in these galaxies is low. 
We derive unrealistic results for IC 5063 and NGC 2992, for which 
$N_{H}^{X-ray} < N_H^{LOS}$. This could be due to the known intrinsic X-ray variability of these galaxies \citep{Polleta96,Turner97,Yaqoob97,Gilli00}.
In the case of variable sources, we cannot directly compare
the column density predictions from IR data modelling and 
X-ray measurements, since the data have not been taken simultaneously.

\begin{deluxetable*}{lcccrrc}
\tablewidth{0pt}
\tablecaption{Silicate Feature Predictions}
\tablehead{
\colhead{Galaxy} & \colhead{$\tau_{10\micron}^{app}$} & \multicolumn{1}{c}{$A_{V}^{app}$} & \multicolumn{1}{c}{$A^{LOS}_{V}$} & 
\multicolumn{1}{c}{$N_H^{LOS}$} & \multicolumn{1}{c}{$N_H^{X-rays}$} & \colhead{Reference(s)} \\ 
 & & (mag) & (mag) & (cm$^{-2}$) & (cm$^{-2}$) }
\startdata
Centaurus A &   -0.67   &\nodata& 260   & 4.9$\times10^{23}$   &	1.5$\times10^{23}$    &  a \\
Circinus    &    0.86   & 20    & 295   & 5.6$\times10^{23}$   &	4.0$\times10^{24}$    &  b \\
IC5063	    &    0.94   & 22    & 580   & 1.1$\times10^{24}$   &	2.6$\times10^{23}$    &  c \\
Mrk 573	    &    0.44   & 10    & 150   & 2.8$\times10^{23}$   &     $>$1.0$\times10^{24}$    &  d \\
NGC 1386    &    0.41   & 10    & 410   & 7.8$\times10^{23}$   &	1.5$\times10^{24}$    &  e \\  
NGC 1808    &   -0.21   &\nodata&  10   & 1.9$\times10^{22}$   &	3.1$\times10^{22}$    &  f \\
NGC 3081    &    0.91   & 21    & 120   & 2.3$\times10^{23}$   &	6.3$\times10^{23}$    &  d \\
NGC 3281    &    0.21   & 5     &  50   & 9.5$\times10^{22}$   &	2.0$\times10^{24}$    &  g \\  
NGC 4388    &    0.51   & 12    &  50   & 9.5$\times10^{22}$   &	2.6$\times10^{23}$    &  h \\
NGC 7172    & \nodata   &\nodata&  50   & 9.5$\times10^{22}$   &	8.3$\times10^{22}$    &  i \\  
NGC 7582    &   -0.61   &\nodata&   1   & 1.9$\times10^{21}$   &	1.4$\times10^{23}$    &  j \\  
\hline
NGC 1365    &    0.00   &\nodata&  10   & 1.9$\times10^{22}$   &	4.8$\times10^{23}$    &  k \\  
NGC 2992    &    0.74   & 17    & 130   & 2.5$\times10^{23}$   &	8.0$\times10^{21}$    &  l \\  
NGC 5506    &   -0.21   &\nodata&  30   & 5.7$\times10^{22}$   &	3.6$\times10^{22}$    &  m \\  
\hline
NGC 3227    &   -0.09   &\nodata&  10   & 1.9$\times10^{22}$   &	5.3$\times10^{22}$    &  n \\  
NGC 4151    &   -0.31   &\nodata&  10   & 1.9$\times10^{22}$   &	6.9$\times10^{22}$    &  o \\
\enddata
\tablecomments
{\footnotesize{Silicate strength $\tau_{10\micron}^{app}$ measured
from the fitted models; negative values indicate a silicate feature in
emission.  It was not possible to properly measure $\tau_{10\micron}^{app}$
for NGC 7172 because the feature is in extremelly weak emission, and
self-absorbed. 
Corresponding apparent optical extinction, $A_{V}^{app}$, assumes
$A_{V}^{app}/\tau_{10\micron}^{app} = 23.6$, 
and $A^{LOS}_{V}$ comes from the modelling. $N_H^{LOS}$ is
calculated from $N_H/ A_V^{LOS} = 1.9\times10^{21} 
\mathrm{cm^{-2}~mag^{-1}}$ \citep{Bohlin78}. $N_H^{X-ray}$ values are
taken from the literature.}}  
\tablerefs{\footnotesize{(a)
\citet{Markowitz07}; (b) \citet{Soldi05}; (c) \citet{Turner97};
(d) This work; (e) \citet{Levenson06}; (f) \citet{Jimenez05}; (g)
\citet{Vignali02}; (h) \citet{Elvis04}; (i) \citet{Awaki06}; (j)
\citet{Turner00}; (k) \citet{Risaliti05}; (l) \citet{Yaqoob97};
(m) \citet{Lamer00}; (n) \citet{Lamer03}; (o)
\citet{Beckmann05}.}}
\label{silicate}
\end{deluxetable*}

\subsection{AGN Luminosities}

The clumpy model fits yield the bolometric luminosity of the intrinsic
AGN in each case.  Combining this value with the torus luminosity, we derive the
reprocessing efficiency of the torus.  The vertical shift parameter
scales with AGN luminosity, which we report in Table \ref{lum} (L$_{bol}^{AGN}$; column 1) for the
median models. 
Integrating the corresponding model torus emission yields the torus luminosity,
(L$_{bol}^{tor}$; column 2).

The Sy2 tori in our sample are efficient reprocessers, absorbing and re-emitting the
majority of the intrinsic AGN luminosity in the IR (with mean
efficiency $0.6\pm0.3$).  Among the Sy2, only Centaurus A, Mrk 573, and the
unreliable results of NGC 1808 and NGC 7582 indicate efficiencies
below 50\% (the mean efficiency is $0.7\pm0.3$, excluding NGC 1808 and NGC 7582).  
With the limited sample size, 
the efficiencies of the intermediate types ($0.4 \pm 0.3$) are formally 
comparable to those of Sy2, but the measurements point out that
possibly Type-2 tori reprocess radiation more efficiently than Type-1 tori, 
although the measured differences between both Seyfert types are not 
significant.
A larger subset of Sy1 and intermediate-type Seyferts is needed 
to clarify if tori of different Seyfert types are genuinely different.
The reprocessing fraction depends primarily on the
total number of clouds available to absorb the incident radiation,
which increases with the model parameters $N_0$ and $\sigma$.  If all
Seyfert nuclei are identical, only the viewing angle would change with
the classification, not the properties of the torus itself.  While
these results are limited, they suggest instead
that classification is biased.  Specifically, when the torus covers a
smaller fraction of the sky viewed from the central engine (which
results in a lower efficiency), the galaxy is less likely to be
classified as a Sy2 (because the AGN is incompletely blocked from more
lines of sight).  Such a selection effect would also extend to Sy1; IR
observations of these galaxies would provide useful tests of this
suggestion against strong forms of AGN unification.

\begin{deluxetable*}{lcccccc}
\tabletypesize{\footnotesize}
\tablewidth{0pt}
\tablecaption{Bolometric Luminosity Predictions}
\tablehead{
\colhead{Galaxy}  & \multicolumn{1}{c}{L$_{bol}^{AGN}$} & \multicolumn{1}{c}{L$_{bol}^{tor}$} & \colhead{L$_{bol}^{tor} / L_{bol}^{AGN}$} & 
\colhead{$R_{o}$ (pc)} &  \colhead{L$_{X bol}^{AGN}$} &\colhead{L$_{X bol}^{AGN}$ / L$_{bol}^{AGN}$}  \\
 & (erg~s$^{-1}$) & (erg~s$^{-1}$) & & &  (erg~s$^{-1}$) & }
\startdata
Centaurus A & 5.5  $\times$ 10$^{42}$  & 1.6 $\times$ 10$^{42}$    &  0.29  &    0.44  & 1.3 $\times$ 10$^{43}$ &   2	 \\
Circinus    & 1.0  $\times$ 10$^{43}$  & 8.1 $\times$ 10$^{42}$    &  0.81  &    0.60  & 1.2 $\times$ 10$^{43}$ &   1.2  \\
IC 5063     & 2.4  $\times$ 10$^{44}$  & 2.1 $\times$ 10$^{44}$    &  0.87  &    2.94  & 1.7 $\times$ 10$^{44}$ &   0.7  \\
Mrk 573	    & 4.3  $\times$ 10$^{44}$  & 6.6 $\times$ 10$^{43}$    &  0.15  &    3.93  & 4.4 $\times$ 10$^{44}$ &   1.0  \\
NGC 1386    & 3.4  $\times$ 10$^{42}$  & 2.0 $\times$ 10$^{42}$    &  0.59  &    0.35  & 1.3 $\times$ 10$^{43}$ &   4	 \\
NGC 1808    & 6.8  $\times$ 10$^{42}$  & 3.1 $\times$ 10$^{42}$    &  0.45  &    0.49  & 2.2 $\times$ 10$^{41}$ &   0.03 \\
NGC 3081    & 1.0  $\times$ 10$^{43}$  & 8.9 $\times$ 10$^{42}$    &  0.89  &    0.60  & 1.0 $\times$ 10$^{44}$ &   10	 \\
NGC 3281    & 7.4  $\times$ 10$^{43}$  & 5.3 $\times$ 10$^{43}$    &  0.72  &    1.63  & 3.0 $\times$ 10$^{44}$ &   4	 \\
NGC 4388    & 2.1  $\times$ 10$^{43}$  & 1.7 $\times$ 10$^{43}$    &  0.81  &    0.87  & 1.5 $\times$ 10$^{44}$ &   7	 \\
NGC 7172    & 8.5  $\times$ 10$^{42}$  & 6.7 $\times$ 10$^{42}$    &  0.79  &    0.55  & 1.1 $\times$ 10$^{44}$ &   13	 \\
NGC 7582    & 1.1  $\times$ 10$^{44}$  & 1.1 $\times$ 10$^{43}$    &  0.10  &    1.99  & 9.8 $\times$ 10$^{43}$ &   0.9  \\  
\hline	      	 		   	    	       
NGC 1365    & 1.7  $\times$ 10$^{43}$  & 1.1 $\times$ 10$^{43}$    &  0.65  &    0.78  & 3.0 $\times$ 10$^{43}$ &   1.8  \\
NGC 2992    & 2.8  $\times$ 10$^{43}$  & 1.5 $\times$ 10$^{43}$    &  0.53  &    1.00  & 3.0 $\times$ 10$^{43}$ &   1.1  \\
NGC 5506    & 7.3  $\times$ 10$^{44}$  & 5.1 $\times$ 10$^{43}$    &  0.07  &    5.13  & 2.2 $\times$ 10$^{44}$ &   0.3  \\
\hline	      	 		   	    	       
NGC 3227    & 2.0  $\times$ 10$^{43}$  & 1.1 $\times$ 10$^{43}$    &  0.55  &    0.85  & 3.8 $\times$ 10$^{43}$ &   1.9  \\
NGC 4151    & 1.4  $\times$ 10$^{44}$  & 2.2 $\times$ 10$^{43}$    &  0.16  &    2.24  & 1.7 $\times$ 10$^{44}$ &   1.2  \\
\enddata     
\tablecomments{\footnotesize{
    Bolometric luminosities corresponding to AGN luminosity and
    integrated flux of the model emission (i.e., reprocessed
    luminosity). Luminosities are good to a factor of 2.  
    Columns 4 and 5 correspond to the fraction of energy emitted by the AGN
    that is reprocessed by the torus (L$_{bol}^{tor} / L_{bol}^{AGN}$)
    and outer radius of the torus calculated using L$_{bol}^{AGN}$.
    Absorption-corrected 2--10 keV X-ray luminosities are taken from
    the literature (references  in Table \ref{silicate}).
    L$_{X bol}^{AGN}$ is derived from 20$\times$L$_{X}^{AGN}$.}}
\label{lum}
\end{deluxetable*}

The outer size of the torus scales with the AGN bolometric luminosity:
$R_{o} = Y R_{d}$, so assuming a dust sublimation temperature of 1500
K, $R_o= 6 (L_{bol}^{AGN}/10^{45})^{0.5}$ pc (with fixed $Y = 15$).  We
find that all tori in our sample have outer extents less than 5 pc
(Table \ref{lum}), in agreement with recent mid-IR
direct imaging of nearby Seyferts \citep{Packham05,Radomski08} and 
also interferometric observations \citep{Jaffe04,Tristram07,Meisenheimer07,Raban09}.  
Specifically, the estimated outer radii
for Circinus ($R_{o} = 0.6$ pc) and Centaurus A ($R_{o} = 0.4$ pc)
are close to the upper limits reported by
\cite{Tristram07} and \cite{Meisenheimer07}, of 1 and 0.3 pc,
respectively.  The uncertainty of these radial sizes is large, mostly
because the unknown value of $Y$ enters the calculation linearly,
but the consistency with other measurements further supports the
assumption of $Y=15$.

The bolometric luminosity of the intrinsic AGN derived from the fits
(Table \ref{lum}; column 1) can be directly compared with the
bolometric luminosities derived from the
absorption-corrected 2-10 keV luminosities
compiled from
the literature (Table \ref{lum}; column 5). To get L$_{X bol}^{AGN}$ from the intrinsic 
2-10 keV luminosities we applied
a conversion factor of 20 \citep{Elvis94}. In general, we find
comparable values of L$_{Xbol}^{AGN}$ and L$_{bol}^{AGN}$ within the
considered errors for L$_{bol}^{AGN}$. Exceptions are NGC 3081, NGC 4388, 
and NGC7172, for which L$_{Xbol}^{AGN} \sim 10 \times
L_{bol}^{AGN}$, and the unreliable NGC 1808, in the opposite direction. 

\section{Conclusions}

We report subarcsecond resolution mid-IR fluxes for eighteen Seyfert galaxies. 
These nuclear fluxes, in combination 
with published near-IR measurements at comparable resolution
are used to construct spectral energy 
distributions corresponding to torus emission. 
We analyze and fit these SEDs with the clumpy dusty torus models 
of \citet{Nenkova02,Nenkova08a,Nenkova08b}. The main results are summarized in the following.

\begin{itemize}

\item
The high spatial resolution mid-IR nuclear fluxes reported in this work provide a spectral shape of the individual SEDs 
that is different from that of large aperture data SEDs (on scales of a few arcseconds).  

\item 
The shape of the average Sy2 SED, constructed using only pure Sy2 galaxies and either diffraction- or near-diffraction-limited data,  rises  
steeply towards the mid-IR, with an IR slope (from $\sim$1 \micron~to 18 \micron)  
$\alpha_{IR} = 3.1\pm0.9$. 

\item The individual Sy2 SEDs are typically steep through the IR, with $\alpha_{IR}$ ranging from 1.8 to 3.8. 
Considering separately the near-IR (from $\sim$ 1 to $\sim$ 9 \micron) and the mid-IR ($\sim$ 10 to $\sim$  18 \micron) slopes, 
we find $\alpha_{NIR} \ga \alpha_{MIR}$, with $\alpha_{NIR}=3.6\pm0.8$ and $\alpha_{MIR}=2.0\pm0.2$ measured from the average Sy2 SED.  

\item 
The IR SEDs of the intermediate-type Seyferts in the sample are
flatter ($\alpha_{IR}$ = 2.0$\pm$0.4 for Sy1.8 and 1.9 and $\alpha_{IR}$ = 1.6$\pm$0.3 for Sy1.5) 
and present larger N/Q band ratios than those of Seyfert 2 galaxies. 
The near-IR excess that flattens
the SED of all intermediate-type Seyferts is due to
the contribution of hot dust from the illuminated faces of the clouds,
while direct AGN emission is also important in the case of Sy1.5.

\item 
The clumpy models of \citet{Nenkova08a,Nenkova08b} successfully reproduce
the SEDs of both the Seyfert 2 and the intermediate-type Seyferts.  
The models accommodate the range of $\alpha_{IR}$ observed in the former, 
while providing optically thick obscuration along the line of sight.

\item The IR SEDs do not constrain the
radial extent of the torus, $Y$, and this parameter remains  
uncorrelated with the other
  parameters.   Because the outer torus contains the coolest material,
high angular resolution measurements at wavelengths longer than 15
  \micron~are needed to reveal significant variations in the torus size.

\item In the individual Sy2 fits, we generally find the number of clouds within
  the interval $N_0$ = [5, 15], the width of the angular distribution
  of clouds within $\sigma$ = [50\degr, 75\degr], high values of the
  inclination angle of the torus ($i$=[40\degr, 90\degr]), and optical depths per cloud $\tau_{V}
  < 100$.  The radial density profile $q$ does not show any clear trend
  within the considered interval ([0, 3]).

\item We find small number of clouds in the
individual intermediate-type Seyfert fits ($N_0 = $ [1,7]).
Low values of $\sigma$ are preferred from the fits of Sy1.8 and Sy1.9 
($\sigma$ = [25\degr, 50\degr]) and even lower for Sy1.5 ($\sigma < 35\degr$).
We require direct (though extinguished) emission of the AGN 
to reproduce the near-IR excess observed in the SEDs of Sy1.5.

\item Views of Sy2 are more inclined than those of the Sy1.5 galaxies.
More importantly, the larger values of $N_0$ and $\sigma$ in the Sy2 models
suggest that these central engines are blocked from direct view along more
lines of sight, in contrast to the intermediate-type Seyferts, which present
more clear views of the AGN. Due to the limited size of the analyzed sample, 
these differences are not significant.

\item Although the fits are based on photometric data, 
the models predict the behavior of the  10 \micron~silicate spectral feature.
Most of the individual Sy2 galaxies show shallow silicate absorption.
In all but one of the intermediate-type Seyferts, the feature is either in shallow
emission or absent. 

\item The columns of material responsible for the X-ray absorption 
are larger than those inferred from the model fits for most of the galaxies in the sample,
which is consistent with 
X-ray absorbing gas located within the dust sublimation radius, 
whereas the mid-IR flux arises from an area farther from the accretion disc.

\item The combination of large number of clouds and high values of $\sigma$ 
results in tori that  more efficiently reprocess the incident nuclear radiation
than those with lower values of these parameters.

\item In the models, the outer radial extent of the torus scales with the
AGN luminosity, and we find the tori to be confined to scales less than 5 pc.

\end{itemize}

\newpage

\begin{figure*}[!h]
\centering
{\par
\includegraphics[width=5.7cm,angle=90]{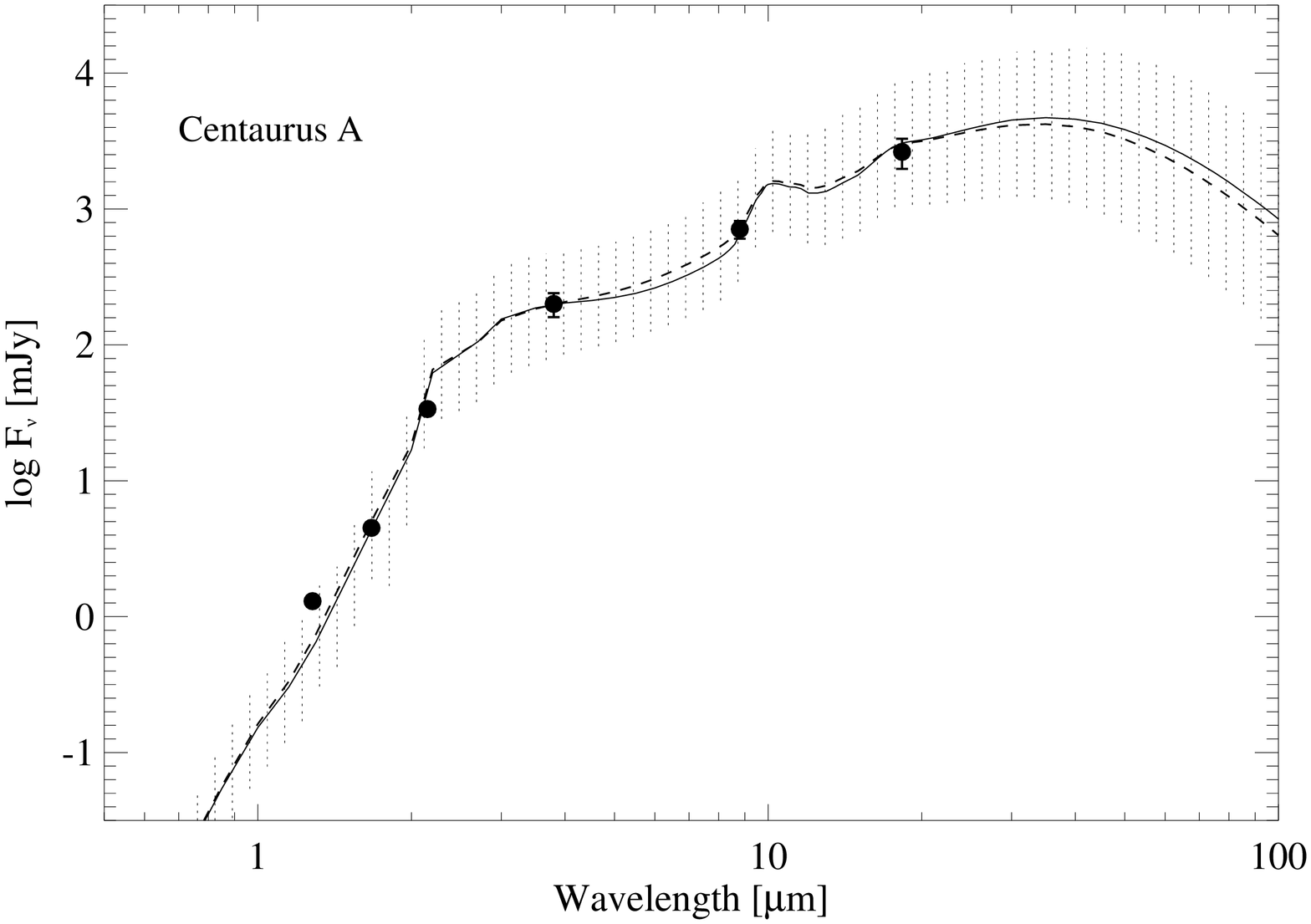}
\includegraphics[width=5.7cm,angle=90]{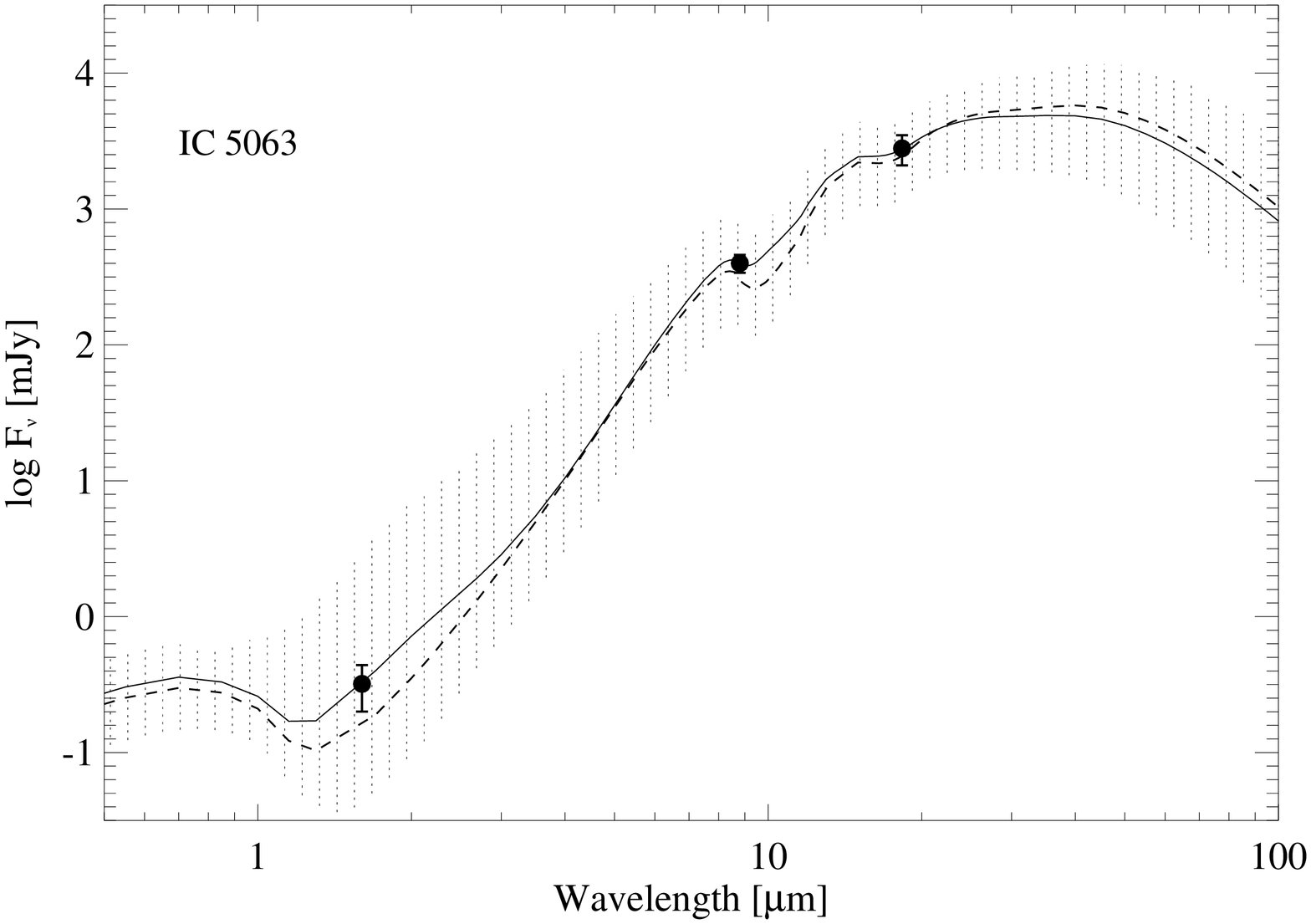}
\includegraphics[width=5.7cm,angle=90]{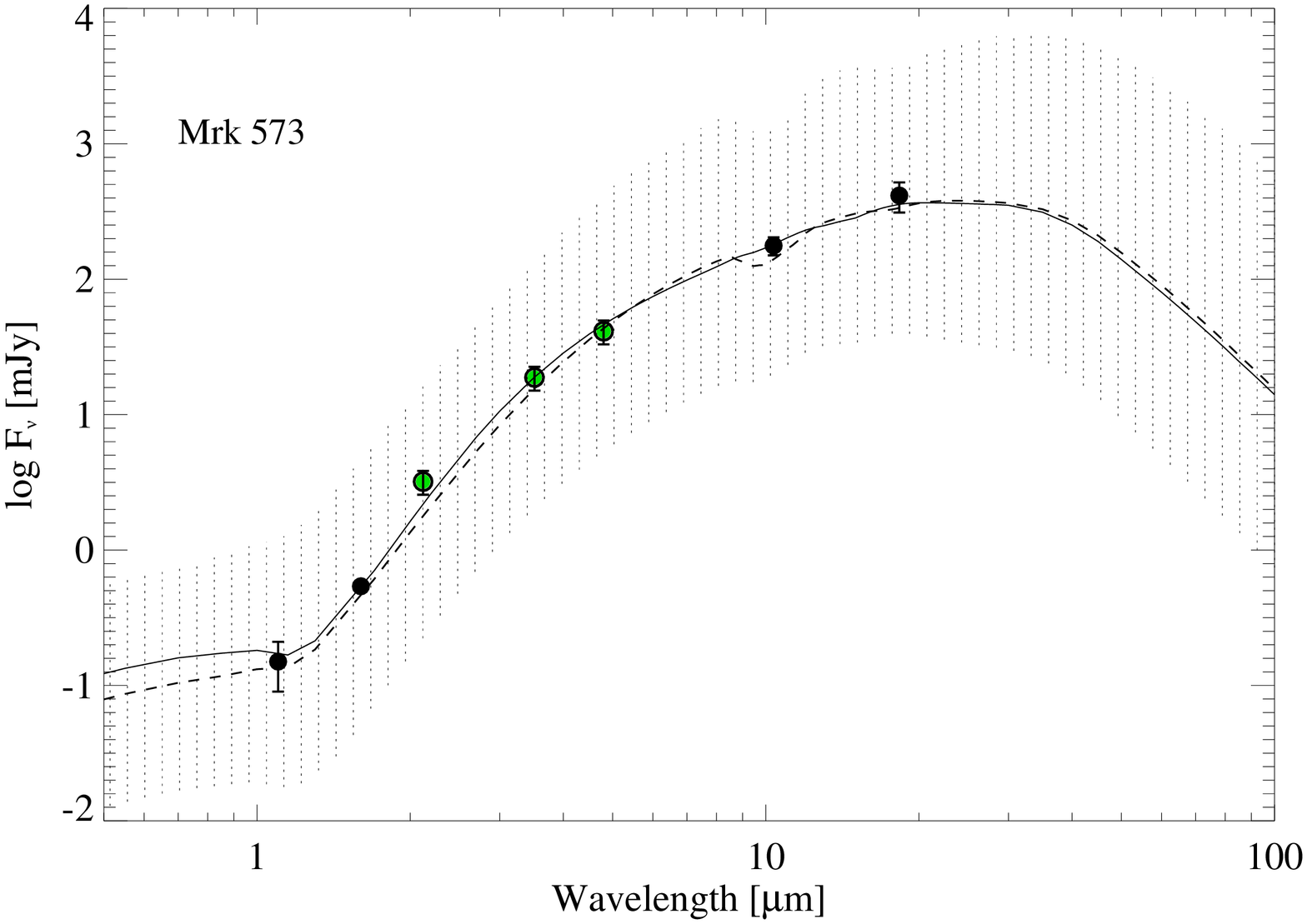}
\includegraphics[width=5.7cm,angle=90]{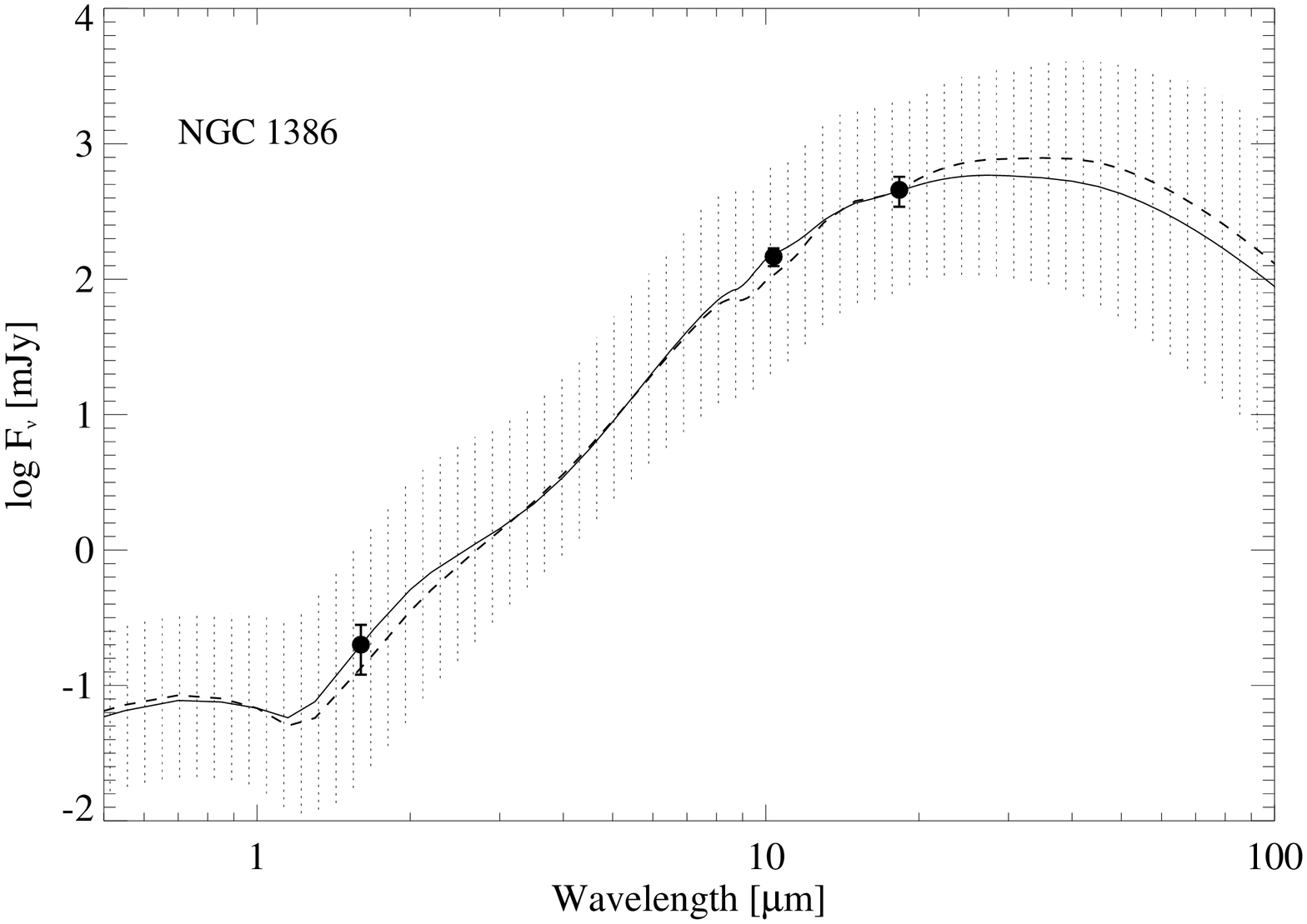}
\includegraphics[width=5.7cm,angle=90]{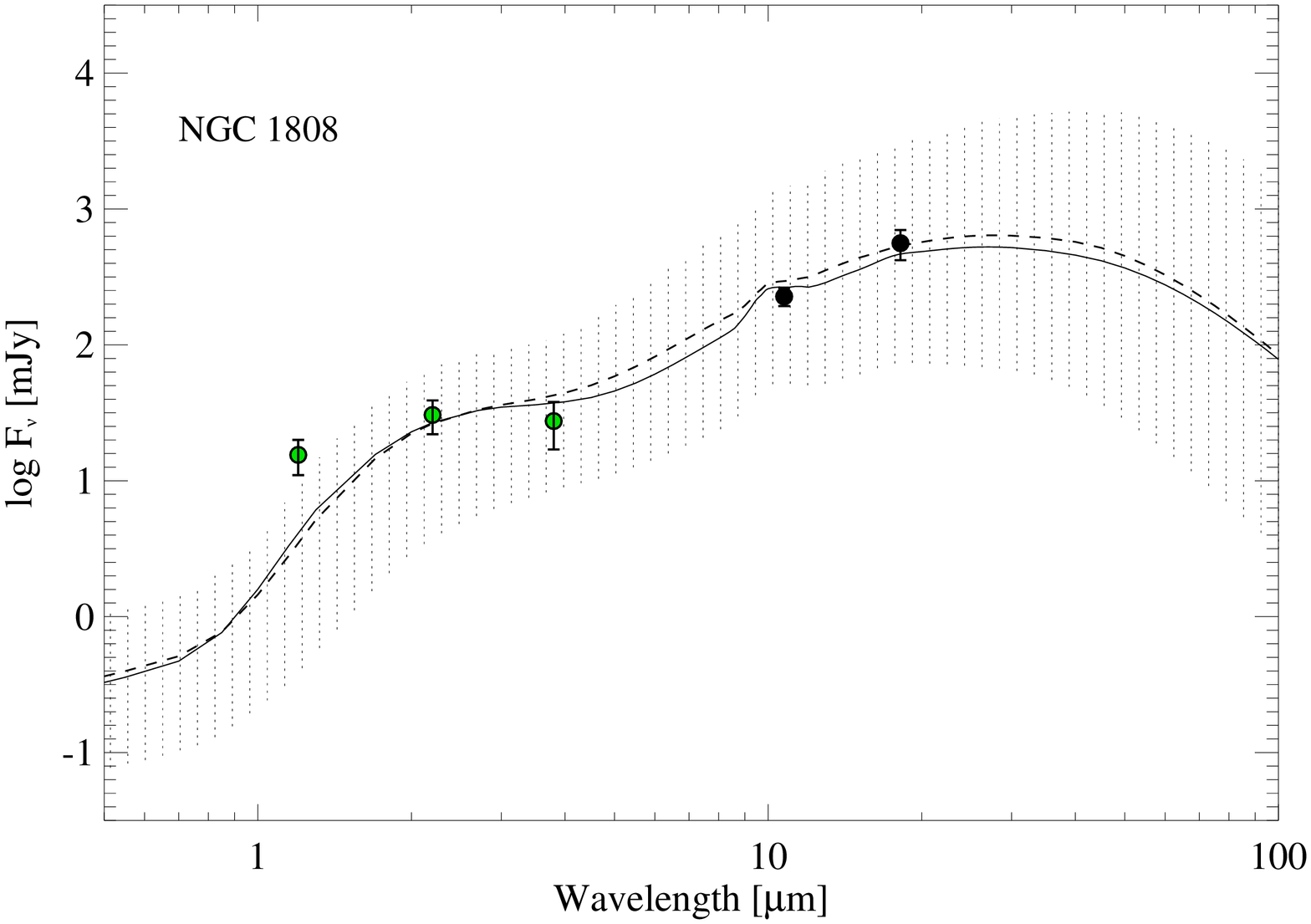}\par}
\label{sy2_1}
\end{figure*}

\begin{figure*}[!h]
\centering
{\par
\includegraphics[width=5.7cm,angle=90]{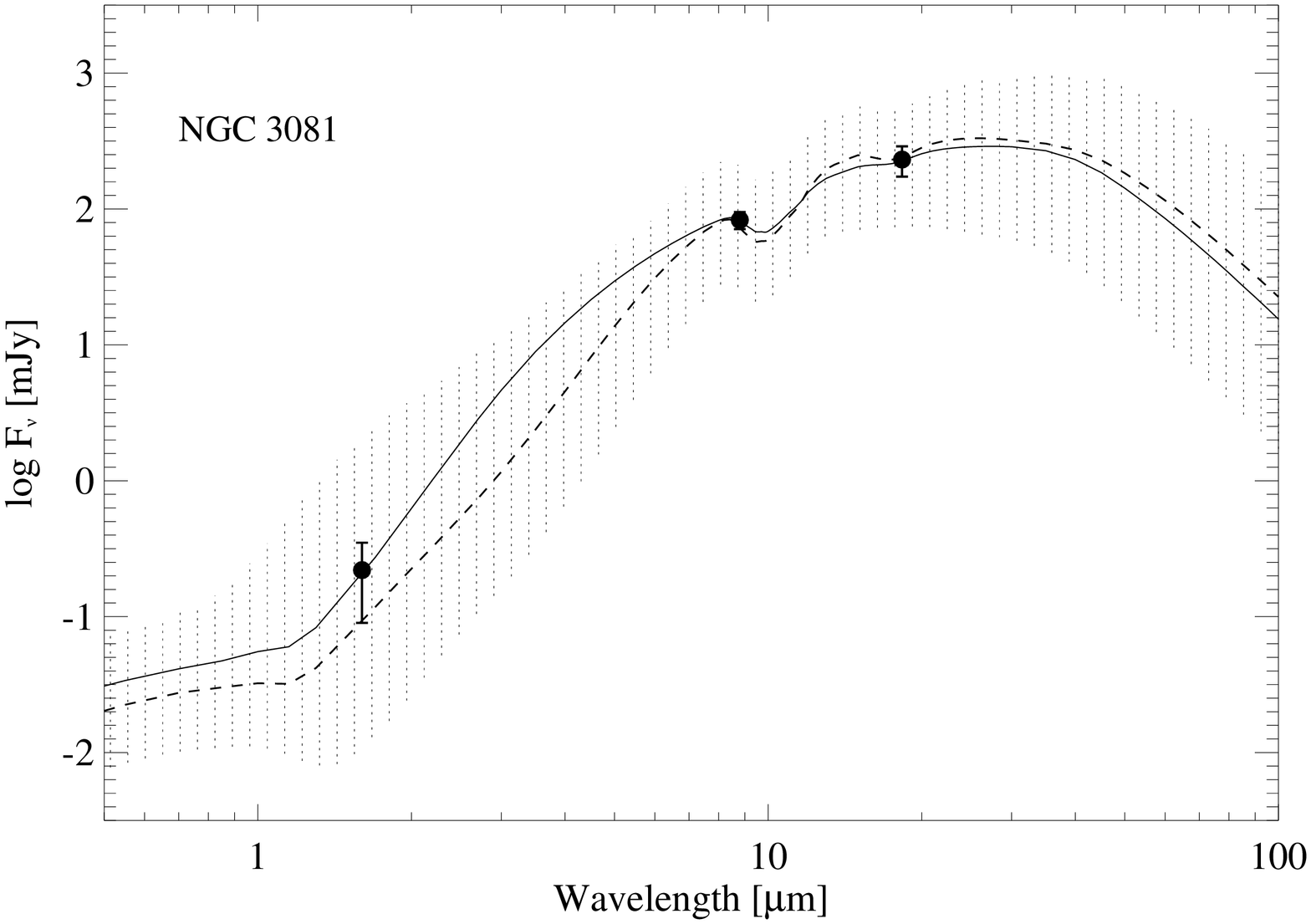}
\includegraphics[width=5.7cm,angle=90]{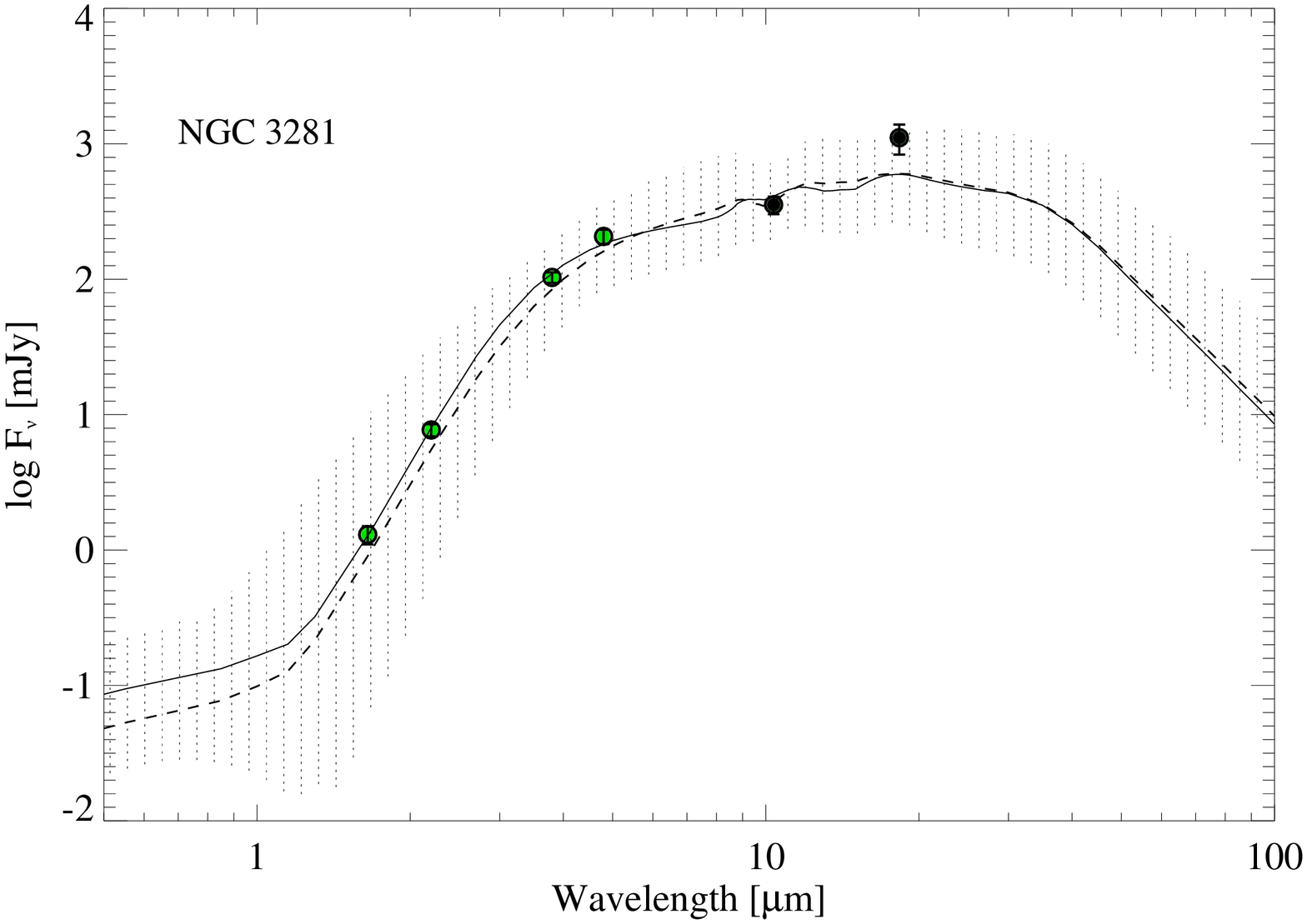}
\includegraphics[width=5.7cm,angle=90]{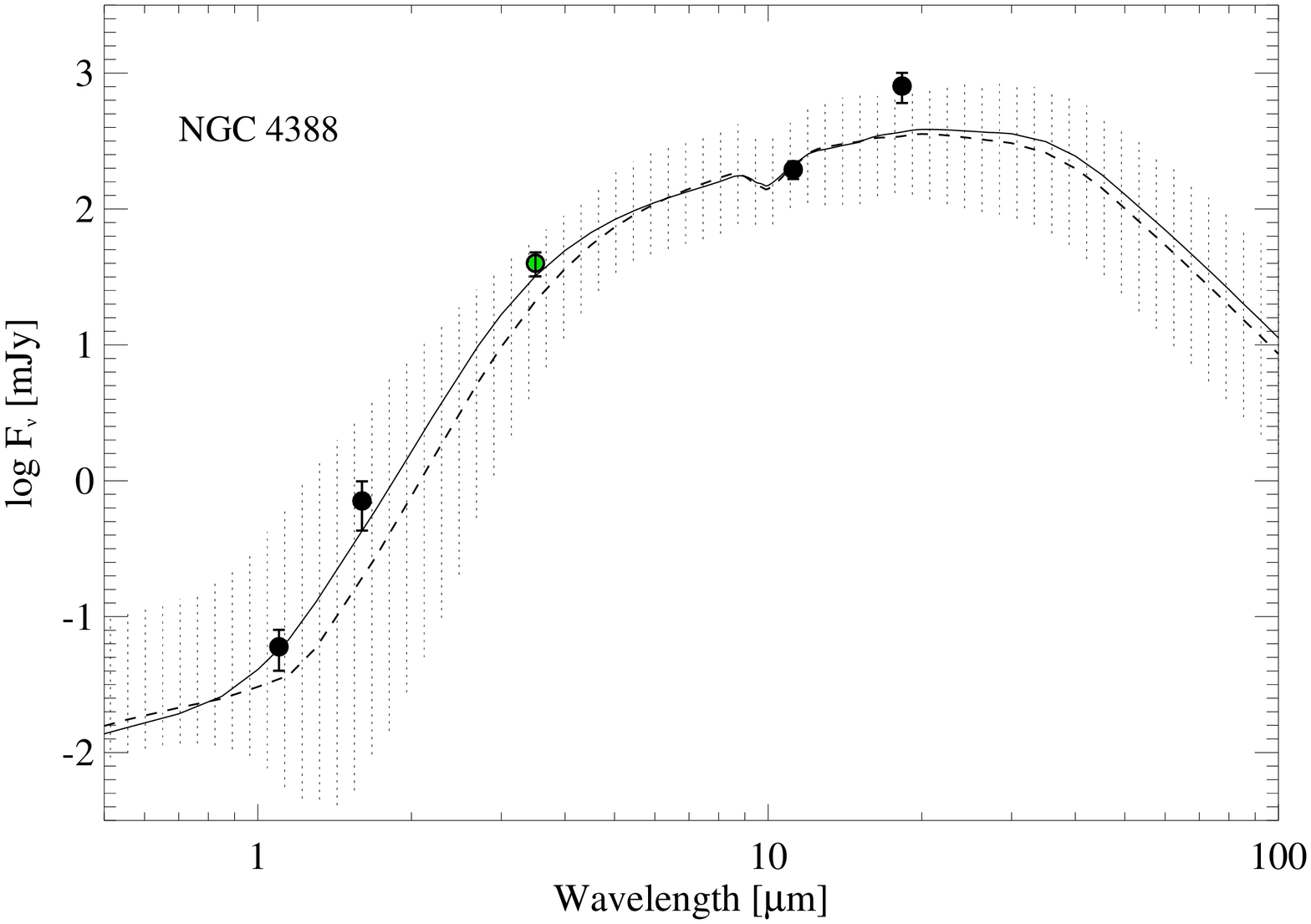}
\includegraphics[width=5.7cm,angle=90]{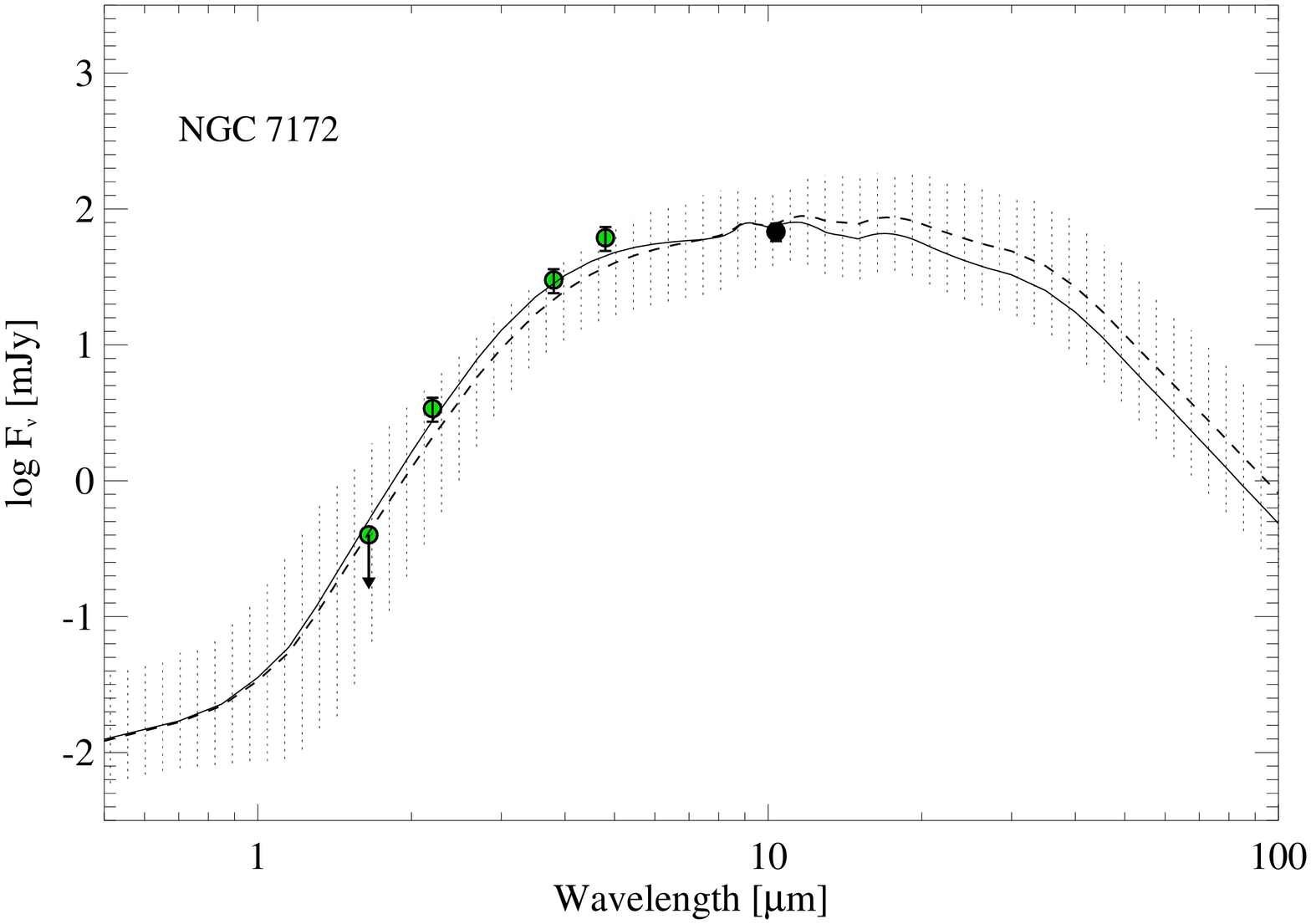}
\includegraphics[width=5.7cm,angle=90]{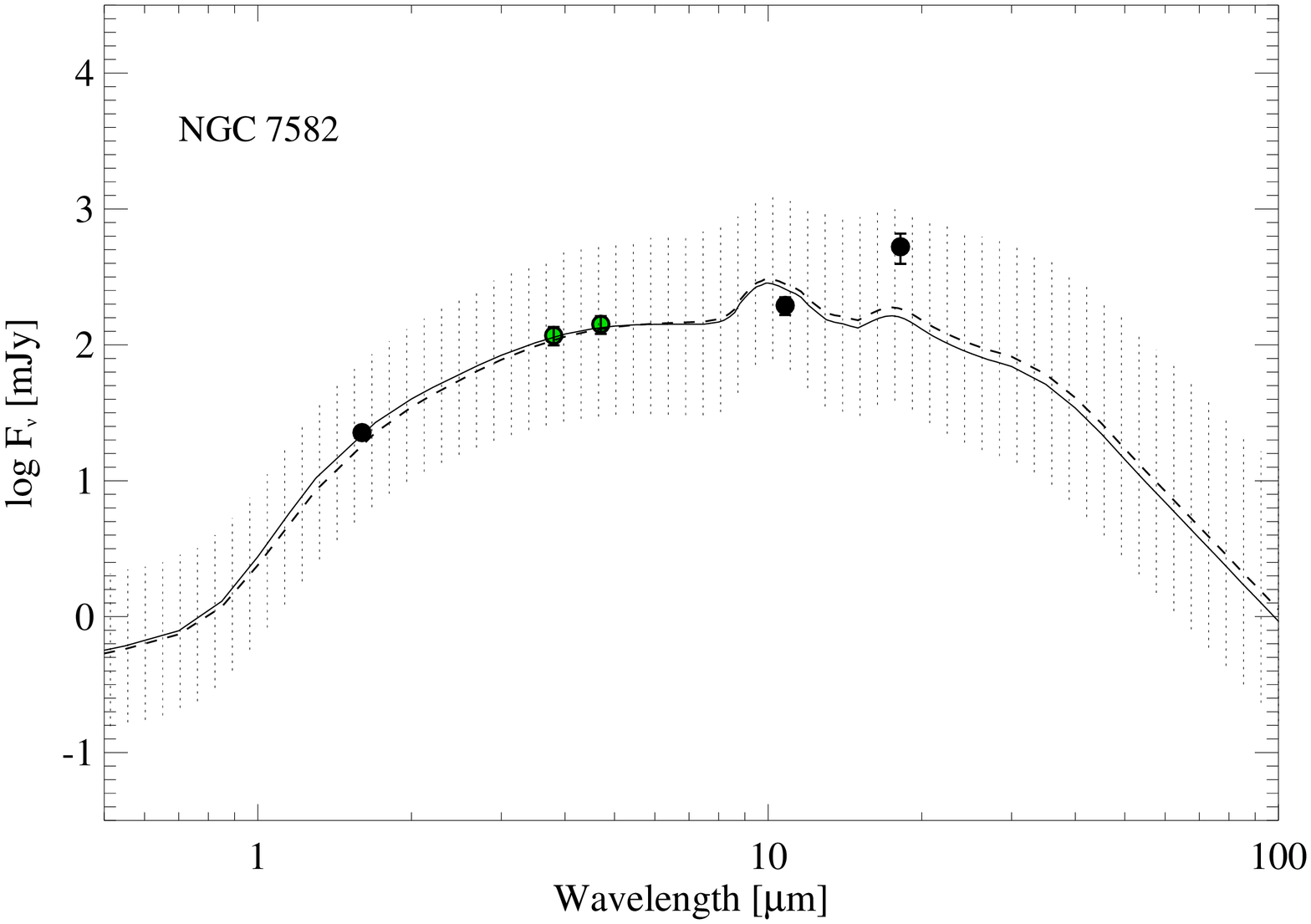}\par}
\caption{\footnotesize{High spatial resolution SEDs of the Sy2 galaxies. 
Solid lines correspond to the model described by the combination
of parameters that maximizes their probability distributions. Dashed lines represent the model computed with
the median value of the probability distribution of each parameter. Shaded regions indicate the range of models compatible
with the 68\% confidence interval for each parameter around the median. Green dots correspond to seeing-limited near-IR data.}
\label{sy2_1}}
\end{figure*}

\newpage

\appendix
\section{Individual Objects}
\label{indiv:sy2}
Below, we comment on the individual fits of the galaxies in our
sample, including Type-2 Seyferts and intermediate types.  
Unless the opposite is indicated, all the reported values from the fits correspond to
or are calculated from the median values. 

\paragraph{Centaurus A.}
Centaurus A (NGC 5128) is a nearby Sy2 nucleus, with its core heavily
obscured by a dust lane. This makes it visible only at wavelengths
longwards $\sim 1$ \micron~\citep{Schreier98,Marconi00}.  Indeed, 
\citet{Hough87}, \citet{Packham96}, and \citet{Marconi00} determined a value for the extinction caused by the dust lane
of A$_V\sim$ 7-8 mag. However, the total obscuration on our LOS to the core 
was estimated by \citet{Meisenheimer07} to be A$_V\sim$ 14 mag, including 
the torus material. Due to the existence of the dust lane, that is probably 
affecting our nuclear fluxes (specially the near-IR ones), we exceptionally 
have taken this foreground extinction into account for the fit of Centaurus A with the 
clumpy models (Figure \ref{sy2_1}), by fixing A$_V$ = 8 mag and applying the \citet{Calzetti00} law to the fitted models.

The clumpy models reproduce the
observed photometry very accurately, but not
the broad absorption silicate feature detected in mid-IR spectroscopy and
interferometry \citep{Siebenmorgen04,Meisenheimer07}.  This mismatch
in the fitted model could be due to the synchrotron radiation contribution 
to the infrared fluxes claimed by \citet{Bailey86}, \citet{Turner92}, \citet{Chiaberge01}, and \citet{Meisenheimer07}.
Indeed, \citet{Meisenheimer07} find from mid-IR interferometric observations that
the contribution of synchrotron emission to the nuclear emission of Centaurus A is
of $\sim$80\% at $\sim$8 $\micron$ and $\sim$60\% at $\sim$13 $\micron$. However, 
the results of \citet{Radomski08}, including the lack of variability at $\sim$10 $\micron$ (see \citealt{Radomski08} and references therein)
and zero polarization at $\sim$1 mm (although low millimeter polarization may be associated with low-luminosity sources; \citealt{Packham96})
are inconsistent with a synchrotron source as that suggested by \citet{Meisenheimer07}.
A more detailed analysis of the SED of this galaxy, e.g., including an optically thin power-law in the fits
to account for the synchrotron radiation can be done. However, this kind of detailed analysis of an individual 
galaxy is beyond the scope of this work.

The probability
distributions derived from the fit of the Centaurus A SED with the clumpy models clearly 
constrain the number of clouds $N_0$,
the optical depth of each cloud $\tau_{V}$, and the width 
of the angular distribution $\sigma$. $N_0$
resembles a Gaussian distribution centered at the median
value $N_0 = 2\pm$1.  The $\tau_{V}$ histogram shows an
asymmetric shape, with a tail towards low values and of median
$\tau_{V}$=176 at a 68\% confidence
level. High-$\sigma$ values are more
probable than others (median value of $58\degr\pm_{15}^{10}$), and the
same for the inclination angle of the torus ($i > 85\degr$). On the
contrary, low-values of $q$ are preferred ($q < 0.2$).  The radial
thickness of the torus $Y$ has been introduced to the code as a
Gaussian prior centered in 15 with a width of 2.5, as explained in
Section \ref{bayesclumpy}. Indeed, the preliminary fit of Centaurus A
without restrictions in any of the model parameters was the only one
that constrained the radial thickness of the torus, with a median
value of its probability distribution of $Y \sim$ 15. The calculated
median value of the optical extinction produced by the torus is $A_{V}^{LOS}$ =
300$\pm_{80}^{100}$ mag for this galaxy.

\paragraph{Circinus.}
The Circinus galaxy is a nearby Sy2 with a prominent cone-shaped
region revealed in the [O III] images \citep{Wilson00}. 
Both the accuracy of the 8 high spatial resolution flux measurements and the proximity of the galaxy
make the resulting fit excellent.
Water vapor
megamasers \citep{Greenhill03} constrain the viewing angle to $i \sim
90 \degr$.  
Thus, in addition to the Gaussian prior for the $Y$
parameter, we introduce a prior on $i$,  a Gaussian centered in
85\degr, with a width of 2\degr\footnote{We choose this value of $i$
to allow for a narrow Gaussian distribution within the considered
interval [0\degr,90\degr].}. 
The resulting fit constrains most model parameters.  The optical depth
per cloud results in a narrow Gaussian centered 
at the median value
$\tau_{V}$ = 30$\pm$2. The number of clouds results in a
Gaussian-like distribution (median
value of $N_0 = 10\pm$2), and the width of the angular
distribution has a median value of $\sigma = 61\degr\pm$8. Finally,
we establish a lower limit of $q < 2.3$ at a 68\% confidence level.
The optical obscuration produced by the torus would be  
$A_{V}^{LOS}$ = 320$\pm_{55}^{80}$ mag.
\citet{Roche06} found the core of Circinus compact and obscured by a substantial column of silicate dust 
from T-ReCS/Gemini-South spectroscopy. They obtained a value for the silicate absorption depth 
of $\tau_{10\micron}$ = 1.6. Our fitted models reproduce the silicate feature in absorption (see Table \ref{silicate}), 
in qualitative agreement with the observations, but shallower ($\tau_{10\micron}^{app}$ = 0.86) than that reported by \citet{Roche06}.
We find that the absorption column density inferred from X-ray measurements ($N_H^{X-rays} = 4\times10^{24}~\mathrm{cm}^{-2}$) 
is larger than that responsible for the mid-IR absorption ($N_H^{LOS} \sim 6\times10^{23}~\mathrm{cm}^{-2}$). 
This is consistent with the X-ray obscuring 
region being closer to the central engine than the mid-IR emitting dust.

\paragraph{IC 5063.}
This Sy2 nucleus is hosted by a merger remnant classified as an elliptical galaxy \citep{Colina91}. 
An obscuring dust lane  partially covers the eastern side of the galaxy. A peculiarity of this source is its 
strong radio luminosity, which is two orders of magnitude larger than the typical values for Seyfert 
galaxies \citep{Colina91}. 
From an  N-band adquisition image of IC 5063 obtained with T-ReCS/Gemini-South, \citet{Young07}
tentatively found the nucleus of this galaxy slightly resolved. However, from our T-ReCS Si2 and Qa band imaging 
the nucleus appears unresolved, and consequently, we subtracted the corresponding PSF at the 100\% level to derive the nuclear fluxes. 
From the fit with the clumpy models, we establish lower limits for three of the parameters
($\sigma > 57\degr$, $N_0 > 11$, and $i > 65\degr$) and an upper limit for the index of the radial density profile 
($q < 1.5$). 
The only Gaussian-like distribution resulting from the fit corresponds to $\tau_{V}$ (median value of 
70$\pm_{23}^{31}$). $A_{V}^{LOS}$ is calculated to be 780$\pm_{345}^{485}$ mag.
The galaxy presents one of the highest $L_{bol}^{AGN}$ of the sample (2.4x10$^{44}$ erg~s$^{-1}$, see Table \ref{lum}), 
and also a high reprocessing efficiency (87\%), consistent with the large values of $N_0$ and $\sigma$ we find.
X-ray observations with the {\it ASCA} satellite allow to infer an intrinsic 2-10 keV luminosity of $\sim$10$^{43}$ erg~s$^{-1}$,
that is within the range expected for Seyfert 1 galaxies \citep{Turner97}.
The silicate feature appears in absorption in the fitted models, in qualitative agreement with mid-IR spectroscopic observations 
obtained with T-ReCS/Gemini-South \citep{Young07}. However, the absorption band appears deeper in our models ($\tau_{10\micron}^{app}=0.94$) than
in the mid-IR spectrum ($\tau_{10\micron}=0.33$).

\paragraph{Mrk 573.}
Mrk 573 is optically classified as a classical Sy2 nucleus
\citep{Tsvetanov92}, and \citet{Ramos08} recently re-classify it as an
obscured NLSy1 galaxy, based on near-IR spectroscopy and X-ray
archival data.  
The X-ray observations of Mrk 573 do not straightforwardly reveal the
intrinsic luminosity of this AGN because it suffers from Compton thick
obscuration \citep{Guainazzi05}. The strong soft X-ray emission is likely
photoionized line emission.  It is not the AGN continuum in any case,
so previous simplified fits of power law models over a broad energy
range \citep[e.g.,][]{Guainazzi05,Shu07} fail to accurately measure the AGN.
We determine the intrinsic AGN luminosity from the XMM-Newton
observation of Mrk 573 with the pn detector, which offers a high
signal/noise ratio.  We use the reduced data and associated
background and calibration files from XAssist pipeline processing
\citep{Ptak03}.  We fit only the data at energies greater than 3 keV, to avoid
confusion from the separate soft emission sources.  The observed continuum
is flat and the Fe K$\alpha$ line equivalent width (EW) is large, which
are characteristic of a purely reflected (not direct) AGN continuum.
We fix the  photon index $\Gamma = 0$ and find
EW $= 2.0\pm 0.7$ keV.  The results are similar if a typical
AGN continuum slope is adopted ($\Gamma = 1.9$).
We follow the technique of \citet{Levenson06}, using the Fe line luminosity
and EW and to determine the intrinsic AGN luminosity, finding
$L_{X}^{AGN} = 2.2 \times 10^{43} \mathrm{\, erg\, s^{-1}}$ 
in the 2--10 keV bandpass.  We note that the fitted hard continuum
directly corresponds to an intrinsic luminosity that is a factor of
100 smaller. 

The IR SED shape of Mrk 573 is practically
the same as those of the rest of Sy2 galaxies in the sample, so we
consider it equivalent to the other Sy2.  This galaxy presents a
strong symmetric double radio lobe \citep{Nagar99}, suggesting that it
contains an edge-on torus; the radio lobes would be asymmetric if the
jets were out of the plane of the sky.  The SED fit reproduces all 7
data points and also constrains the majority of the parameters.   As for the case of
Circinus, we have introduced the inclination angle of the torus as a
Gaussian prior, centered in 85\degr~with a width of 2\degr.
The probability distribution of $N_0$ present a Gaussian
shape, with median value $N_0 = 6\pm_{2}^{5}$ (there is a tail towards 
large values that enlarges the error). 
The $\sigma$ and $q$ distributions allow to establish upper limits of 
$\sigma  < 37\degr$ and $q < 1.4$. 
The optical depth of each cloud shows a Gaussian 
distribution with median value of $\tau_{V}$ = 30$\pm_{8}^{10}$. 
The optical extinction produced by this clumpy torus is $A_{V}^{LOS}$ = 185$\pm_{75}^{120}$ mag. 
Mrk~573 has the highest $L_{bol}^{AGN}$ of the Sy2 subset
($4.3 \times 10^{44}$ erg~s$^{-1}$; Table \ref{lum}), but due to 
the low values of $N_0$ and $\sigma$ derived from its fit, its
reprocessing efficiency is low ($\sim$15\%).
The 10 \micron~silicate feature appears in shallow absorption in the
median fitted model ($\tau_{10\micron}^{app}$ = 0.44; see Table \ref{silicate}), producing an estimated
value of the optical extinction $A_V^{app}$ = 11 mag. The hydrogen column density obtained from 
the modelling along the LOS ($N_H^{LOS} \sim 3 \times 10^{23} cm^{-2}$) is much lower than 
the one that obscures the X-ray emission ($N_H^{X-rays} > 1 \times 10^{24} cm^{-2}$).

\paragraph{NGC 1386.}
NGC 1386 is one of the nearest Sy2 galaxies and therefore extensively
studied.  \citet{Mauder92} resolved its NLR on a 0.3\arcsec~scale
($\sim 15$ pc) using speckle interferometry, detecting individual NLR
clouds and claiming a clumpy structure. The detection
of a water vapor megamaser in NGC 1386 \citep{Braatz97}
constrains the inclination angle of the torus to be $i \sim 90\degr$,
as in the case of Circinus and Mrk~573.  The fit with the clumpy models reproduces
the observed SED; we find 
$N_0 = 11\pm3$, $\sigma = 50\degr\pm_{19}^{16}$, 
$\tau_{V} = 95\pm_{51}^{66}$, and $q$ = 1.5$\pm_{1.0}^{0.9}$ at a 68\%
confidence level.  The optical extinction produced by the clumpy torus is $A_V^{LOS} <
1460$ mag.
The fitted models show a shallow silicate absorption feature ($\tau_{10\micron}^{app}$ = 0.41) 
from which we derived an optical extinction of $A_V^{app}=11$ mag.
As for most of the galaxies in the sample, we find $N_H^{LOS} < N_H^{X-rays}$, indicating 
that the X-ray obscuring 
region must be closer to the central engine than the torus material.

\paragraph{NGC 1808.}
The Sy2 galaxy NGC 1808 is undergoing an intense episode of star formation in its 
central 750 pc \citep{Tacconi96,Tacconi05,Galliano08}. 
The near-IR fluxes were obtained using ISAAC/VLT, and we measured the mid-IR flux from
10.7 and 18.2 \micron~OSCIR/CTIO images. 
Consequently,  NGC 1808 has relatively
low spatial resolution for our sample (the
angular resolution of the ISAAC images is $\sim0.6\arcsec$ and that of the OSCIR image at CTIO is $\sim 0.9-1\arcsec$). 
The near-IR SED shape is flat and peculiar, and 
star formation may contaminate these measurements 
(according to the argument of \S\ref{stellar_contamination}), making the fit not very reliable.
Low values of $\sigma$ and $i$ are preferred 
($< 35\degr$ and $37\degr\pm_{23}^{19}$, respectively), contrary to the general trend
for the Sy2 galaxies in our sample. The median value of the number of clouds is $N_0 = 8\pm_{4}^{5}$, 
the optical depth per cloud  $\tau_{V} > 122$, and $q$ = 1.2$\pm$0.9. 
The calculated optical extinction produced by the torus is $A_{V}^{LOS} < 140$ mag.
From the bolometric luminosites derived from the fit, we derive the reprocessing 
efficiency of the modelled torus, that results on $\sim$45\%. 
The 10 \micron~silicate feature appears in weak emission in the fitted spectrum ($\tau_{10\micron}^{app}=-0.21$). 
Summarizing, due to the intense stellar formation that is taking place in the nuclear region of this galaxy, 
and to the lower spatial resolution near- and mid-IR measurements  that we have for it (all from 3 or 4 m telescopes), 
the infrared SED of NGC 1808 is very likely contaminated with starlight. 
This is probably producing its peculiar shape and the different trends in the fitted parameters, in comparison with the rest of Sy2 in the sample.
For these reasons, we consider the NGC 1808 SED fit unreliable.

\paragraph{NGC 3081.}
NGC 3081 is a Sy2 galaxy with intense stellar formation located in a
series of nested ringlike features \citep{Buta04}.  
As in the case of
NGC 1386, the SED of NGC 3081 is formed with only three data
points, and the model results are similar, with
the probability distributions of the parameters
presenting the same shapes.  
The $N_0$ and $\tau_{V}$ histograms have median values of $N_0$ = 10$\pm$3 and
$\tau_{V}$ = 47$\pm_{23}^{42}$, respectively. 
We determine lower limits of $\sigma > 52\degr$, $q > 1.2$, and $i > 42\degr$.
$A_{V}^{LOS}$ is estimated to be lower
than 450 mag.  According to the calculated luminosities for this
galaxy, the fraction of reprocessed radiation
(L$_{bol}^{tor}/L_{bol}^{AGN}$) is among highest in the sample
($\sim$90\%).
The 10 \micron~silicate feature is reproduced in absorption by the models ($\tau_10\micron^{app}=0.91$), 
from which we inferred an optical obscuration $A_V^{app}\sim23$ mag.

\paragraph{NGC 3281.}
This Sy2 galaxy suffers large extinction ($A_{V} = 22\pm 11$ mag to
the IR-emitting and $N_H = 2\times 10^{24}\mathrm{\, cm^{-2}}$ to the
X-ray emitting regions; \citealt{Simpson98,Vignali02}).
The clumpy models  fit the data well, except for the 18.3
\micron~point, which the models underestimate slightly.
The probability distributions resulting from the fit show
that large values of $\sigma$ and $i$ are more probable than others ($> 68\degr$
and $> 49\degr$, respectively)
and establish an upper limit for the optical depth
per cloud of $\tau_{V} < 12$ at a 68\% confidence level. 
The number of clumps $N_0$ presents a Gaussian distribution 
with a tail towards large
values, with median value $N_0 = 6\pm_{2}^{4}$, as the $q$ histogram (median value of $q$ = 1.3$\pm_{0.8}^{1.1}$). 
The optical extinction produced by the clumpy torus corresponds to a Gaussian
distribution of median value $A_{V}^{LOS} = 55 \pm_{15}^{18}$ mag.
The models predict the 10 \micron~silicate feature in shallow absorption ($\tau_10\micron^{app}=0.21$), 
resulting in an apparent optical obscuration of $A_V^{app}=5$ mag.
$N_H^{LOS} < N_H^{X-rays}$, indicating a closer X-ray obscuring 
region to the central engine than the torus material.

\paragraph{NGC 4388.}
This highly inclined disk galaxy is spectroscopically classified as a
Sy2 nucleus \citep{Phillips82}.  However, \citet{Shields96} report the
detection of weak, broad H$\alpha$ emission.  The ground-based L-band
point introduces a bump in the SED around 3-4 \micron, which likely
represents starlight contamination (\S\ref{stellar_contamination}).
The galaxy appear very extended in the L band \citep{Alonso03}, 
making difficult to isolate of the nuclear emission.  
Thus, the fit correctly reproduces the near-IR and the N-band points but
severely underestimates the Q-band measurement. 
The probability distributions establish lower limits of $\sigma > 53\degr$ 
and $i > 38\degr$, and upper limits of $q < 1.6$ and $\tau_{V} < 14$ at a 68\% confidence level.
The median value of the number of clouds is $N_0$ = 9$\pm_{3}^{4}$.
Along the line of sight, $A_{V} < 80$ mag.
The silicate feature appears in shallow absorption in the fitted models ($\tau_10\micron^{app}=0.51$), allowing
to estimate an optical obscuration of $A_V^{app}=13$ mag.
As for most of the galaxies in the sample, we found $N_H^{LOS} < N_H^{X-rays}$.

\paragraph{NGC 7172.}
NGC 7172 is a Type-2 Seyfert nucleus in a nearly edge-on spiral
galaxy. A prominent dust lane in the east-western direction crosses
the galaxy.   Similar to the case of NGC
4388, we observe a near-IR bump of emission  in the SED of NGC 7172,
which stellar contamination of the seeing-limited fluxes likely produces
(\S\ref{stellar_contamination}). 
The $N_0$ histogram shows a Gaussian shape, with median value $N_0 =
5\pm_{1}^{3}$. High values of
$i$, $\sigma$, and $q$ are more probable ($i > 45\degr$, $\sigma > 54\degr$, and $q > 1.7$, respectively),
while low values of $\tau_{V}$ are preferred ($\tau_{V} < 12$).
The torus produces optical extinction $A_{V}^{LOS}$ = 50$\pm$20 mag.
The reprocessing efficiency of this galaxy is high ($\sim$80\%), according to 
the derived values of the AGN and torus bolometric luminosities. 
T-ReCS mid-IR spectroscopy shows a substantial silicate absorption feature \citep{Roche07} that 
is not reproduced by the fitted models, that predict it in very extremely  emission (and self-absorbed). This could be due to 
the above mentioned stellar contamination of the near-IR fluxes, that in addition to the lack of Q-band data point for this galaxy, make it 
difficult the correct reproduction of the silicate feature in absorption.

\paragraph{NGC 7582.}
NGC 7582 is a Sy2 galaxy that harbors a nuclear starburst surrounding the active nucleus 
\citep{Cid01,Sosa01,Wold06,Bianchi07} and presents a nuclear outflow \citep{Morris85}. 
The fit with the clumpy models shown in Figure \ref{sy2_1} does not reproduce the 18.2
\micron~point, and the resulting probability distributions of the
parameters are very different compared with the rest of 
Sy2 galaxies discussed here (except for NGC 1808). 
Low values of $\sigma$ are more probable ($< 29\degr$), 
as are extremely small numbers of clouds ($N_0 <$ 2). 
High values of $q$ ($> 2.5$) and low values of the $\tau_V$ ($< 27$) are preferred. The $i$
probability distribution has a Gaussian shape which median value is $i = 41\degr\pm_{28}^{19}$. 
The optical obscuration produced by the torus results in an unrealistic value of $A_{V}^{LOS} < 6$ mag.
The reprocessing efficiency fraction of this galaxy is among the lowest in the sample ($\sim$10\%),
a consequence of the low values of $N_0$ and $\sigma$.
The 10 \micron~silicate feature appears in emission in the fitted models, 
in contradiction with TIMMI2 mid-IR observations \citep{Siebenmorgen04} that
show a strong absorption band characterized by $\tau_{10\micron}=1.1$.
All these results are probably due to stellar contamination of our mid-IR data (obtained at the 4 m CTIO).
The intense circumnuclear star formation present in this galaxy makes it more difficult to isolate the torus emission from that of 
the host galaxy. Consequently, ss for the case of NGC 1808, we consider the model fit of NGC 7582 unreliable.

\paragraph{NGC 1365.}
This Sy1.8 nucleus is hosted in a galaxy with intense nuclear star
formation. 
The X-ray emission of NGC 1365 is also remarkable, showing
 the most dramatic spectral changes observed in an AGN
\citep{Risaliti05,Risaliti07}. 
The rapid X-ray variability is attributable to variations in the 
LOS density, which  is compatible with a 
low number of clouds along the LOS. 
The  clumpy models fit the SED of NGC 1365 well, but the model parameters
are not well-constrained.  The number of clouds has a median value of
$7\pm_{4}^{5}$ (which is consistent with the rapid X-ray variability),
$\sigma = 35\degr\pm_{13}^{20}$, and $\tau_{V} = 111\pm_{56}^{58}$.
Low values of $q$ and $i$ ($q < 1.7$ and $i < 42\degr$) are more probable. 
The optical obscuration due to the torus is $A_{V}^{LOS} < 170$ mag.
The silicate feature is completely absent in the fitted models. 

\paragraph{NGC 2992.}
This galaxy is optically classified as a Sy1.9 nucleus, based on early
published spectra \citep{Ward80}, which showed a weak
broad H$\alpha$ component, but no broad H$\beta$. In a later
observation \citep{Allen99}, the broad component of H$\alpha$
dissapeared, leading to a new classification of the galaxy as a
Sy2. \citet{Gilli00} reported that the spectrum regained the broad
wings of H$\alpha$ coinciding with a period of intense X-ray activity,
contrary to the low-stage that was taking place during Allen's 
observations. \citet{Trippe08} report
re-classification of NGC 2992 as a Sy2 nucleus again, as of 2006.
The near-IR
ground-based data were taken in 1998 April, coinciding with the epoch
of Gilli's observations, and thus with the classification
as intermediate-type Seyfert. On the other hand, our mid-IR
measurements from Michelle were obtained in 2006 May, coinciding with
the Sy2 state. Thus, our IR SED includes fluxes taken during diferent periods
of activity of NGC 2992. This can explain its similarity with a
typical Sy2 SED.  
The clumpy models that include only torus emission (and no direct
AGN component) fit the observed data well.
The probability distributions are Gaussians for $\sigma$, $N_0$, and $\tau_{V}$, 
with median values $\sigma = 45\degr\pm_{21}^{18}$,
$N_0 = 7\pm_{3}^{5}$, and $\tau_V$ = 36$\pm_{11}^{14}$. High values of $i$ ($> 53\degr$) and low
values of $q$ ($< 1.0$) are preferred. The
calculated optical extinction due to the torus is $A_{V} < 160$ mag.
NGC 2992 is the only intermediate-type Seyfert with the 10 \micron~silicate feature
in absorption in the fitted models ($\tau_{10\micron}^{app}=0.74$). The derived apparent optical extinction from the measurement 
of the band is $A_V^{app}\sim19$ mag. The  hydrogen column density along the LOS results in $N_H^{LOS} = 2.5\times10^{23}~cm^{-2}$.
This value results unrealistic when compared with that inferred from X-ray measurements ($N_H^{X-rays} = 8\times10^{21}~cm^{-2}$).
This could be due to the intrinsic X-ray variability observed in NGC 2992 \citep{Trippe08}.

\paragraph{NGC 5506.}
The nucleus of this galaxy is classified as a Sy1.9 based on the
detection of broad wings of the Pa$\beta$ profile in the near-IR
\citep{Blanco90}.  However, more recently, \citet{Nagar02} presented
evidence that NGC 5506 is an obscured Narrow Line Sy1, as a result of
the finding of the permitted O I$\lambda$1.129 \micron~line, together
with a broad pedestal of Pa$\beta$ and  rapid X-ray variability. 
The clumpy models reproduce
the near-IR and the N-band data points well, whilst slightly underestimating the 
Q-band flux.
The discrepancy between the spectral observations and the model fit along with
the poor mid-IR fit together suggest either that additional emission
contaminates the near-IR measurements or that the models cannot produce this observed
SED shape  (\S\ref{stellar_contamination}). Based on
the detection of a water vapor megamaser \citep{Raluy98}, we have
introduced the inclination angle to the code as a Gaussian-prior
centered in 85\degr~with a width of 2\degr. 
In the resulting 
probability distributions, we find $\sigma=25\degr\pm_{7}^{9}$, $q$ = 2.5$\pm_{0.4}^{0.3}$, 
and the average number of clouds along an equatorial ray to be less than 2. The
optical depth per cloud $\tau_{V} < 68$, and the calculated
torus optical extinction is $A_{V} < 90$ mag.
NGC 5506 has the highest $L_{bol}^{AGN}$ of the sample
($7.3\times 10^{44}$ erg~s$^{-1}$; Table \ref{lum}). 
As a consequence of
the low values of $N_0$ and $\sigma$ derived from its fit, the
reprocessing efficiency is low (7\%).
The galaxy nucleus is very compact in the mid-IR, with an apparent optical depth of $\tau_{10 \micron}^{app} \sim 1.4$
\citep{Roche91,Roche07,Siebenmorgen04b}, although
NGC 5506 also shows variations in its silicate absorption depth on parsec scales \citep{Roche07}. 
Unfortunately, the fitted models reproduce the silicate feature in emission ($\tau_{10\micron}^{app}=-0.21$), contrary to the spectroscopic observations.
This galaxy has comparable X-ray and mid-IR the absorbing columns, 
indicating that the dust-free absorption in this galaxy is lower than for the Sy2 reported here.

\paragraph{NGC 3227.}
\label{ngc3227}
This active nucleus is usually classified as a Sy1.5. XMM-Newton
observations show short term hard X-ray variability likely related to
variation in the intrinsic continuum emission \citep{Gondoin03}.  The
clumpy models
reproduce the SED
well, including the direct though extinguished AGN emission (using the \citealt{Calzetti00} law), as in the case
of NGC 4151.  
The resulting histograms show broad Gaussian distributions of
the $\sigma$, $N_0$, and $\tau_{V}$ parameters, with median values of
$33\degr\pm_{12}^{20}$, $6\pm_{4}^{5}$, and $116\pm_{55}^{56}$,
respectively.  Low values of $i$ ($< 48\degr$)
are more  probable than others, and the $q$ parameter results unconstrained.
The optical depth calculated using the fitted model parameters is $A_V^{LOS} < 210$ mag. 
On the other hand, the extinction that obscures the direct AGN emission is
$A_{V} = 2.7\pm_{1.5}^{2.0}$ mag (non related with the torus). 
This value is intermediate among those reported in the literature
of $A_{V} = 1.2$--1.7 mag \citep{Cohen83,Gonzalez97} and $A_{V} =
4.5$--4.9 mag \citep{Mundell95}.  
The 10 \micron~silicate feature is practically absent from the fitted
models ($\tau_{10\micron}^{app}=-0.1$).  

\paragraph{NGC 4151.}
\label{ngc4151}
The nucleus of this well-known Sy1.5 galaxy shows flux variability in
a wide wavelength range, with timescales ranging from a few hours in
the hard X-rays \citep{Yaqoob93} to several months in the IR
\citep{Oknyanskij99}. \citet{Shapovalova08} reported changes in the
spectral type of the nucleus from a Sy1.5 to a Sy1.8, coinciding with
maximum and minimum activity states, respectively.  The SED of NGC
4151 is one of the flattest in our sample, with 
the near-IR excess likely due to direct AGN
emission, which is absent in Sy2 galaxies.
We include this power-law contribution and allow for some extinction
to account for the observed reddening of the 
active nucleus. The fit correctly reproduces the observed SED,
especially in the near-IR.  The probability distributions
establish a median value of $q$ = 1.7$\pm_{0.9}^{0.8}$, $i$ = 41\degr$\pm_{28}^{23}$, 
and $\tau_V$ = 120$\pm_{48}^{55}$.
Low values for the number of clouds along equatorial rays 
($N_0 < 3$) and for the width of the angular distribution ($\sigma < 32\degr$)
are more probable. 
The optical extinction derived from the fitted model parameters is $A_{V}^{LOS} < 65$ mag. 
On the other hand, 
the optical obscuration of the direct AGN emission (not produced by
the torus) is another free parameter of the fit, and we find  $A_{V} =
2.5\pm_{1.6}^{1.4}$ mag.
The 10 \micron~silicate feature appears in weak emission ($\tau_{10\micron}^{app}=-0.31$) in the fitted models, in 
qualitative agreement with Spitzer observations of this galaxy \citep{Weedman05}. 

\acknowledgments

C.R.A. and J.R.E. acknowledge the Spanish Ministry of Education and Science through projects 
PN AYA2007-67965-C03-01 and Consolider-Ingenio 2010 Program grant CSD2006-00070: First Science with the GTC 
(http://www.iac.es/consolider-ingenio-gtc/).
N.A.L. acknowledges work supported by the NSF under Grant
0237291 and thanks the University of Florida Astronomy Department and
the IAC for their hospitality during this project. 
A.A.H acknowledges support from the Spanish Plan Nacional del Espacio under grant
ESP2007-65475-C02-01. 
A.A.R. acknowledges the Spanish Ministry of Education and Science through project AYA2007-63881.

This work is based on observations obtained at the Gemini Observatory, which is operated by the
Association of Universities for Research in Astronomy, Inc., under a cooperative agreement
with the NSF on behalf of the Gemini partnership: the National Science Foundation (United
States), the Science and Technology Facilities Council (United Kingdom), the
National Research Council (Canada), CONICYT (Chile), the Australian Research Council
(Australia), Minist\'{e}rio da Ci\^{e}ncia e Tecnologia (Brazil), and Ministerio de Ciencia, Tecnolog\'{i}a e Innovaci\'{o}n Productiva (Argentina). 
The Gemini programs under which the data were obtained are GS-2003B-DD-4, GS-2004A-DD-4, GS-2005B-DD, GS-2005A-Q-6, GN-2006A-Q-11, and GS-2006A-Q-30.

This work is based on observations made with the NASA/ESA Hubble Space Telescope, and obtained 
from the Hubble Legacy Archive, which is a collaboration between the Space Telescope Science Institute 
(STScI/NASA), the Space Telescope European Coordinating Facility (ST-ECF/ESA) and the Canadian Astronomy 
Data Centre (CADC/NRC/CSA).

This research has made use of the NASA/IPAC Extragalactic Database (NED) which is 
operated by the Jet Propulsion Laboratory, California Institute of Technology, under 
contract with the National Aeronautics and Space Administration.

This research has made use of data obtained from the High Energy
Astrophysics Science Archive Research Center (HEASARC), provided by
NASA's Goddard Space Flight Center. 

The authors acknowledge Ana M. P\'{e}rez Garc\'\i a, Rachel Mason, Itziar Aretxaga, and Tanio D\'\i az Santos for their valuable help.

\end{document}